\documentclass[3p,number]{elsarticle}

\usepackage{amsmath}
\usepackage{amssymb}
\usepackage{amsthm}
\usepackage{color}
\usepackage{subfigure}
\usepackage{graphicx}

\def\p{\partial}    %Partial derivative
\def\i{{\textrm i}} %Unit imaginary

\def\eqdef{\stackrel{\mathrm{def}}{=}}
\def\epsilon{\varepsilon}
\def\phi{\varphi}

\def\hlf{\tfrac{1}{2}}
\def\qtr{\tfrac{1}{4}}

\def\bc{\begin{center}}
\def\ec{\end{center}}
\def\bi{\begin{itemize}}
\def\ei{\end{itemize}}
\def\be{\begin{enumerate}}
\def\ee{\end{enumerate}}
\def\beq{\begin{equation}}
\def\eeq{\end{equation}}
\def\ba{\begin{eqnarray}}
\def\ea{\end{eqnarray}}

\newcommand{\braket}[1]{\langle #1 \rangle}

\def\delplus{\Delta^{\!+}}
\def\deltimes{\Delta^{\!\times}}

\def\fix{\mathrm{Fix}}

\def\Z{\mathbb{Z}}

\def\AUTO{{\sc auto}}
\def\mumx{\mu_{\rm mx}}
\def\reflection{\rho}
\def\translation{\tau}
\def\updown{\kappa}

\title{Snaking and isolas of localised states in bistable discrete lattices}
\author[damtp]{Chris Taylor\fnref{fn1}}
\ead{C.R.N.Taylor@damtp.cam.ac.uk}

\author[bath]{Jonathan H.P. Dawes\fnref{fn2}}
\ead{J.H.P.Dawes@bath.ac.uk}

\address[damtp]{Department of Applied Mathematics and Theoretical Physics,
Centre for Mathematical Sciences, University of Cambridge, Wilberforce Road,
Cambridge CB3 0WA, UK}
\address[bath]{Department of Mathematical Sciences, University of Bath,
Claverton Down, Bath BA2 7AY, UK}

\fntext[fn1]{CRNT acknowledges financial support from the EPSRC.}
\fntext[fn2]{JHPD holds a University Research
Fellowship from the Royal Society.}

\begin{document}

\begin{abstract}
We consider localised states in a discrete bistable Allen-Cahn
equation. This model equation combines
bistability and local cell-to-cell coupling in the simplest
possible way. The existence of stable localised states
is made possible by pinning to the underlying lattice; they do not
exist in the equivalent continuum equation.

In particular we address the existence of `isolas': closed curves
of solutions in the bifurcation diagram.
Isolas appear for some non-periodic boundary conditions in one
spatial dimension but seem to appear
generically in two dimensions. We point out how features of the
bifurcation diagram in 1D help to explain
some (unintuitive) features of the bifurcation diagram in 2D.
\end{abstract}


\maketitle

\section{Introduction}

The general term `pattern formation' refers to the spontaneous appearance of
structure in a dissipative dynamical system in a
continuum or spatially discrete medium,
when a spatially homogeneous state undergoes an instability of some kind
\cite{CH93,H06,P06}.
In many physical situations it is appropriate to use mathematical models for
which the spatial dynamics are modelled discretely,
rather than as a continuum. Such 
lattice models arise in many physical applications: in particular we
mention recent work in nonlinear optics \cite{YCS08,YC09,CC09}
and mathematical modelling of processes
in developmental biology \cite{CMML96,OSW00,Webb04}. In many
applications the existence of domain-filling (almost) regular
solutions exhibiting spatial periodicity
is of interest. Mathematical work on periodic patterns in regular
lattices has also been the subject of considerable recent
interest \cite{WG05,ADGW05}.

In addition, there has been a renewed interest in the study
of localised states in
nonlinear dissipative systems in recent years, motivated by a wealth of
experimental and numerical results both in continuum systems
\cite{CRT00,BK06,BK07} and
spatially discrete models \cite{AA05}. Such localised states have been
referred to by a number of different names, including `light bullets'
and `dissipative solitons'.

In continuum systems one recurring mechanism for the
generation of localised states is the existence of
a subcritical Turing (pattern-forming) instability. In the simplest
case, two branches of small-amplitude localised states emerge at
the Turing instability, also in this context referred to as a
Hamiltonian--Hopf bifurcation \cite{IP93}.
Away from the initial instability they
develop into a complicated structure of
fully nonlinear localised states which exist on a pair of
intertwining curves \cite{WC99}; a complete asymptotic description of the
bifurcation structure
involves consideration of exponentially-small effects \cite{KC06,CK09}.
In an infinite domain this `homoclinic snaking'
behaviour continues, and the width of the
localised states increases without bound. In a finite domain the
branches of localised states terminate on the
branches of spatially
periodic patterns which also originate at the initial
Turing instability \cite{BBKM08,D09}.
There is now a pretty complete picture of
the existence and stability
of localised states in canonical 1D model equations, for example the
Swift--Hohenberg model \cite{SH77} with quadratic-cubic
or cubic-quintic nonlinearities, but in contrast, comparatively
little is known about the detailed structure of localised states in
spatially discrete problems.

Motivation for the detailed investigation of discrete models 
comes from a number of sources, but in particular discrete models arise
as natural descriptions of arrays of waveguides
in nonlinear optics \cite{YCS08}. For example,
the discrete bistable nonlinear Schr\"odinger equation
\beq
\i \dot \psi_n + C\Delta \psi_n + s\psi_n |\psi_n|^2 - \psi_n|\psi_n|^4 = 0,
\label{eqn:DNLS}
\eeq
for an array of complex fields $\psi_n(t)$ naturally describes
properties of discrete optical
`dissipative solitons' in a $k$-dimensional square
lattice whose sites are indexed
by the (multiple) suffix $n \in \Z^k$.
The operator $\Delta$ is an
appropriate sum of nearest neighbour (and possibly next-nearest neighbour)
differences depending on the lattice dimension $k$.
Substituting the ansatz $\psi_n=u_n \exp(-\i \mu t)$ where $u_n$ is
a real stationary field,
we see that $u_n$ must satisfy the real difference equation
\beq
\mu u_n + C \Delta u_n + su_n^3 - u_n^5 = 0. \label{eqn:AC}
\eeq
Hence the possible steady states correspond to those of a discrete Allen--Cahn
equation \cite{CMV96}, albeit with a cubic-quintic nonlinearity rather than the
well-studied purely cubic case.

In this Letter we investigate the existence of
localised states in the canonical nonlinear lattice model~(\ref{eqn:AC})
in one and two
dimensions. Such models are in many senses simpler than the corresponding
spatially continuous equations and we discuss the
similarities and differences that arise.
To aid this discussion we add a time derivative term to~(\ref{eqn:AC}) which
makes stability correspond to that found in the spatially-continuous
Swift--Hohenberg model.
In contrast to previous work on this canonical problem by other authors, our
central focus in this paper is on the appearance of isolas. In particular
we show that, firstly, in 1D and on a finite lattice,
isolas are an immediate consequence of changing the boundary conditions
from periodic to Neumann (or, in fact Dirichlet). Secondly we show that
in 2D isolas exist when periodic boundary conditions are used, and
reconnect and then disconnect again as the cell to cell coupling parameter is
increased.
We also comment on (i) the overall bifurcation
structure in 1D, (ii) the similarities
between the 1D bifurcation structure and that of the
continuum 1D Swift--Hohenberg model and
(iii) the connections between 1D lattice structures
and 2D lattice structures, using the former to explain features
of the bifurcation diagram we obtain for the latter.

The structure of the Letter is as follows. In section~\ref{sec:1d} we
introduce the 1D lattice case
and explore the bifurcation structure of localised states. We discuss
the effects of different boundary conditions when the problem is
restricted to a finite lattice.
In section~\ref{sec:2d} we consider a 2D lattice, highlight the
new features that arise, and use the 1D results to
explain aspects of the snaking curve. We summarise
the main results of the paper in section~\ref{sec:conclusions}.

\section{One-dimensional lattices}
\label{sec:1d}

\begin{figure}[!h]
\centering
\subfigure[]{
	\includegraphics[width=0.45\textwidth]{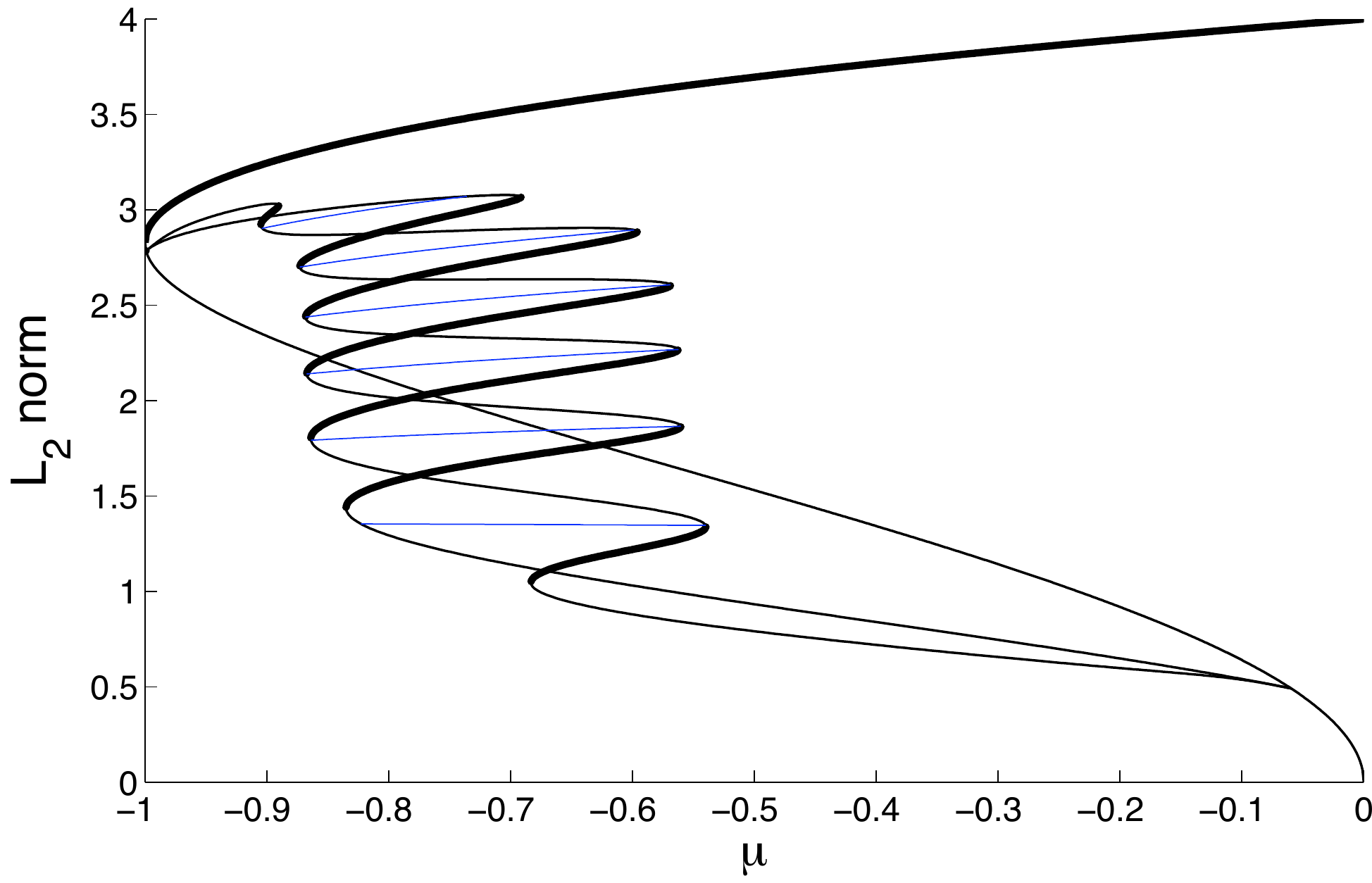}
	}
\subfigure[]{
	\includegraphics[width=0.45\textwidth]{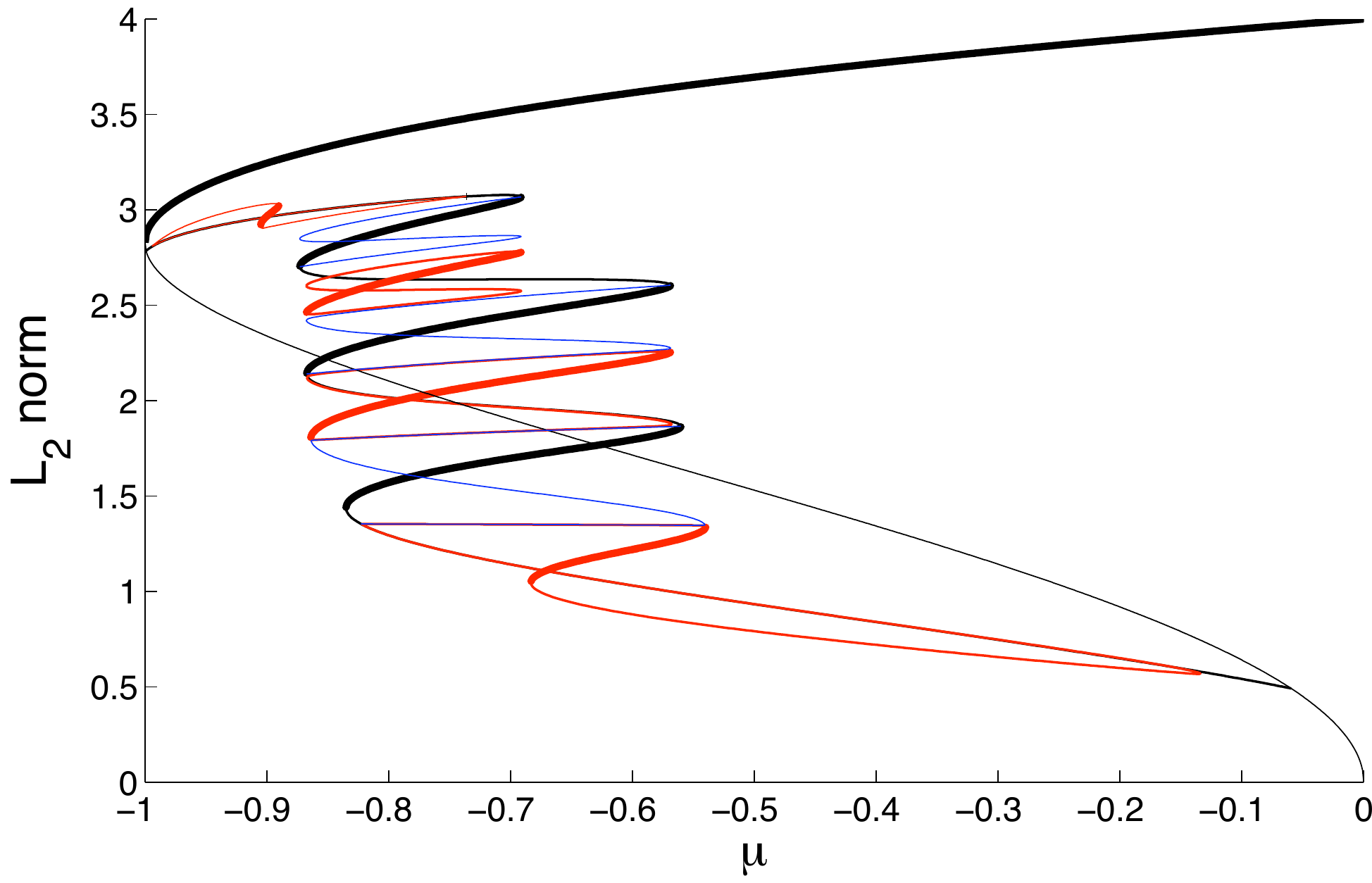}
	}
\caption{(a) Bifurcation diagram in one dimension for an array of $N=8$ cells with PBC and coupling strength $C=0.2$. Thick lines represent stable solutions, thin lines represent unstable solutions. The thin blue horizontal
lines are asymmetric `ladder' solutions.
(b) The bifurcation diagram with NBC.
Red curves are isolas, and blue curves terminate at
branch points on the black snaking curve.
The red isolas and blue bubbles appear alternately:
four red isolas are shown; both the lowest and the highest
parts of the snaking curve form isolas in this case.
Three blue bubbles are shown as thin lines between the isolas.}
\label{fig:1dsnake}
\end{figure}

Motivated by \eqref{eqn:AC}, after applying the rescaling
$u_n = \sqrt{\frac{s}{2}}\tilde u_n$, rescaling $\mu$ and $C$ appropriately, dropping the tilde and adding 
a time derivative term appropriate for strongly dissipative rather than conservative dynamics,
we consider the following equation posed on the integer lattice $\Z$:
\beq \label{eq:1d}
\dot{u}_n = C(u_{n+1}+u_{n-1}-2u_n) + \mu u_n + 2u_n^3 - u_n^5,
\eeq
where $u_n$ is a real scalar variable defined at each lattice site,
$C$ is the strength of linear coupling between adjacent sites,
and $-1<\mu<0$ is the primary (real) bifurcation
parameter. The addition of a time derivative term is, of course, not
necessary for the investigation of the existence of equilibrium states.
However, it is a useful addition to indicate stability and to
provide a contextual similarity to the studies of the Swift--Hohenberg
equation in spatially continuous systems referred to above.
If $C<0$ then we apply the parameter symmetry
given by applying the `staggering' transformation
\beq\label{eq:stagger}
u_n \to (-1)^n u_n, \quad C \to -C, \quad \mu \to \mu - 4C,
\eeq
which leaves the equation unchanged, but switches the sign of $C$.
Thus in this Letter we will only consider positive $C$, noting that
the existence (and stability)
of states $\{u_n\}$ with $C>0$ is equivalent to the existence (and stability)
of states $\{v_n\}=\{(-1)^n u_n\}$ when $C<0$ and $\mu$ is shifted
as in~(\ref{eq:stagger}).

Importantly, and in common with the Swift--Hohenberg equation,
equation~\eqref{eq:1d} is a gradient flow, that is,
we can write $\dot{u}_n=-\p F/\p u_n$ where the potential $F$ is given by
\beq
F(u) = \sum_n \hlf C (u_{n+1}-u_n)^2 - \hlf\mu u_n^2 - \hlf u_n^4 + \tfrac{1}{6} u_n^6.
\eeq
Thus $\dot{F}\leq 0$, so that every solution of~\eqref{eq:1d} flows down
gradients of the potential toward an equilibrium solution, and
every stable equilibrium state is a local minimum of the potential. No
periodic or complex dynamics are therefore possible.

There are five homogeneous equilibria: the trivial 
state $u_n=0$, and four non-trivial equilibria $u_n=\pm u_{\pm}$, where
\beq
u_\pm^2 = 1 \pm \sqrt{1+\mu}.
\eeq
The state $u_n=0$ is linearly stable for $\mu<0$ and unstable
for $\mu>0$. Taking the positive square roots,
the lower uniform
state $u_-=\sqrt{ 1 - \sqrt{1+\mu}}$ exists in $-1<\mu<0$ and is
always unstable,
and the upper uniform state $u_+=\sqrt{ 1 + \sqrt{1+\mu}}$
exists in $\mu>-1$ and is always stable;
there is a saddle-node bifurcation at $\mu=-1$.

The per-cell potential for a homogeneous equilibrium $u_n=u_*$ is given by $F(u_*)=-\frac{1}{2}\mu u_*^2-\frac{1}{4}u_*^4+\frac{1}{6}u_*^6$. Since the system acts to minimize the total potential, the relative values of potential at the homogeneous equilibria are important. The zero state has zero potential, and the potential of the upper state $u_+$ depends only on $\mu$. By looking for a double root of $F(u)=0$ we find that the upper state has zero potential when $\mu=\mumx=-3/4$; this is the so-called {\em Maxwell point} at which
the upper uniform state and the zero state are energetically
equal. When $\mu>\mumx$ the upper state is energetically
more favourable; when $\mu<\mumx$ the zero state is energetically
more favourable.

\subsection{Boundary conditions}

In many physical applications it is appropriate to consider
\eqref{eq:1d} posed on a finite grid $n\in\{1,\dots,N\}$ with
boundary
conditions specified with the aid of `ghost cells' $u_0$ and $u_{N+1}$.
There are three obvious choices:
periodic boundary conditions (PBC) for which $u_0=u_N$ and $u_{N+1}=u_1$,
Neumann boundary conditions (NBC) for which $u_0=u_1$ and $u_{N+1}=u_N$,
and Dirichlet boundary conditions (DBC) for which $u_0=u_{N+1}=0$.

With PBC or NBC the homogeneous state is an equilibrium configuration.
With DBC this is not the case and this obvious difference
leads us to focus in this discussion
only on PBC and NBC: we find there are
substantial differences between these two cases that require
some discussion.

\subsection{Symmetries}

Equation~\eqref{eq:1d} posed on the infinite integer lattice has a
symmetry group generated by the independent operations of
translation $\translation$, reflection
$\reflection$ and an up-down symmetry $\updown$ (since
the nonlinearity is an odd function of $u_n$):
\beq
\translation:u_n\mapsto u_{n+1},\quad
\reflection: u_n\mapsto u_{-n},\quad
\updown:u_n\mapsto -u_n.
\eeq
Thus the full symmetry group of the infinite lattice
problem is $\Z\ltimes\Z_2\times\Z_2$.
Posing the equation on a finite lattice implies changes in the symmetry
group. Consider the following actions of the
elements $\translation$, $\reflection$ and $\updown$:
\beq
\translation:u_n\mapsto u_{n+1\,({\rm mod } N)},\quad
\reflection: u_n\mapsto u_{N+1-n},\quad
\updown:u_n\mapsto -u_n.
\eeq
With PBC the translation subgroup $\Z$, generated by $\translation$,
is simply replaced with the cyclic group $\Z_N$ so that the full
symmetry group is $\Z_N \ltimes \Z_2 \times \Z_2 \cong D_N \times \Z_2$.
With NBC or DBC we lose translation symmetry entirely, so the
symmetry group becomes
$\Z_2\times\Z_2=\braket{\reflection}\times\braket{\updown}$.

We will be particularly interested in localised solutions
to~\eqref{eq:1d} which lie in $\fix(\reflection)$ or 
$\fix(\reflection\translation)$, i.e.
which are symmetric under reflection 
either about a lattice point, or about a point halfway between two
lattice points. We refer to such solutions as {\em site-centred}
and {\em bond-centred}. When $N$ is even, site-centred solutions lie in
$\fix(\reflection\translation)$ and bond-centred solutions lie in 
$\fix(\reflection)$. When $N$ is odd, site-centred solutions lie in
$\fix(\reflection)$ and bond-centred solutions lie in
$\fix(\reflection\translation)$.

\subsection{Linear stability on a finite lattice}

\begin{figure}[!h]
\centering
\includegraphics[width=0.32\textwidth]{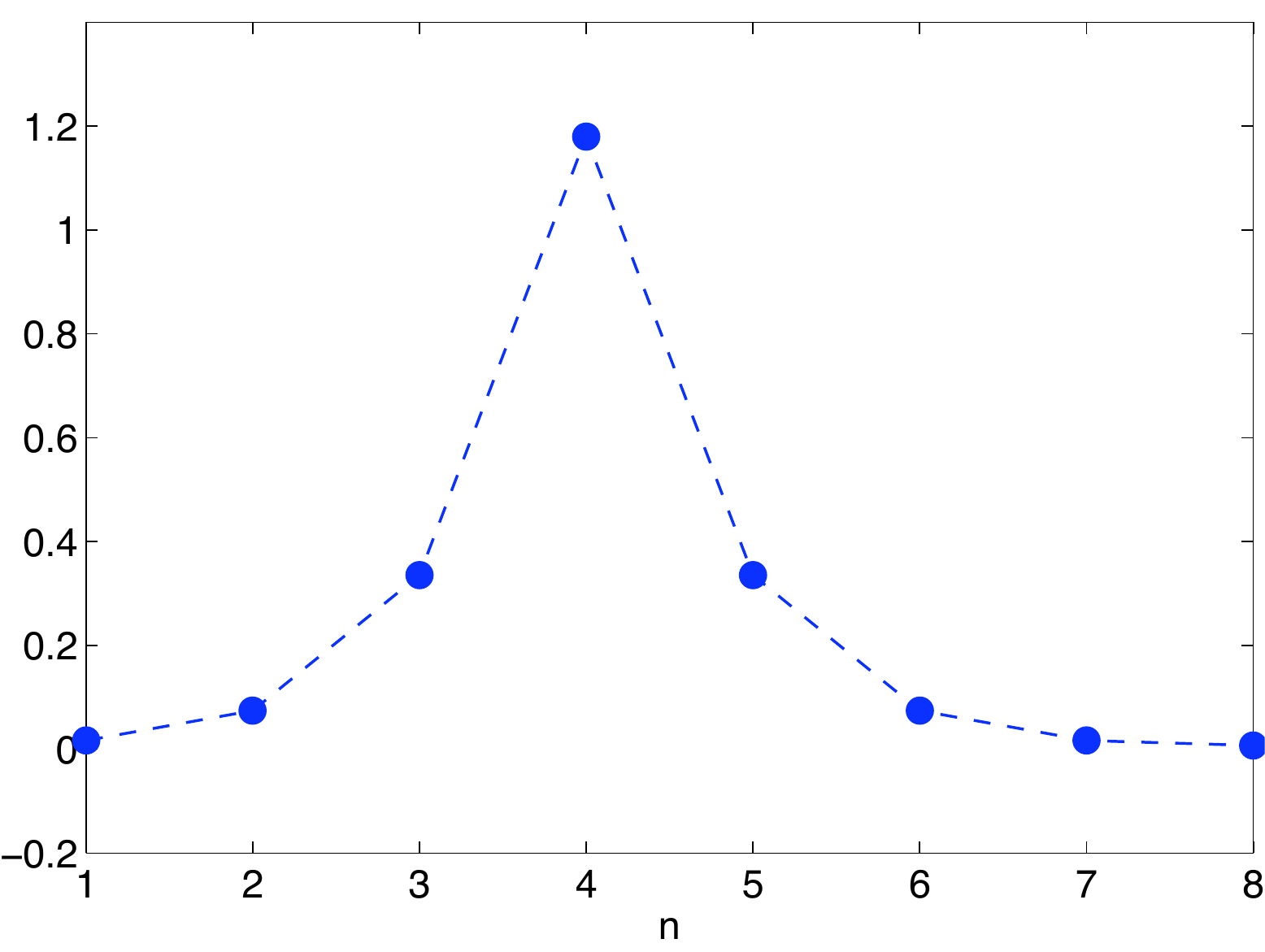}
\includegraphics[width=0.32\textwidth]{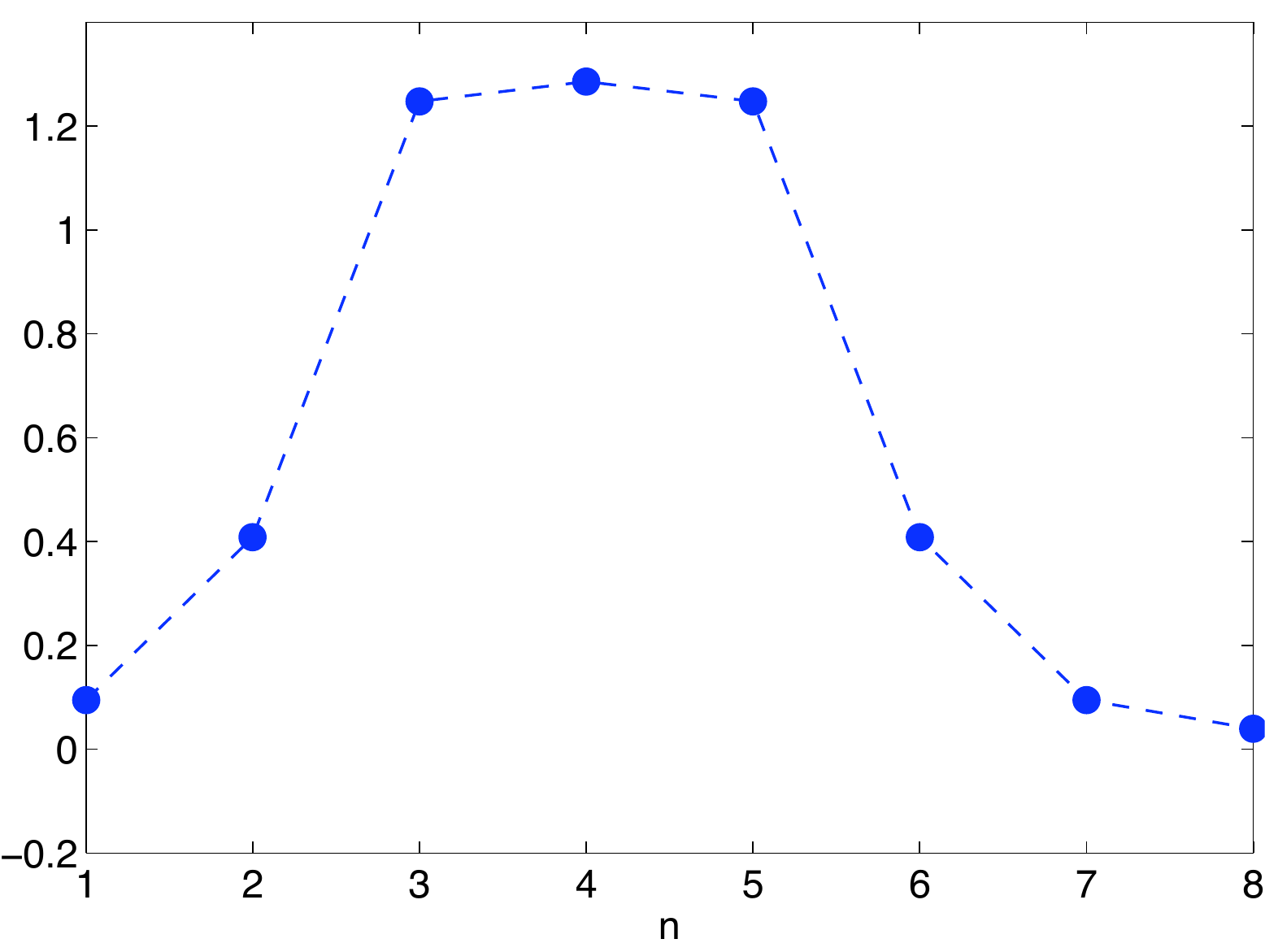}
\includegraphics[width=0.32\textwidth]{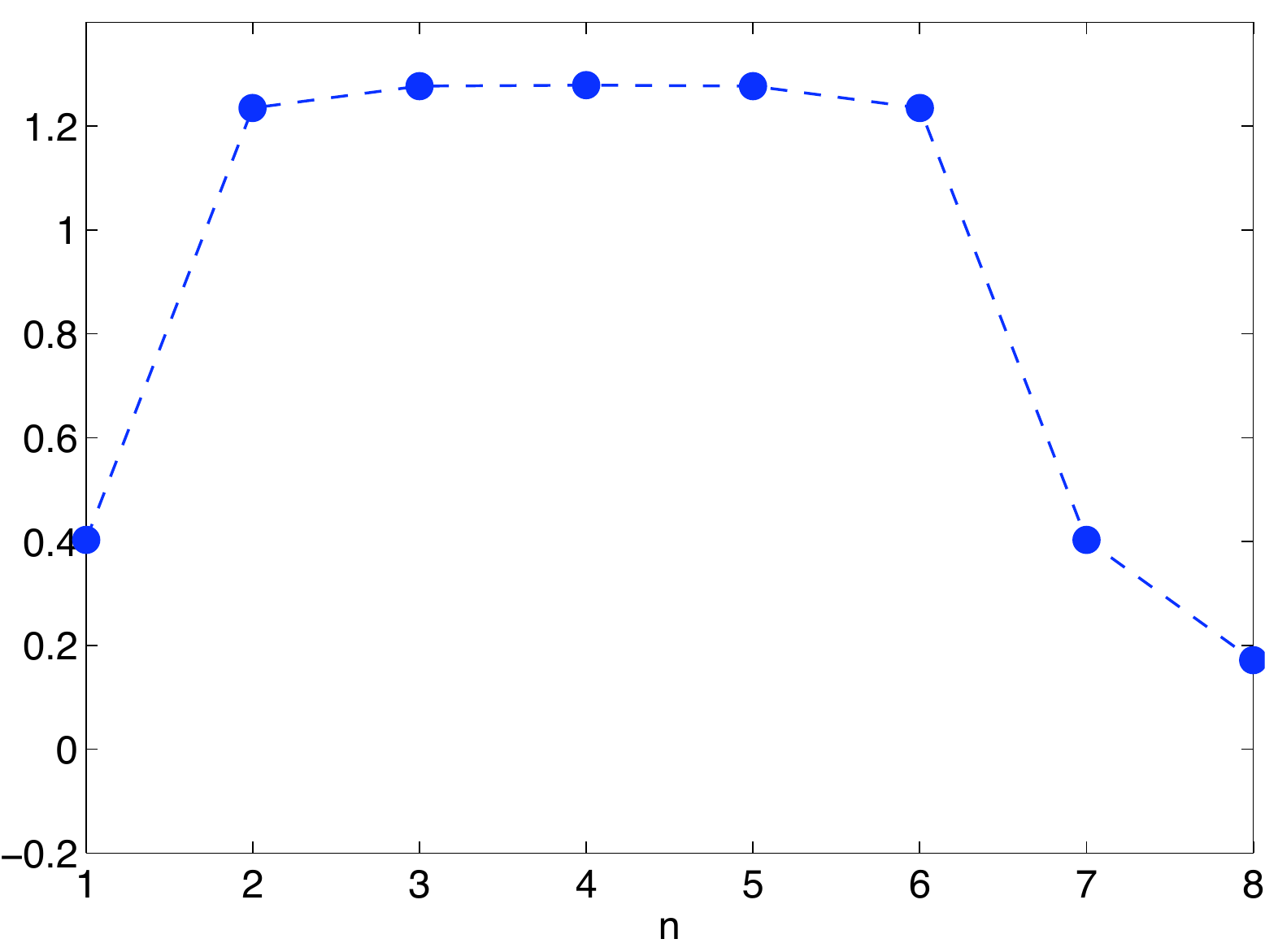}\\
\includegraphics[width=0.32\textwidth]{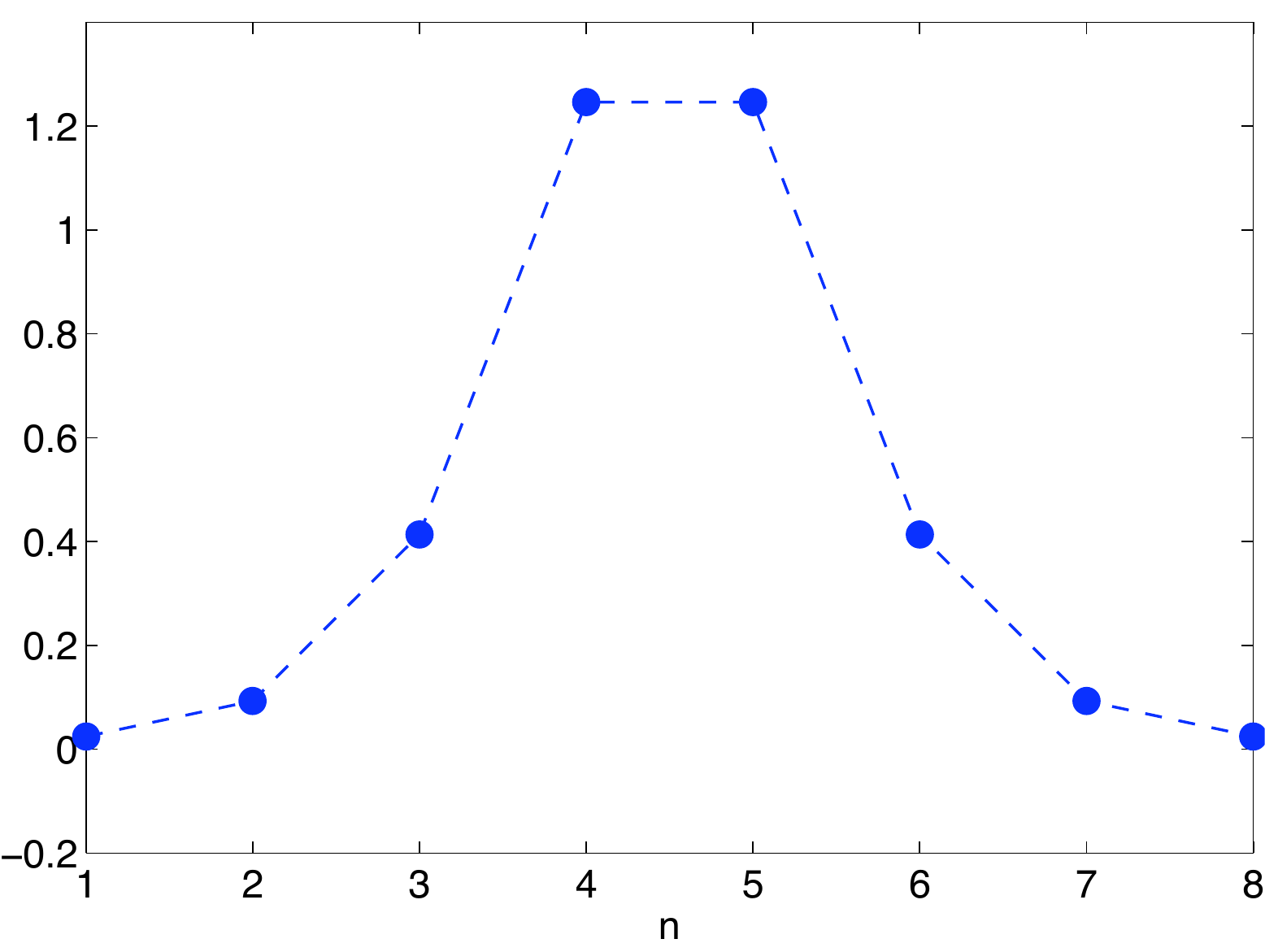}
\includegraphics[width=0.32\textwidth]{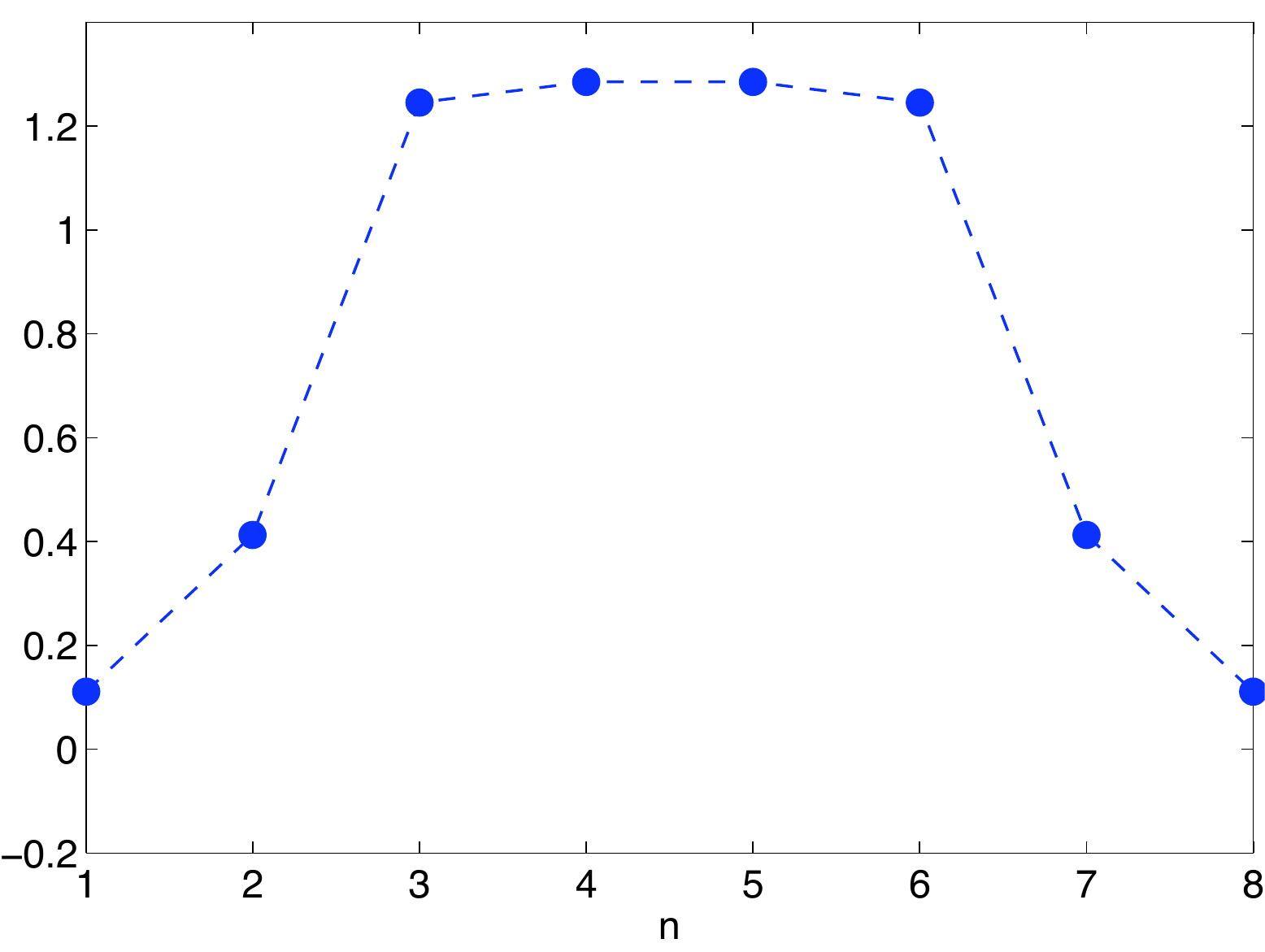}
\includegraphics[width=0.32\textwidth]{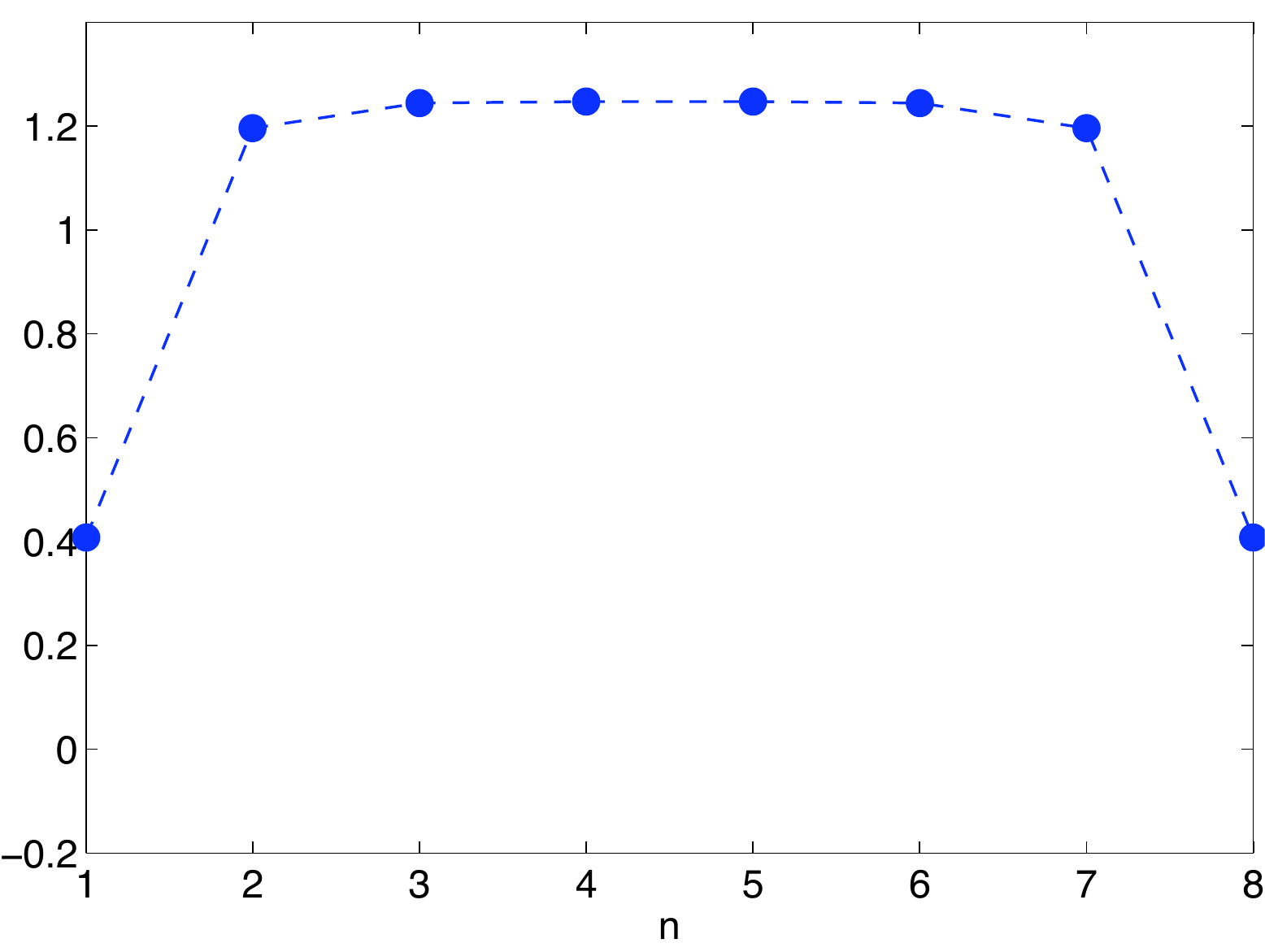}
\caption{Equilibrium solution profiles of~(\ref{eq:1d})
solved on a 1D lattice with $N=8$ using PBC and taking $C=0.2$.
Profiles correspond to the right-hand saddle-node
bifurcation points
of Figure \ref{fig:1dsnake}(a). The upper three solutions
are site-centered and hence are reflection-symmetric about the
lattice point $n=4$, since $N$ is even in this case. Similarly,
the lower three plots are bond-centred and are
reflection-symmetric about the midpoint of sites $n=4$ and $n=5$.}
\label{fig:1dprofiles}
\end{figure}

The eigenfunctions $v_n^{(k)}$ and eigenvalues $\lambda_k$
of the discrete Laplacian, obtained by solving $\Delta v_n^{(k)} \equiv
v_{n+1}^{(k)}-2v_n^{(k)}+v_{n-1}^{(k)}=\lambda v_n^{(k)}$
on an $N$-cell lattice with PBC are
easily computed to be $v_n^{(k)}=\mathrm{e}^{\i \theta_k n}$ and
$\lambda_k = -4\sin^2(\theta_k/2)$ where
$\theta_k=2\pi k/N$ for $k=0,\dots,N-1$. Thus the zero state
undergoes an instability to a mode with `wavenumber' parameter
$\theta_k$ when $\mu = 4C\sin^2(\theta_k/2)$. Since $C>0$ the first
instability encountered when increasing $\mu$ is a bifurcation to the lower
uniform state $u_-$ at $\mu=0$. This branch bifurcates subcritically, turning
around in a saddle-node bifurcation and restabilising at
$\mu=-1$ to form the upper uniform state $u_+$ as shown in
figure~\ref{fig:1dsnake}.

The upper uniform state $u_+$ undergoes no instabilities as $\mu$ is increased.
In contrast, the lower state $u_-$ has many secondary bifurcation points, occuring at
\beq
\mu = \mu_k^\pm \eqdef -\hlf - C\sin^2(\theta_k/2) \pm \sqrt{\qtr - C\sin^2(\theta_k/2)}, \label{eq:instability}
\eeq
for values of $k$ such that the expression under the square root is positive.
The bifurcations are supercritical `double pitchforks', each creating
two new bifurcating branches of `odd' and `even' modulated states which
develop into single- and multi-pulse localised states, each
of which terminates on $u_-$ at both ends. In figure~\ref{fig:1dsnake}
only the single pulse branches are shown for clarity. This behaviour of
secondary bifurcation creating branches of odd and even-symmetric
modulated and localised states is extremely similar to that given
for the continuum Swift--Hohenberg equation in \cite{BBKM08,D09} without
the additional complexity that arises from continuous changes
of wavenumber along the branch which is prohibited in the present
case by the spatial discreteness.

Noting that the instability creating
single-pulse localised states is the $k=1$ branch,
inspection of~(\ref{eq:instability}) shows that
the $k=1$ case is the
last instability to remain as $C$ increases at fixed $N$. Hence
the maximum coupling strength above which
there are no modulational instabilities of $u_-$
is $C_{\rm max} = 1/[4\sin^2(\pi/N)]$. $C_{\rm max}$ therefore
also provides an upper bound on the
range of $C$ over which localised
states asymptotic to zero may exist.

\subsection{Homoclinic snaking}
We use the numerical continuation program \AUTO~\cite{AUTO}
to explore the bifurcation
structure of localised solutions to \eqref{eq:1d} using both
PBC and NBC. Stability results come directly from \AUTO,
since \eqref{eq:1d} is simply a collection of ODEs.

With PBC the bifurcation diagram figure~(\ref{fig:1dsnake}) closely resembles
the homoclinic snaking picture for the Swift--Hohenberg equation
in a finite domain,
with branches of localised solutions emerging from the uniform
branch at $\mu \approx -0.07$ and turning around in a
succession of fold bifurcations
to form two intertwined
snakes of localised structures, one of site-centered
and one of bond-centered solutions, before reconnecting to
the uniform branch at $\mu \approx -0.98$.
For $N$ even, as is the case in figure \ref{fig:1dsnake},
these solutions lie in $\fix(\reflection\translation)$ and
$\fix(\reflection)$ respectively; this is shown clearly in
figure~\ref{fig:1dprofiles}.

\begin{figure}[!h]
\centering
\includegraphics[width=0.8\textwidth]{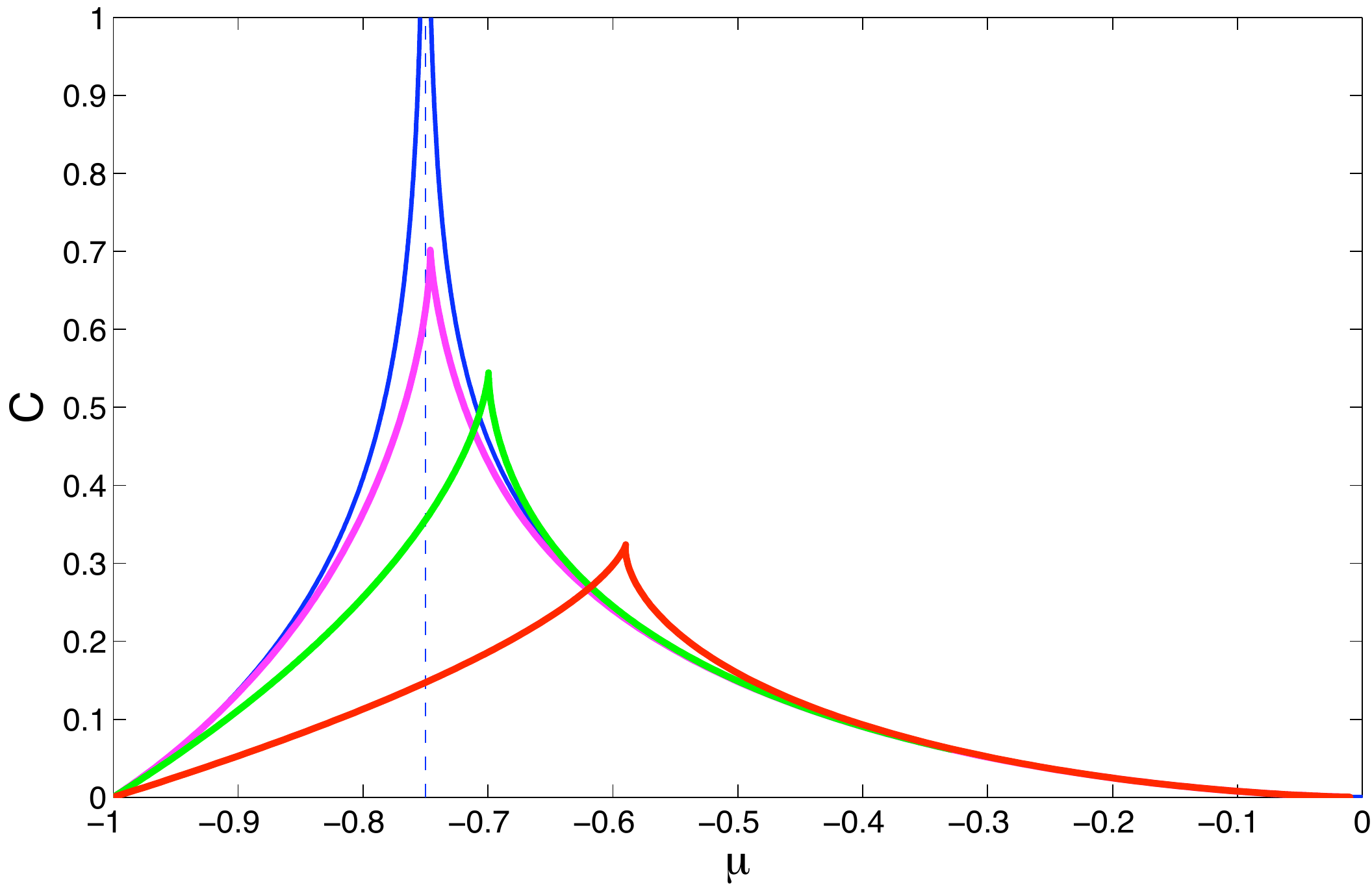}
\caption{Regions of existence of localised states in the $(\mu,C)$ plane:
$1$-cell localised state (red, peak at $C \approx 0.33$), $2$-cell
(green, peak at $C \approx 0.55$) and $3$-cell (magenta,
peak at $C \approx 0.7$).
The front solution is indicated by the blue curves which extend up to
$C=1$: on an infinite lattice
the edges of the snaking region asymptote to these curves as the width
of the localised state increases.
The vertical dashed line is the Maxwell point $\mu=-3/4$ for reference.
Results obtained for $N=100$.}
\label{fig:muC}
\end{figure}

Again, just as in the continuum Swift--Hohenberg equation,
the two snaking branches are connected by `ladders' of asymmetric 
structures: the ladders comprise solutions with no symmetry at all
that are 
always unstable. These branches bifurcate subcritically 
from the main snake, at locations near to the fold bifurcations 
making up the main body of the snake (not shown in figure 
\ref{fig:1dsnake}). Solution profiles at the lowest (in terms of $L^2$ norm)
six right-hand saddle nodes are shown in figure~\ref{fig:1dprofiles}. The
regions of the $(\mu,C)$ plane in which 1-cell, 2-cell and 3-cell
states exist are shown in figure~\ref{fig:muC}.

\subsection{Neumann boundary conditions} \label{sec:nbc}

With PBC the bifurcation structure of the discrete model \eqref{eq:1d}
exactly parallels that of the continuous bistable Swift-Hohenberg equation.
When we switch to NBC one of the snaking branches persists
but the other fragments into a series of isolas (closed curves
of equilibria that no longer connect up into a snake) and `bubbles'
(curves of equilibria that begin and end in bifurcations
from the persistent snaking curve). In figure~\ref{fig:1dsnake} the
persistent snaking branch is indicated in black,
isolas are indicated by the red curves and the bubbles are shown
in blue. For $N$ even, the persistent snaking curve consists of
bond-centred states with symmetry $\fix(\reflection)$. The
two halves of the pitchfork bifurcation where the bond-centred
states bifurcate from the uniform state are no longer
related by symmetry: in figure~\ref{fig:1dsnake}(b) we
show the branch corresponding to states that are localised in the
centre of the lattice rather than at the edges of the lattice.

The snaking curve of site-centred states in
figure~\ref{fig:1dsnake}(a)
breaks up in figure~\ref{fig:1dsnake}(b) into isolas and
bubbles which do not directly interact. In
figure~\ref{fig:1dsnake}(b) the lowest isola
approaches the bifurcation point where the bond-centred branch
connects to the uniform branch but does not bifurcate from either.
As the $L_2$ norm increases, isolas alternate with
bubbles which do bifurcate from the bond-centred snaking branch. The
bubbles bifurcate from points close to where the `ladders' bifurcate in
figure~\ref{fig:1dsnake}(a) and lack symmetry: in fact they comprise
states which evolve smoothly from the exactly bond-centred state
on the snaking curve and become very close to the subspace
$\fix(\reflection\translation)$ before moving back towards
$\fix(\reflection)$ and reconnecting with the snaking branch further
up. Moreover there are, properly speaking, four copies of
each isola and bubble branch of site-centred
asymmetric localised states, related by the 
symmetries $\updown$ and $\reflection$.

This behaviour mimics what has been observed for
the Swift--Hohenberg equation with non-periodic boundary conditions
\cite{BBKM08,D09,HK09}. It is important to note however, that
in contrast with these investigations, the domain size cannot
be used as a continuously varying parameter in the present
context. In fact, we anticipate that increasing $N$ by unity
reverses the roles of the bond-centred and site-centred solutions
when the boundary conditions are left unchanged. For example,
with NBC as we consider here, when $N$ is odd we expect the
site-centred snaking curve to persist and the bond-centred one
to fragment into isolas and bubbles in a manner similar to
that shown in figure~\ref{fig:1dsnake}.

\section{Two-dimensional lattices} \label{sec:2d}

In two dimensions our interpretation of~(\ref{eqn:AC}) must be modified 
since there are additional possibilities for the coupling terms.
On the square lattice $\Z^2$ there are two natural nearest-neighbour difference
operators:
\begin{align}
\delplus u_{nm}  & \equiv u_{n+1,m} + u_{n-1,m} + u_{n,m+1} + u_{n,m-1} - 4u_{nm}, \\
\deltimes u_{nm} & \equiv u_{n+1,m+1} + u_{n-1,m-1} + u_{n-1,m+1} + u_{n+1,m-1} - 4u_{nm}.
\end{align}
This leads to the following 2D generalization of~(\ref{eq:1d})
\beq \label{eq:2d}
\dot{u}_{nm} = C^+\delplus u_{nm} + C^\times \deltimes u_{nm} + \mu u_{nm} + 2u_{nm}^3 - u_{nm}^5,
\eeq
where there are now two coupling parameters $C^+$ and $C^\times$, the 
coefficients of nearest-neighbor (NN) and next-nearest-neighbor 
(NNN) coupling.
As in the 1D case,~(\ref{eq:2d}) is variational.
For this equation our primary interest is in establishing
that even with PBC there are isolas containing stable
localised states. Therefore, we confine our presentation to
localised solutions that are far from the lower and upper
ends of the snake where reconnection to the uniform state occurs
in a finite domain. To further simplify the
discussion we set $C^\times=0$ for the remainder of this
Letter. Clearly it is of interest to investigate how the
snaking structure evolves with $C^\times$ nonzero, and we
will return to this and other issues in subsequent work.

\begin{figure}[!h]
\centering
\includegraphics[width=0.32\textwidth]{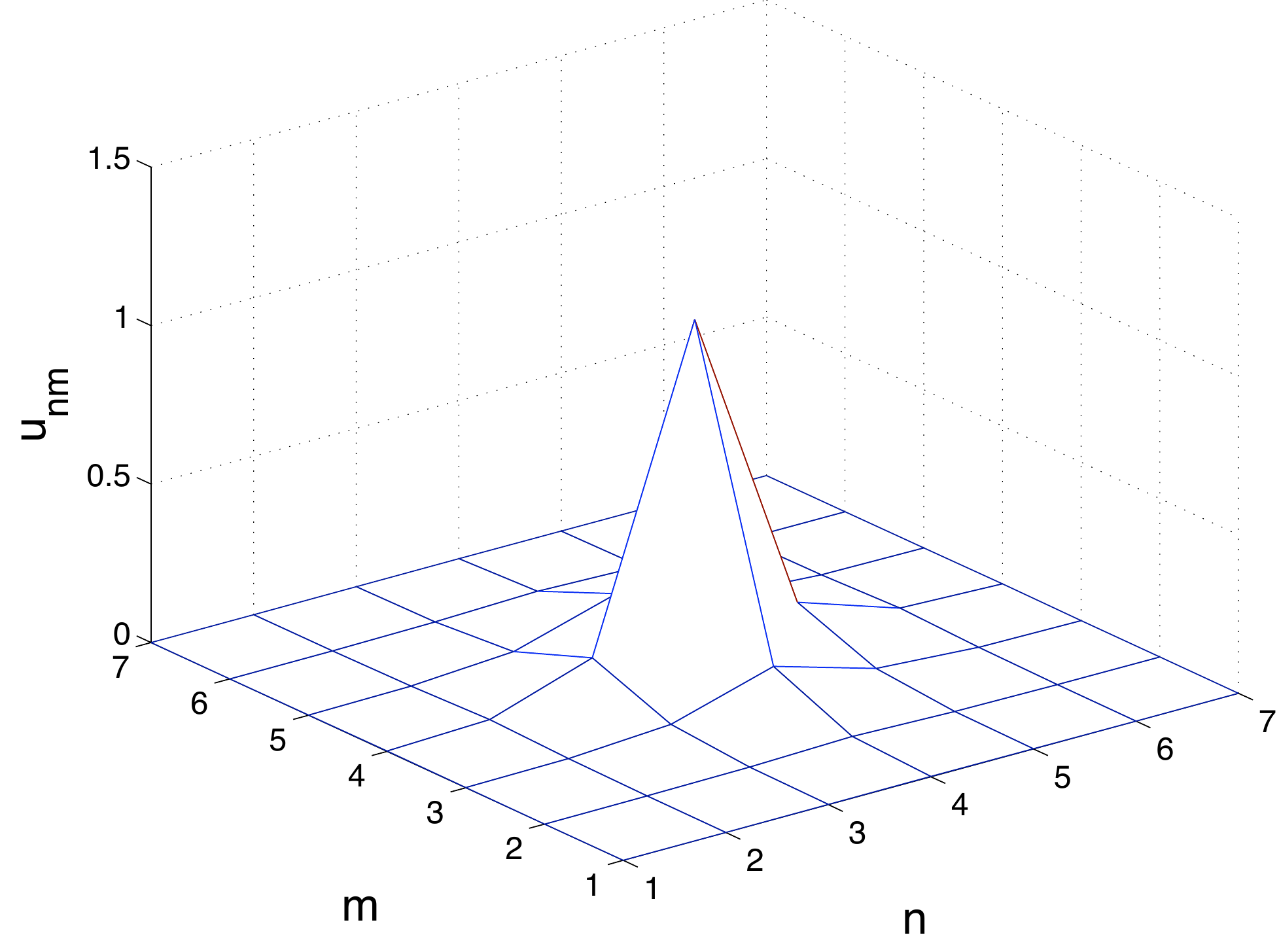}
\includegraphics[width=0.32\textwidth]{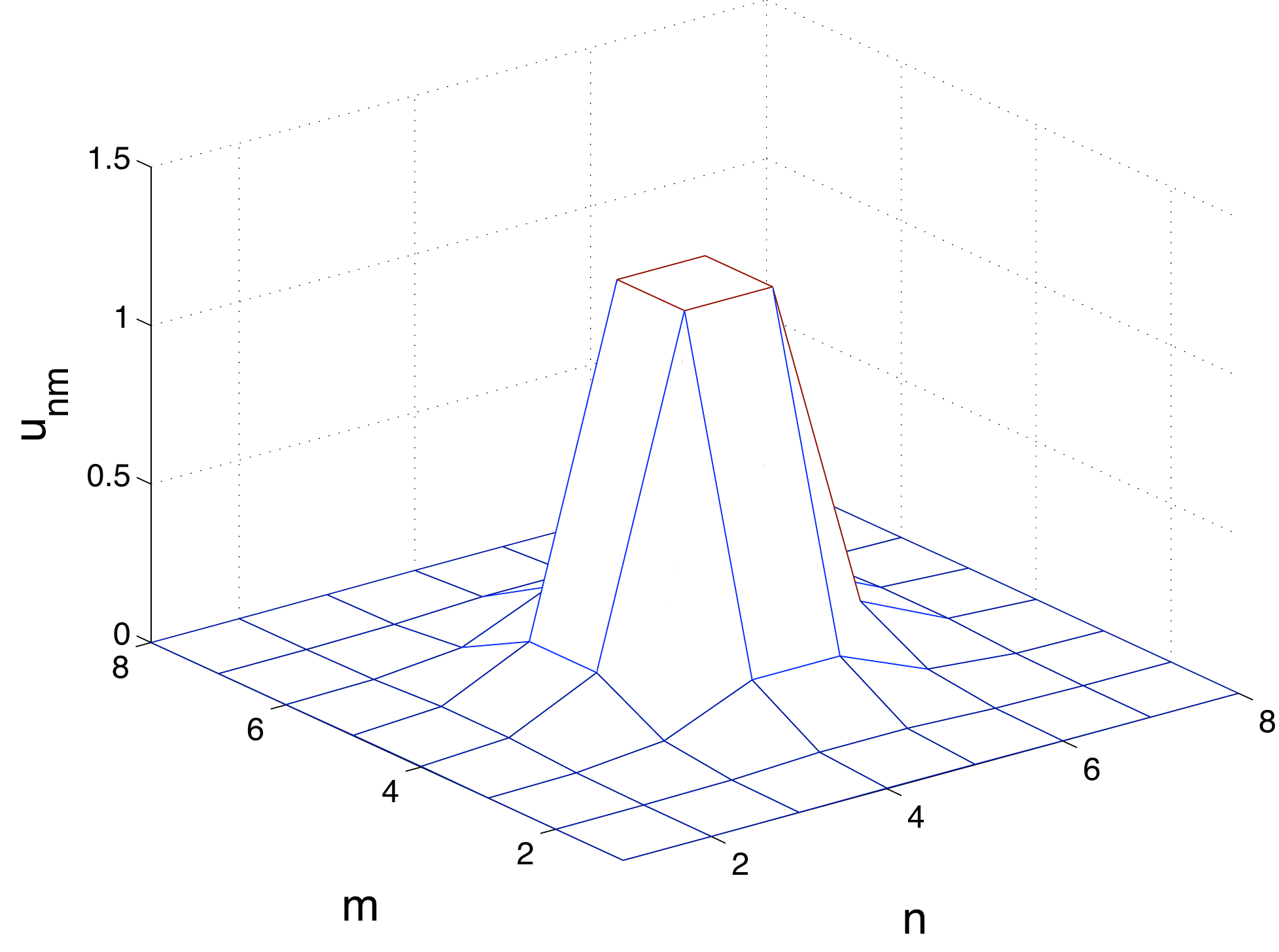}
\includegraphics[width=0.32\textwidth]{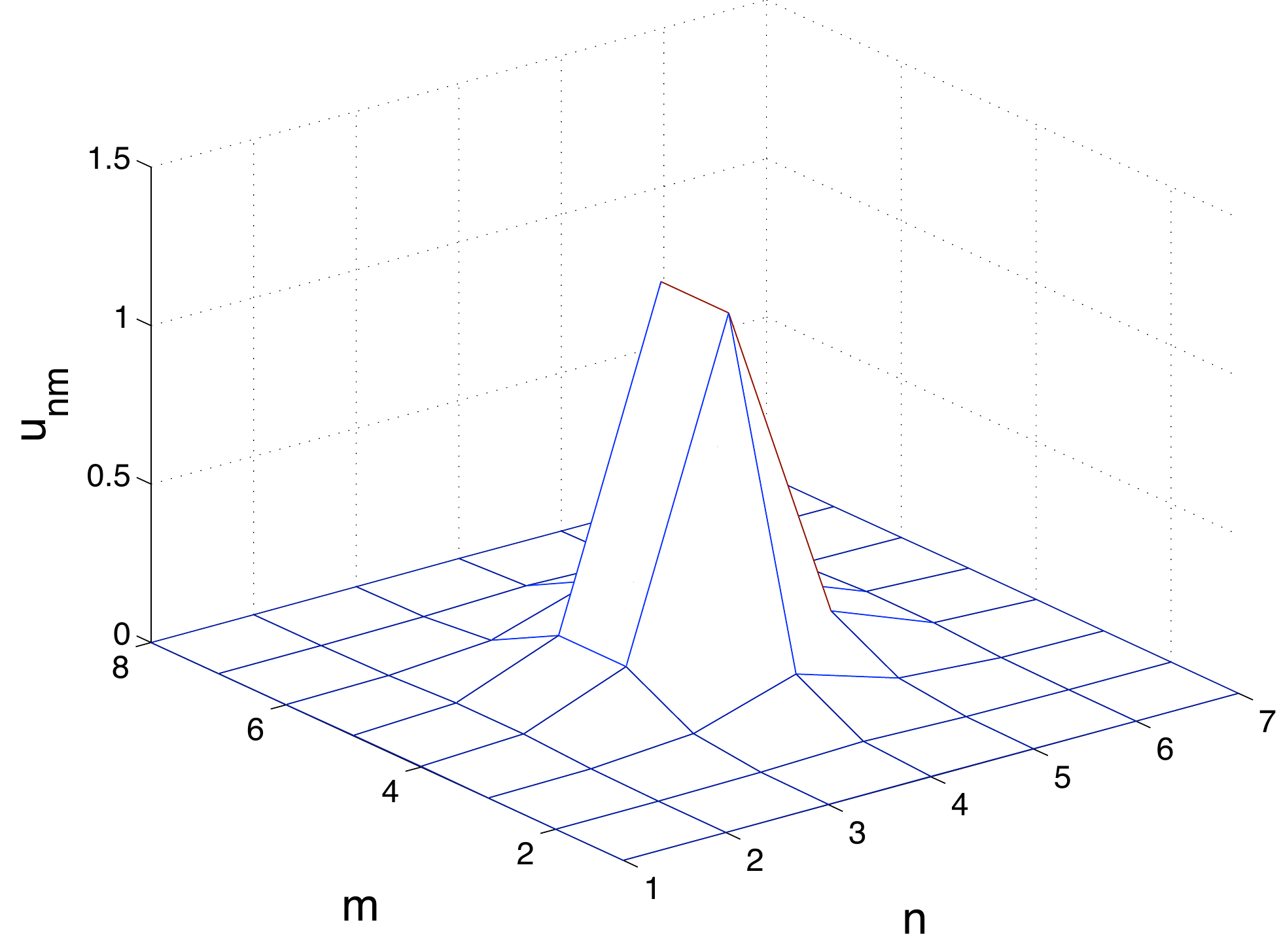}
\caption{Three typical localised solutions to \eqref{eq:2d}
with $C^+=0.1$, $C^\times=0$ and $\mu=-0.6$. The three
solutions are respectively site-centred, bond-centred and hybrid.}
\label{fig:local2d}
\end{figure}

\begin{figure}[!h]
\centering
\includegraphics[width=0.16\textwidth]{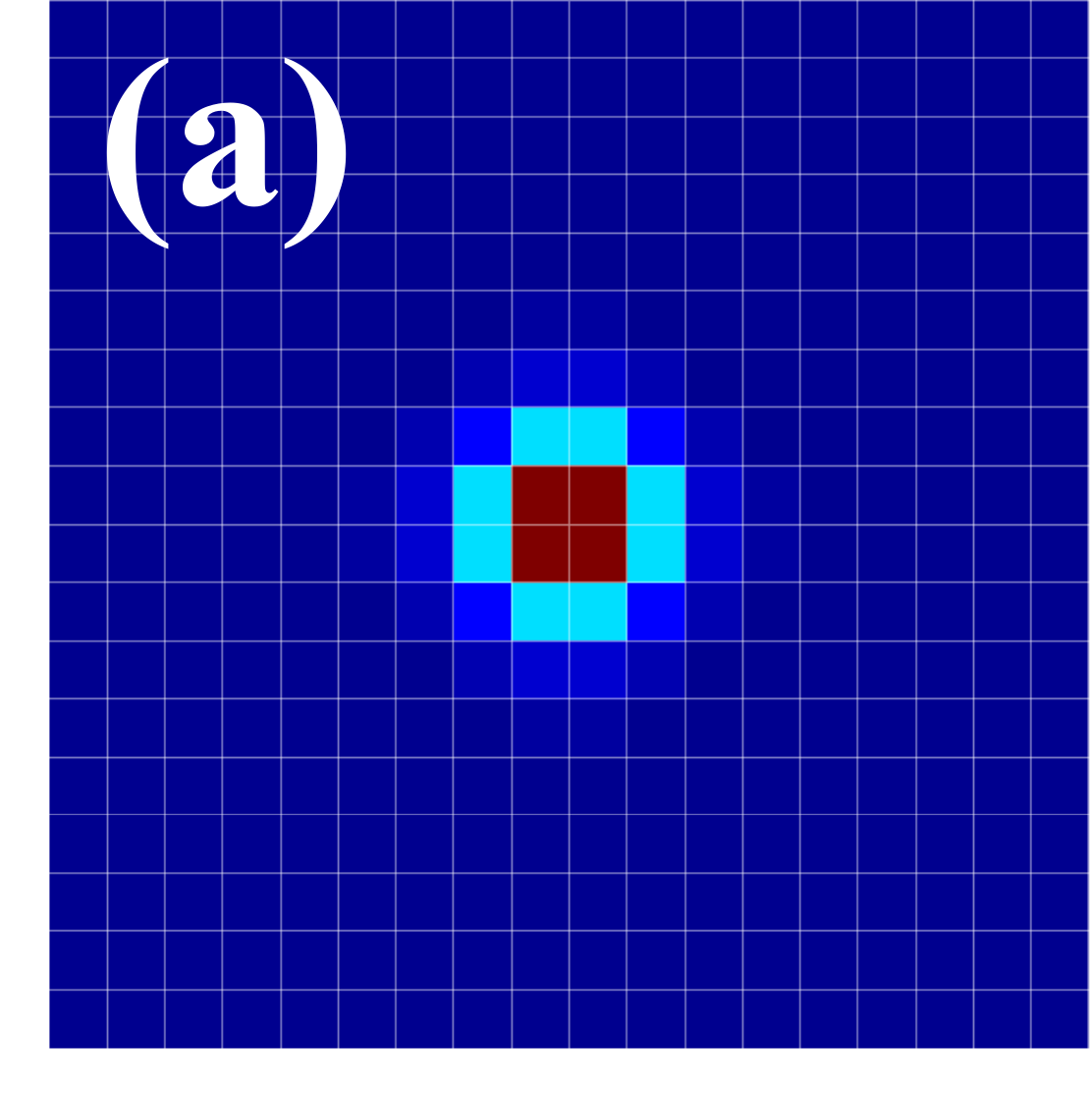}
\includegraphics[width=0.16\textwidth]{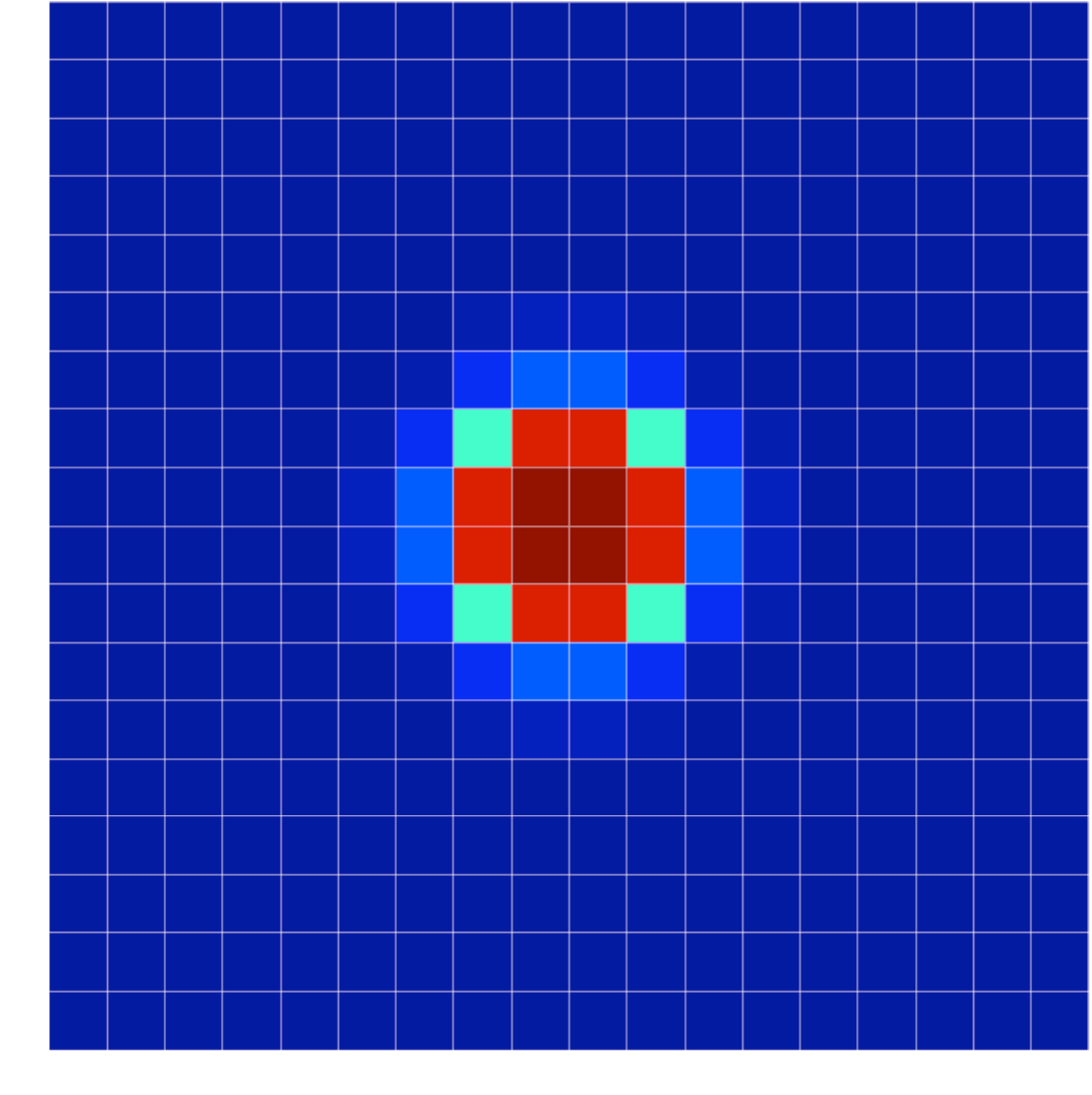}
\includegraphics[width=0.16\textwidth]{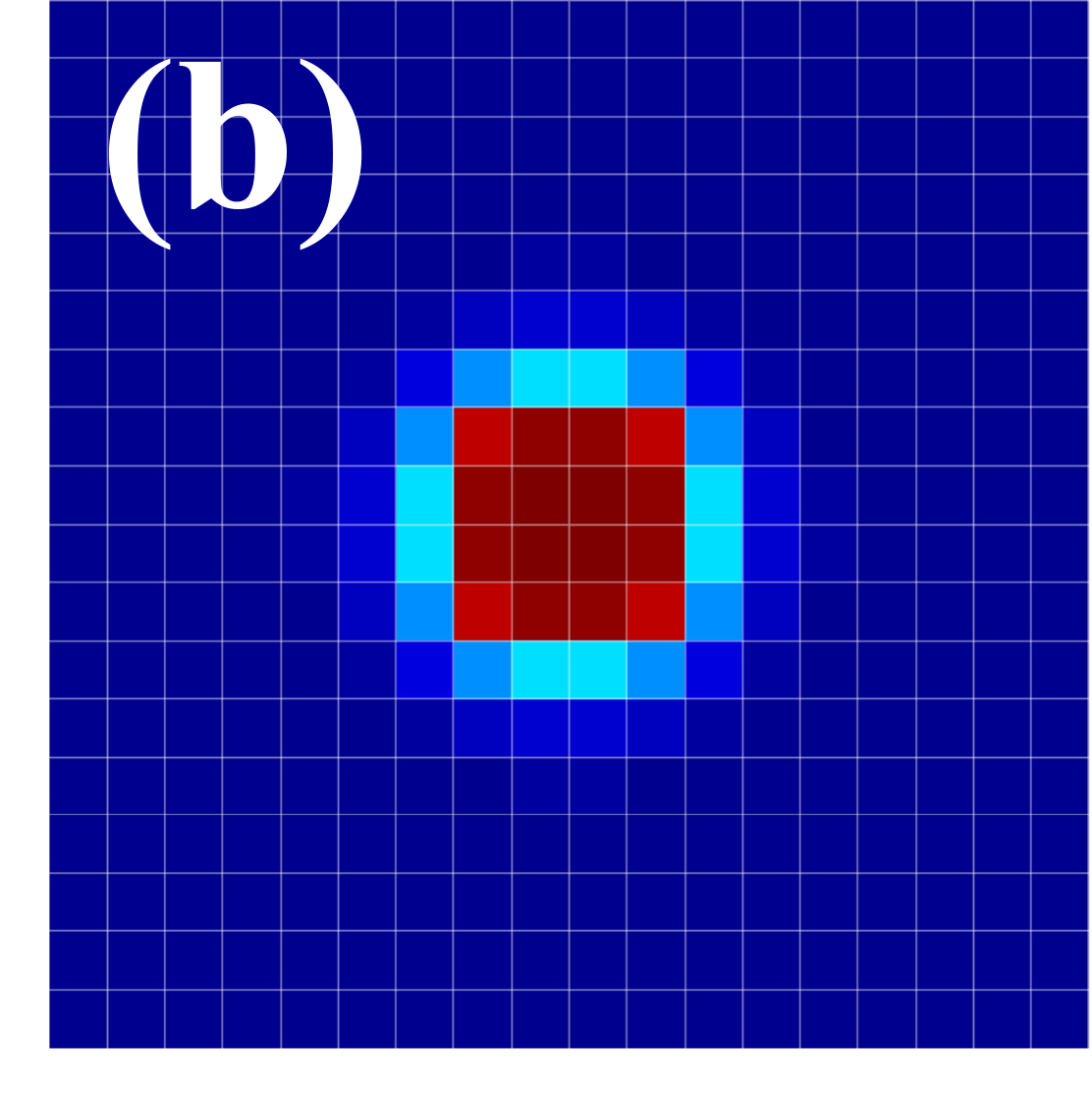}
\includegraphics[width=0.16\textwidth]{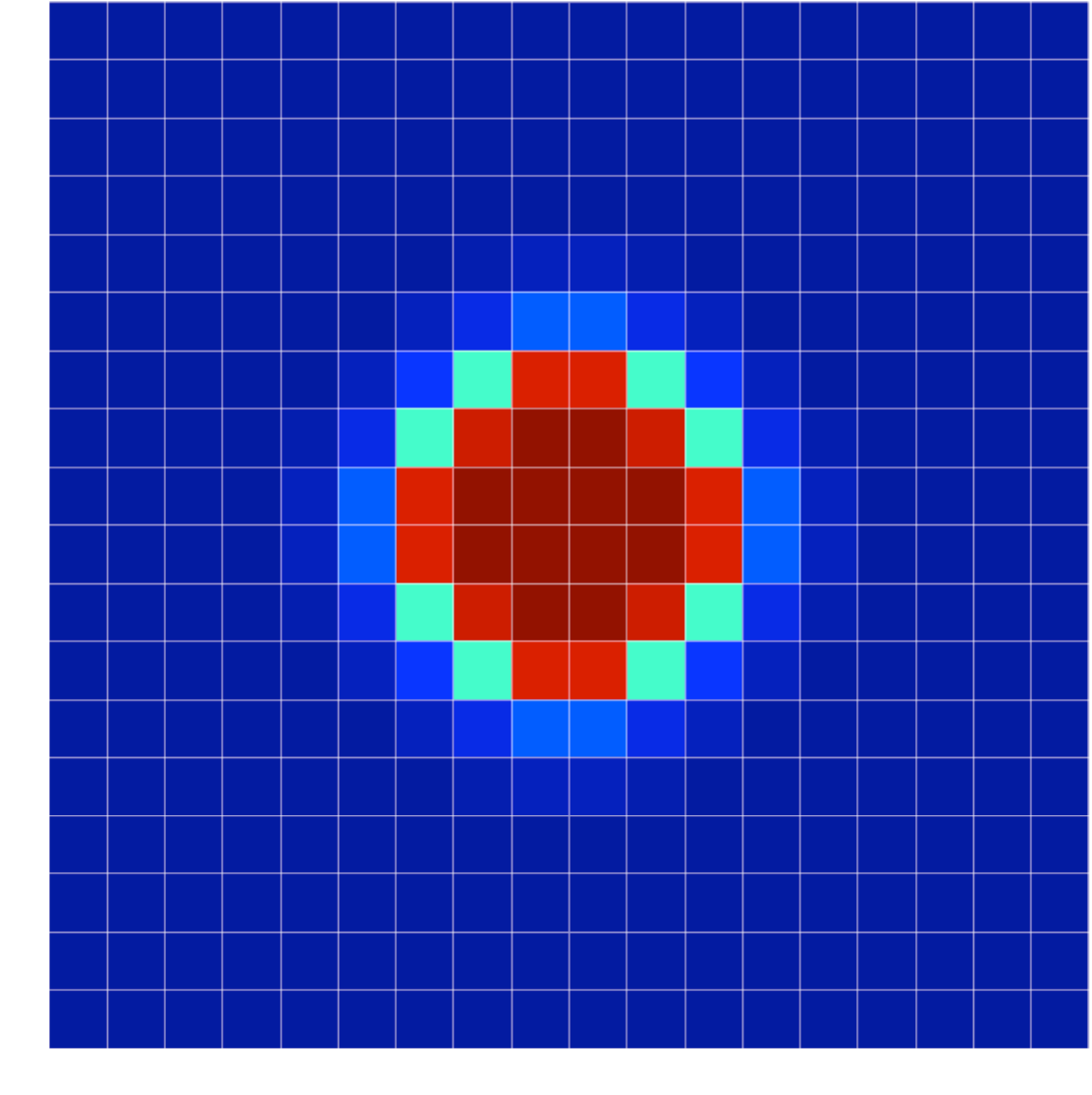}
\includegraphics[width=0.16\textwidth]{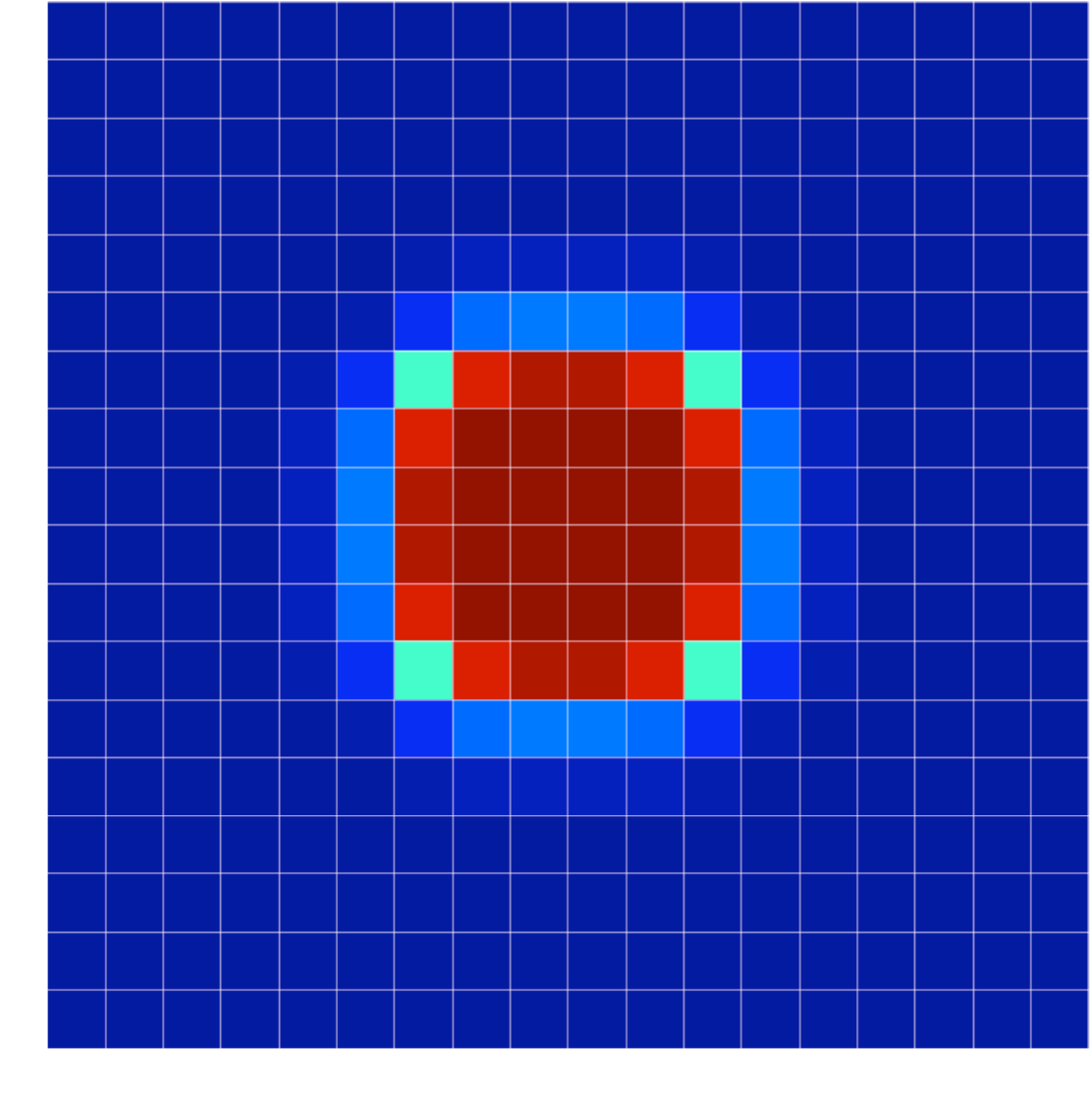}
\includegraphics[width=0.16\textwidth]{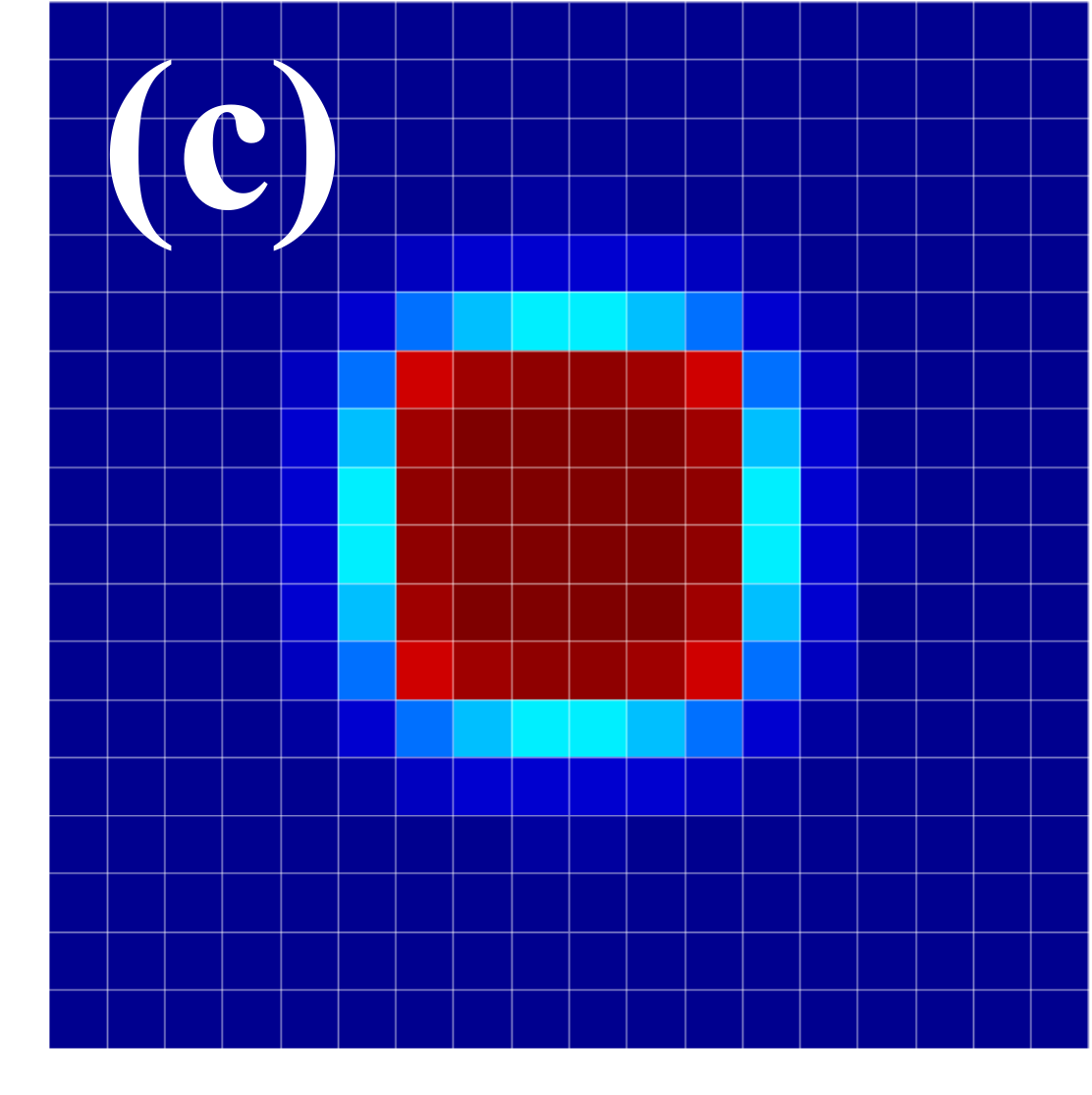}\\
\includegraphics[width=0.16\textwidth]{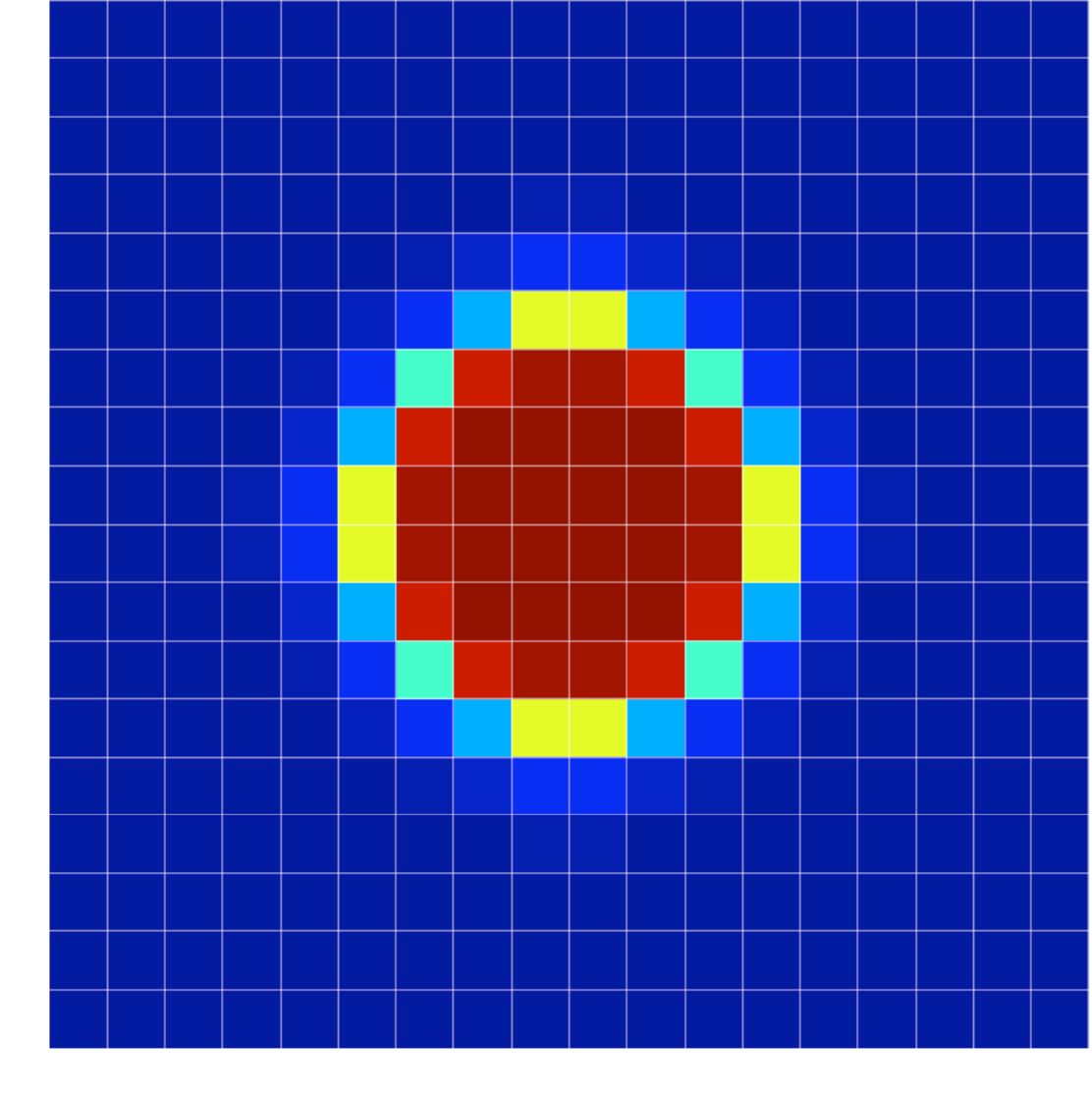}
\includegraphics[width=0.16\textwidth]{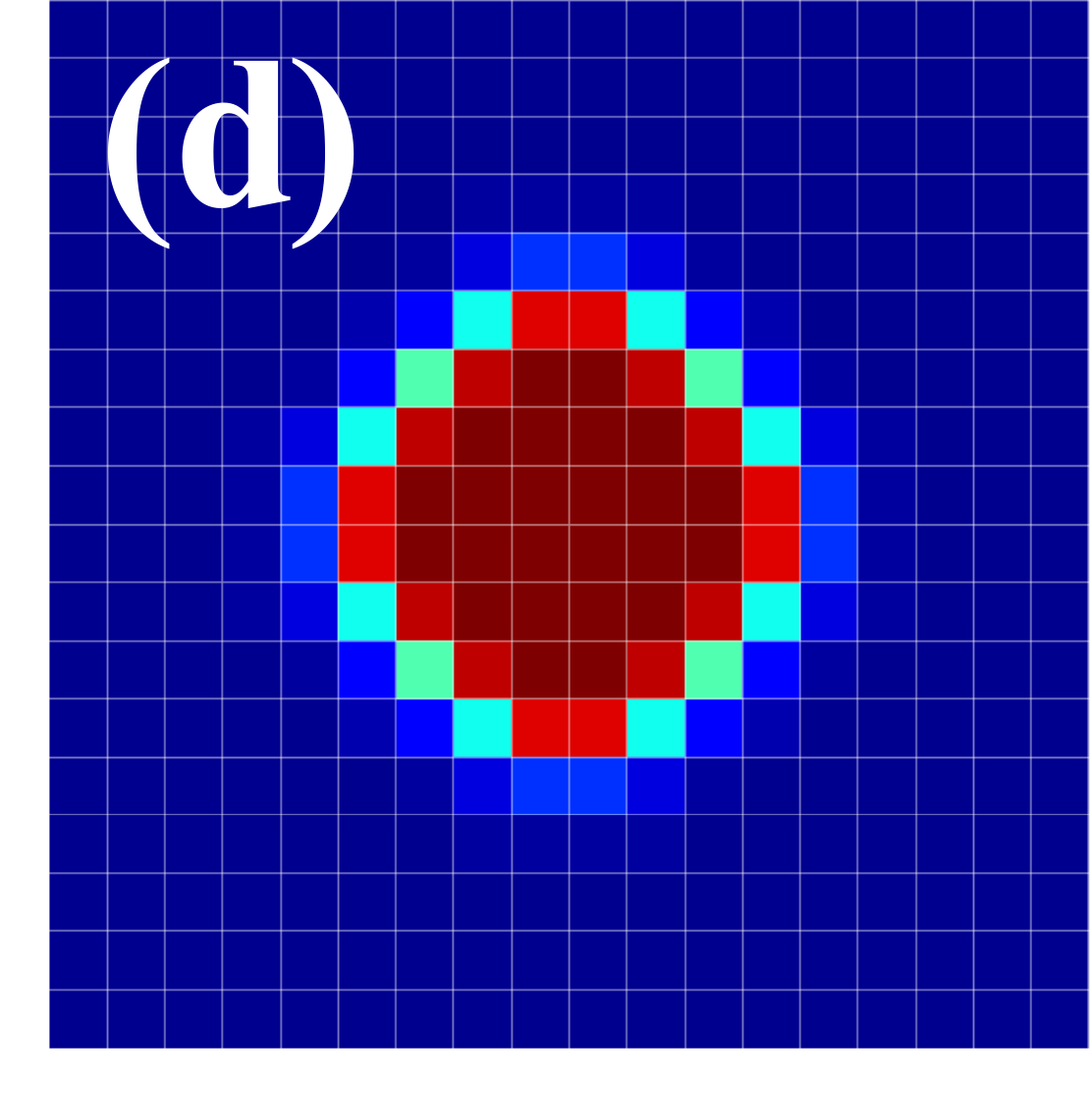}
\includegraphics[width=0.16\textwidth]{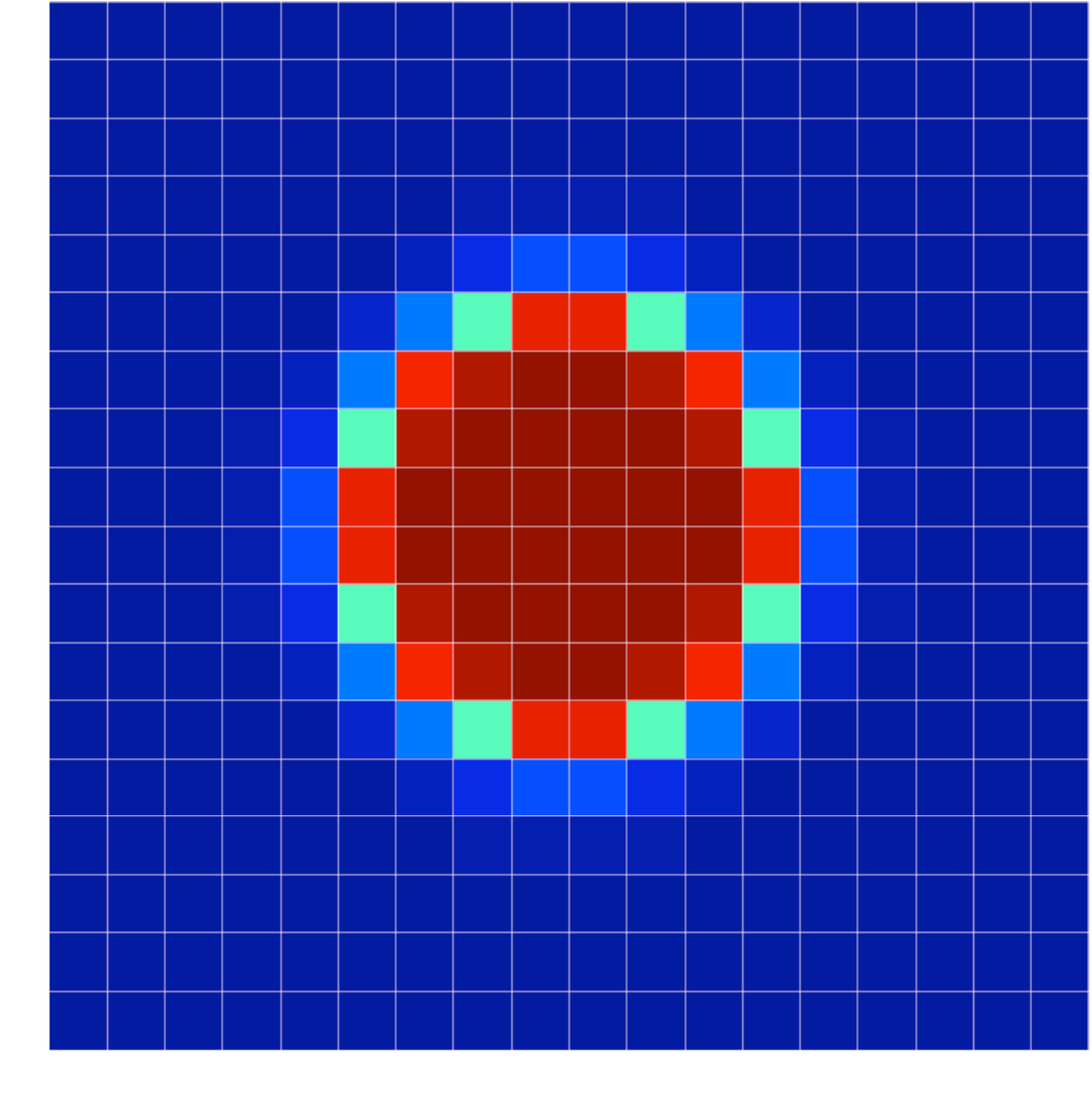}
\includegraphics[width=0.16\textwidth]{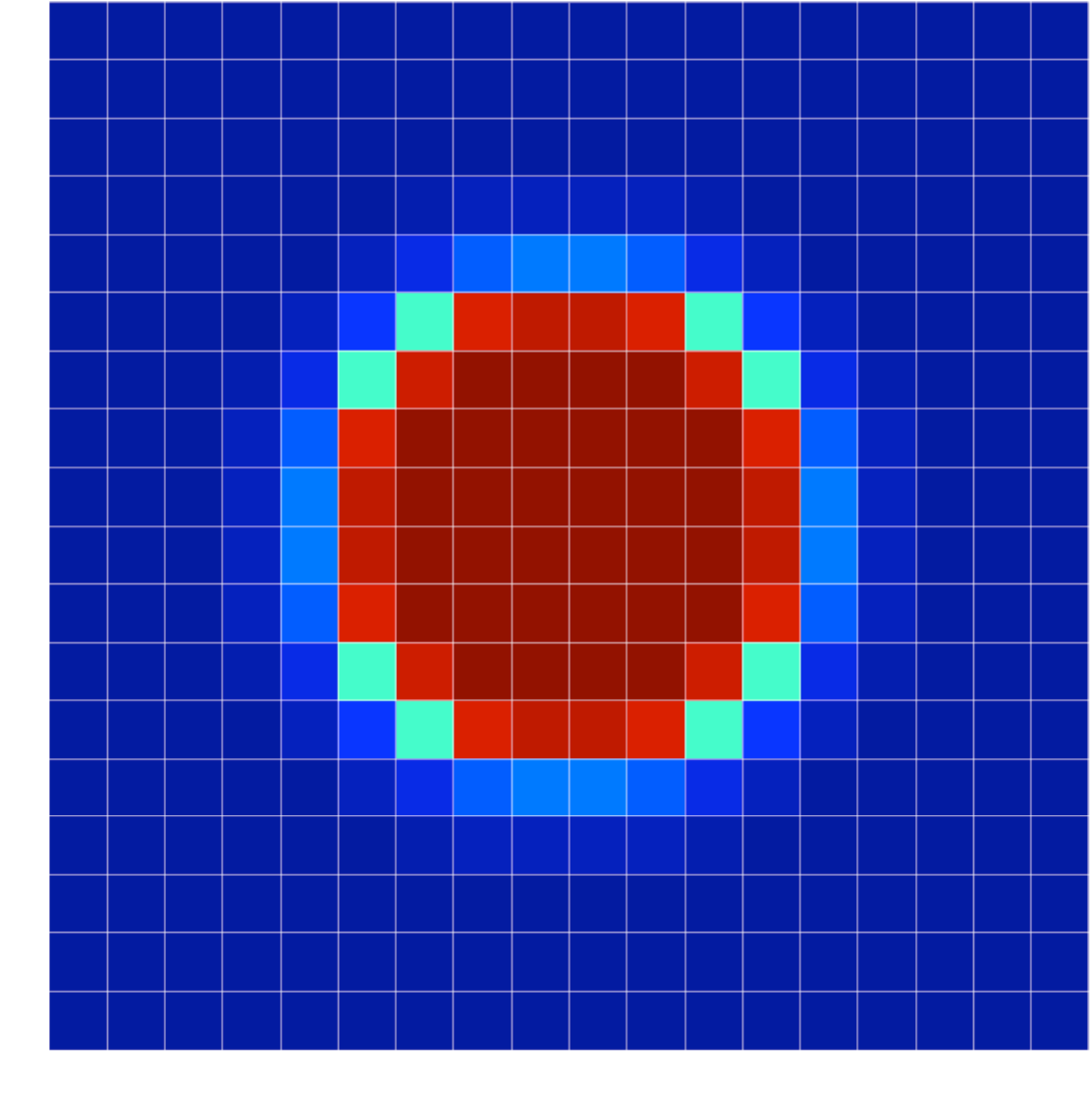}
\includegraphics[width=0.16\textwidth]{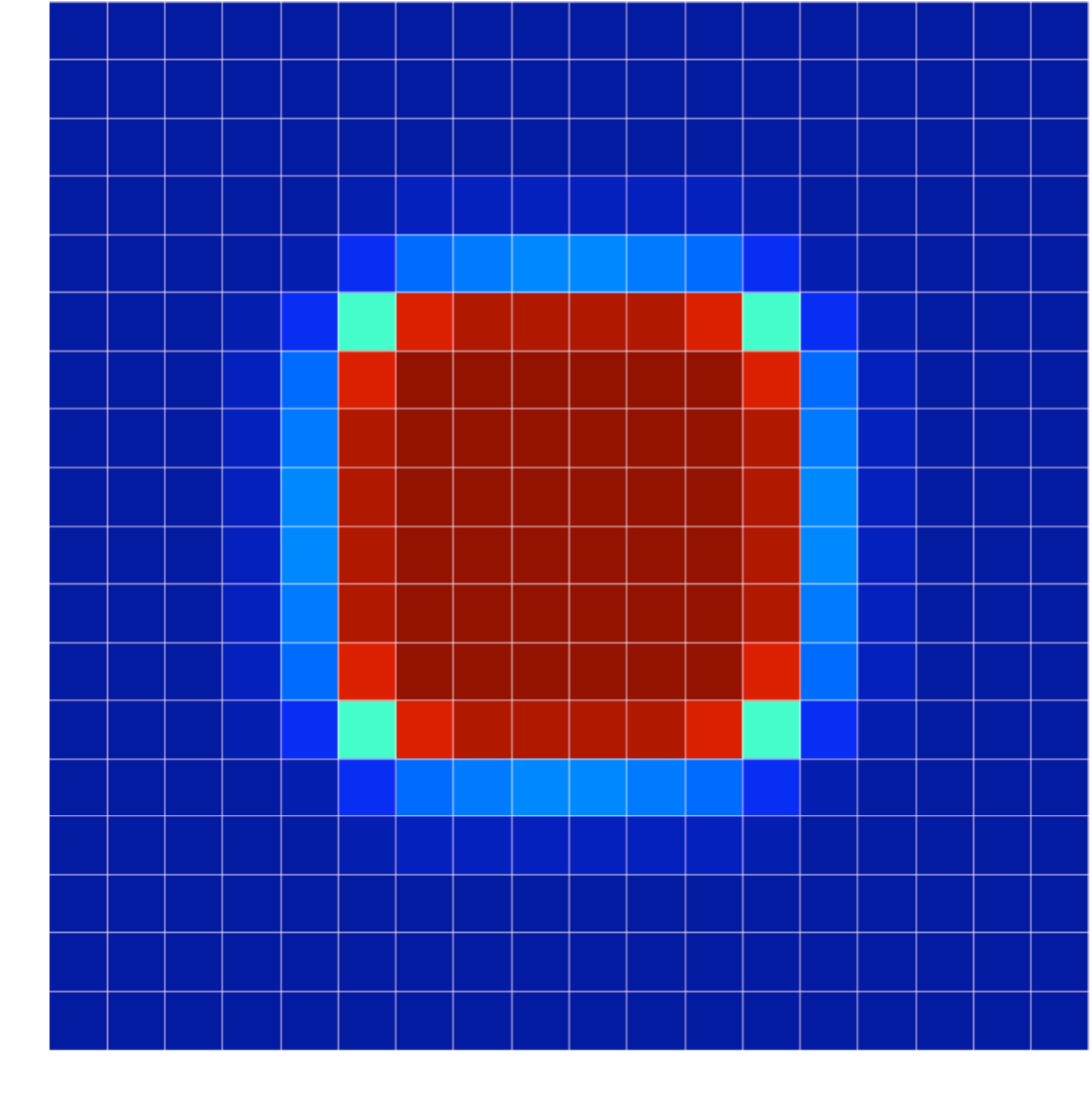}
\includegraphics[width=0.16\textwidth]{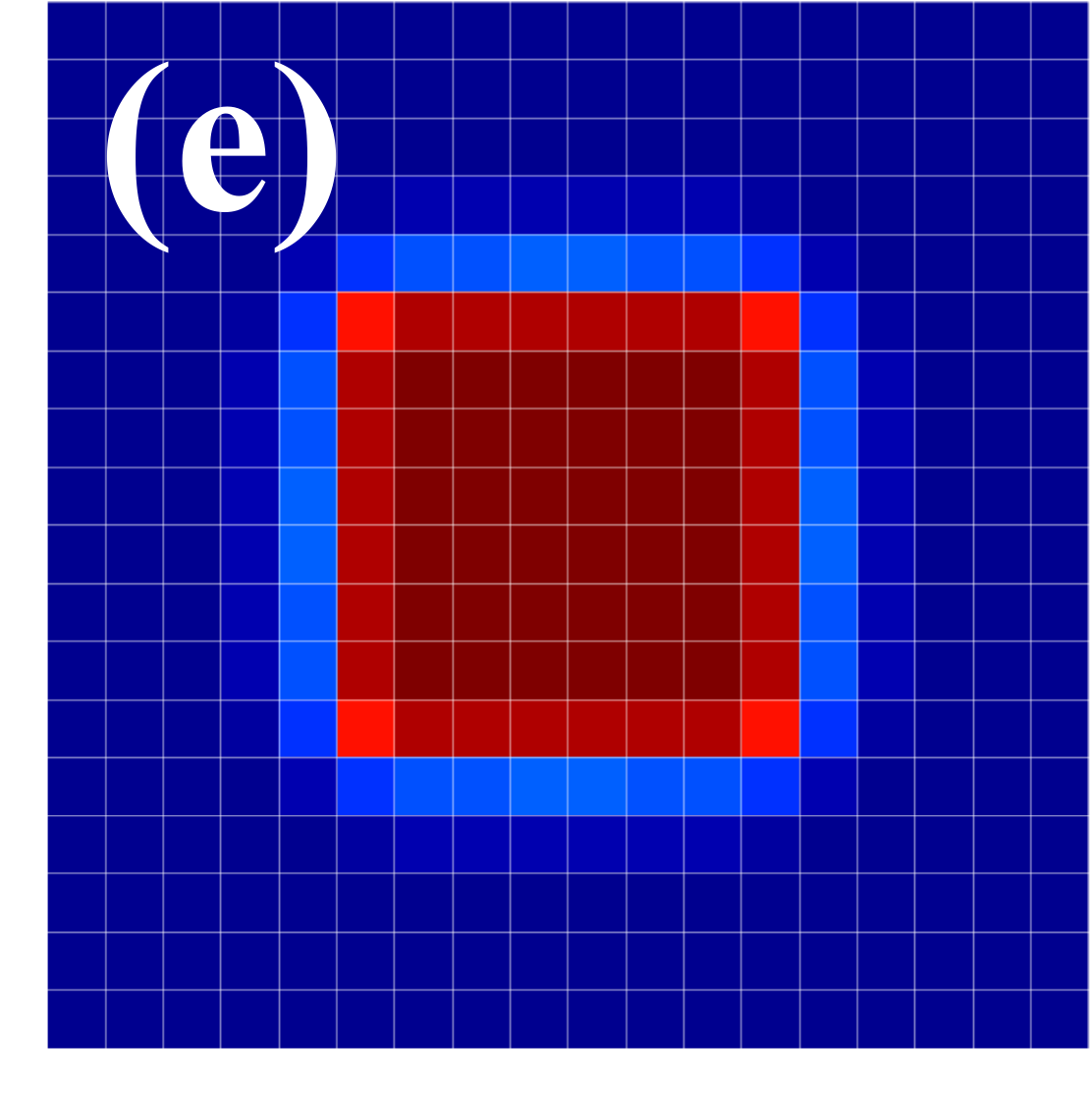}\\
\includegraphics[width=0.16\textwidth]{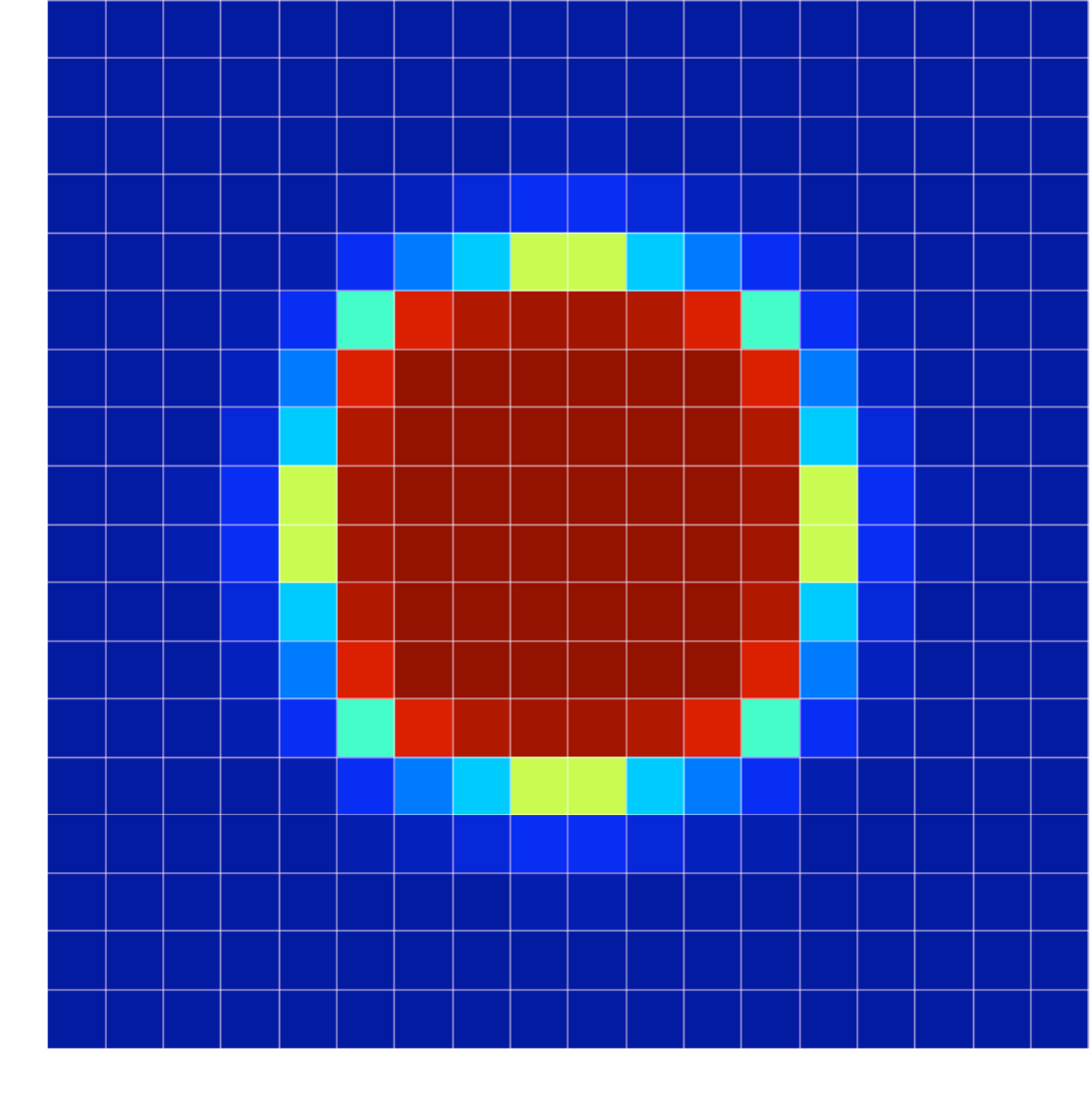}
\includegraphics[width=0.16\textwidth]{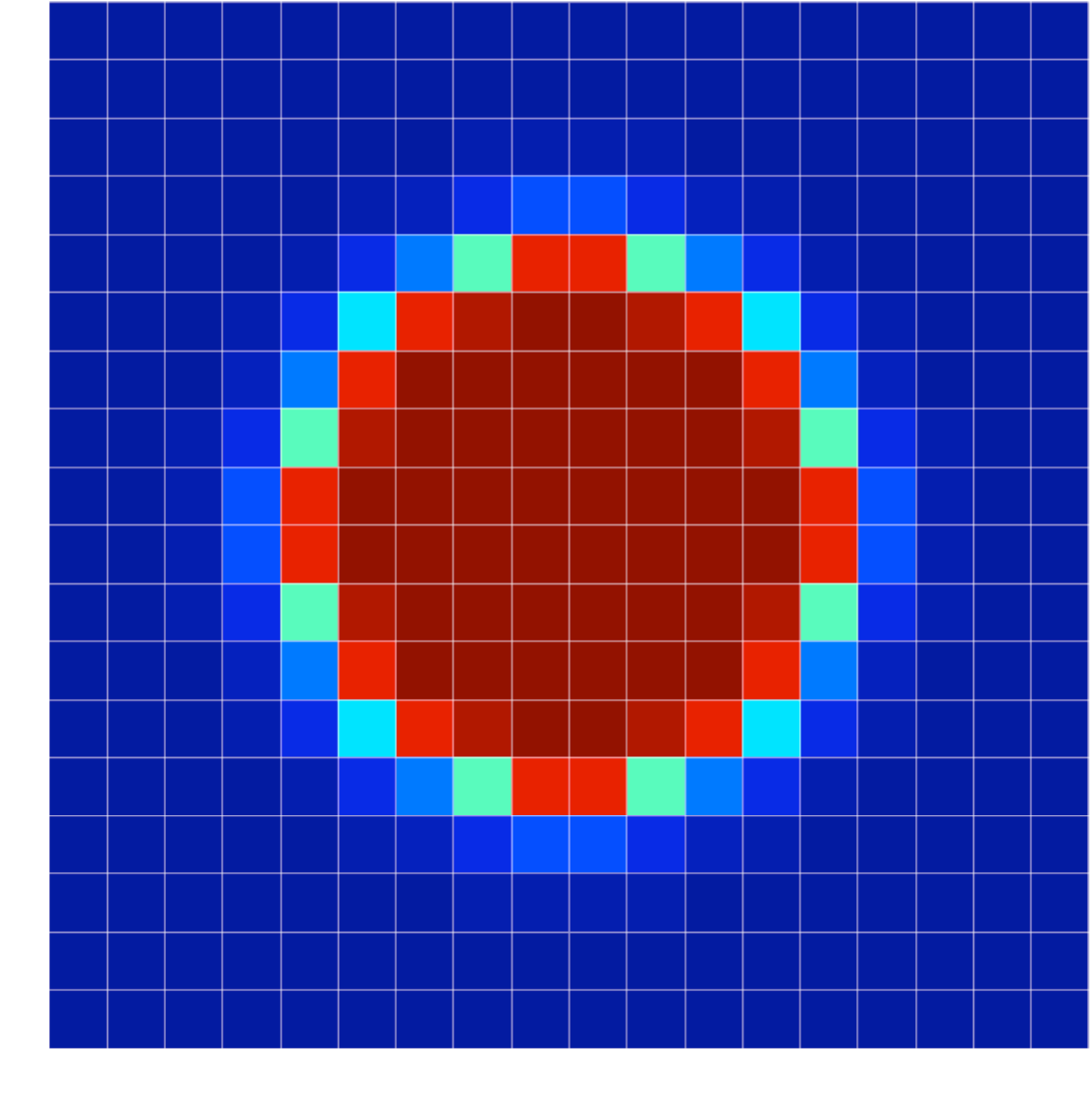}
\includegraphics[width=0.16\textwidth]{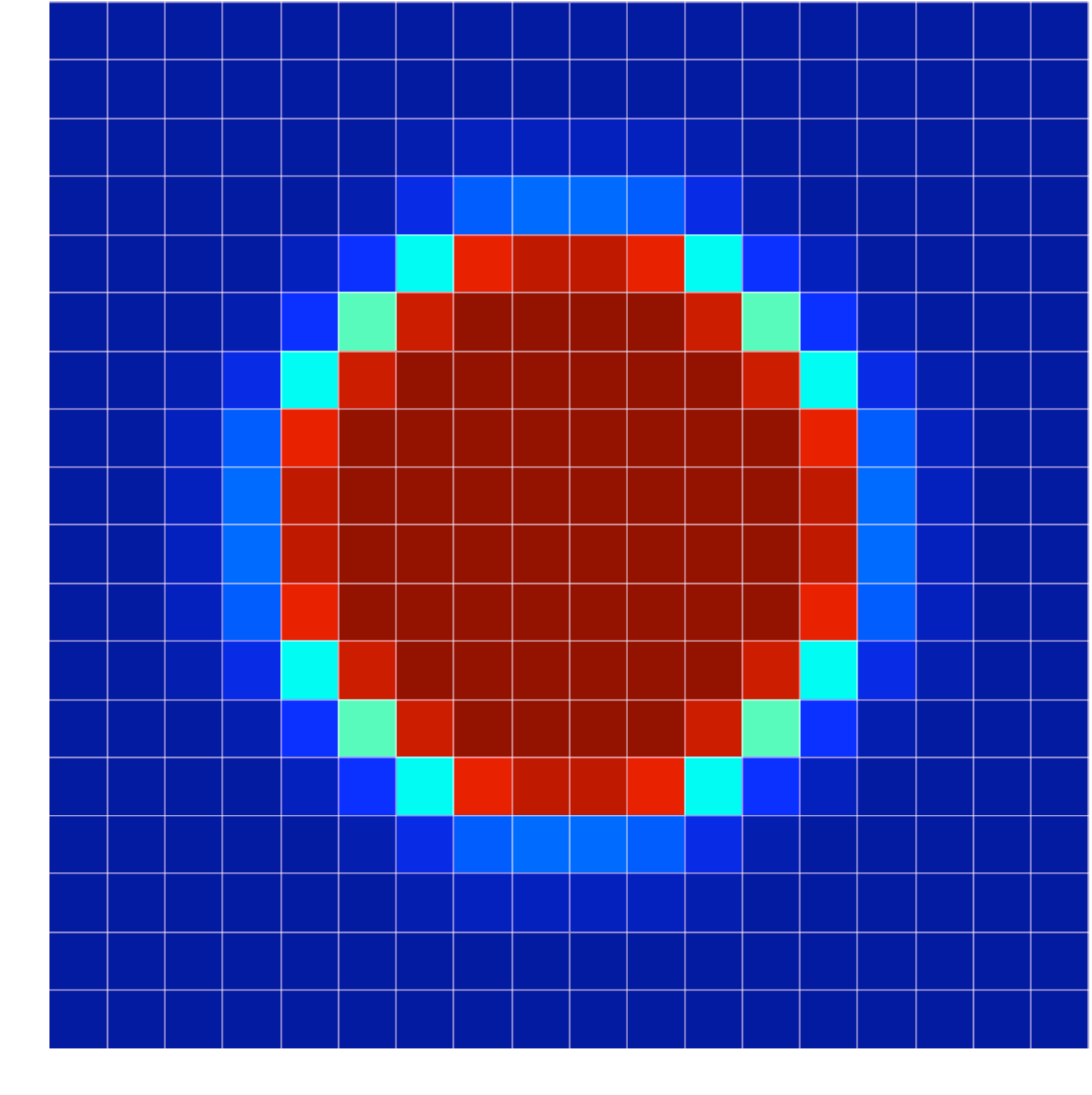}
\includegraphics[width=0.16\textwidth]{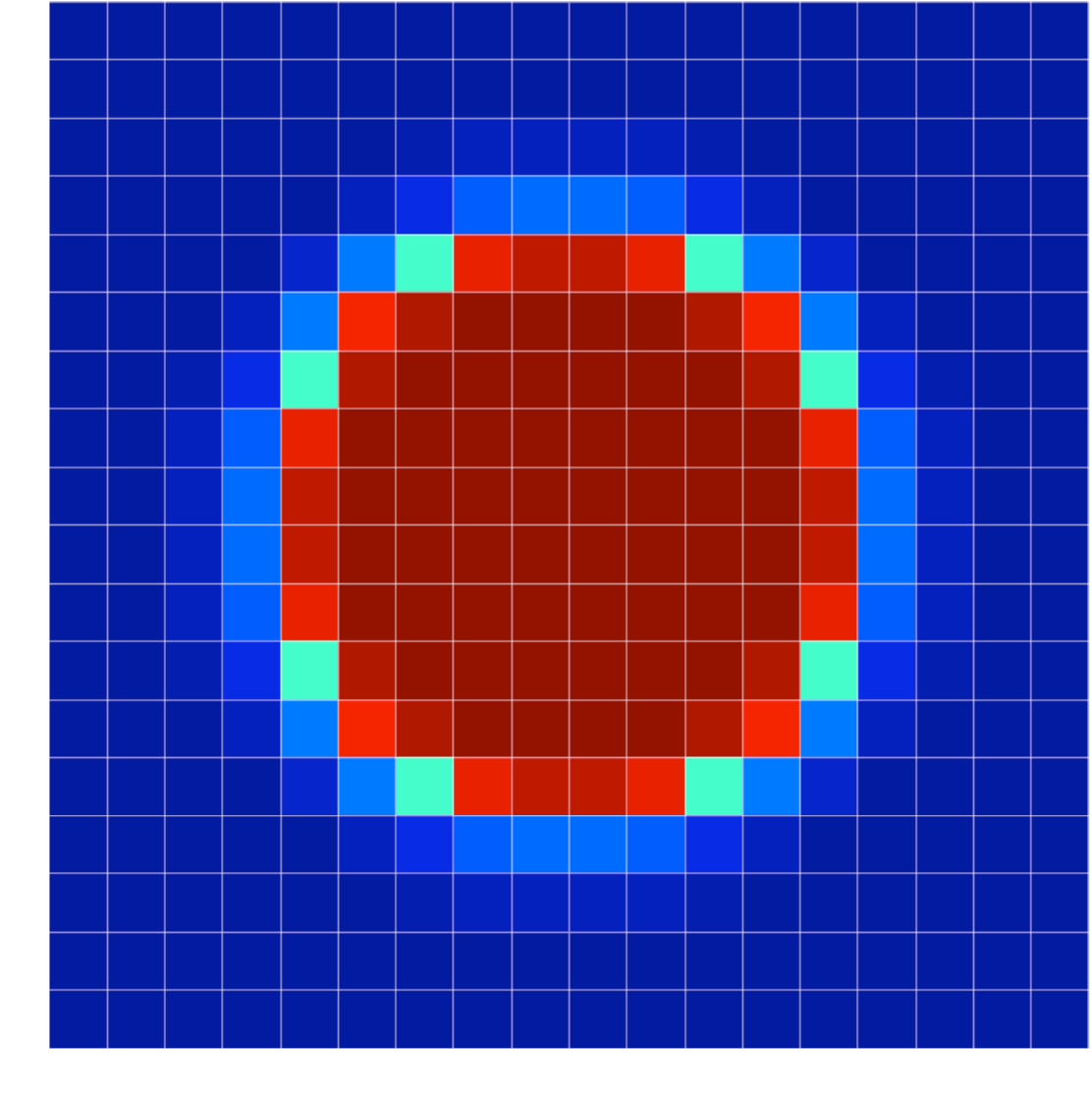}
\includegraphics[width=0.16\textwidth]{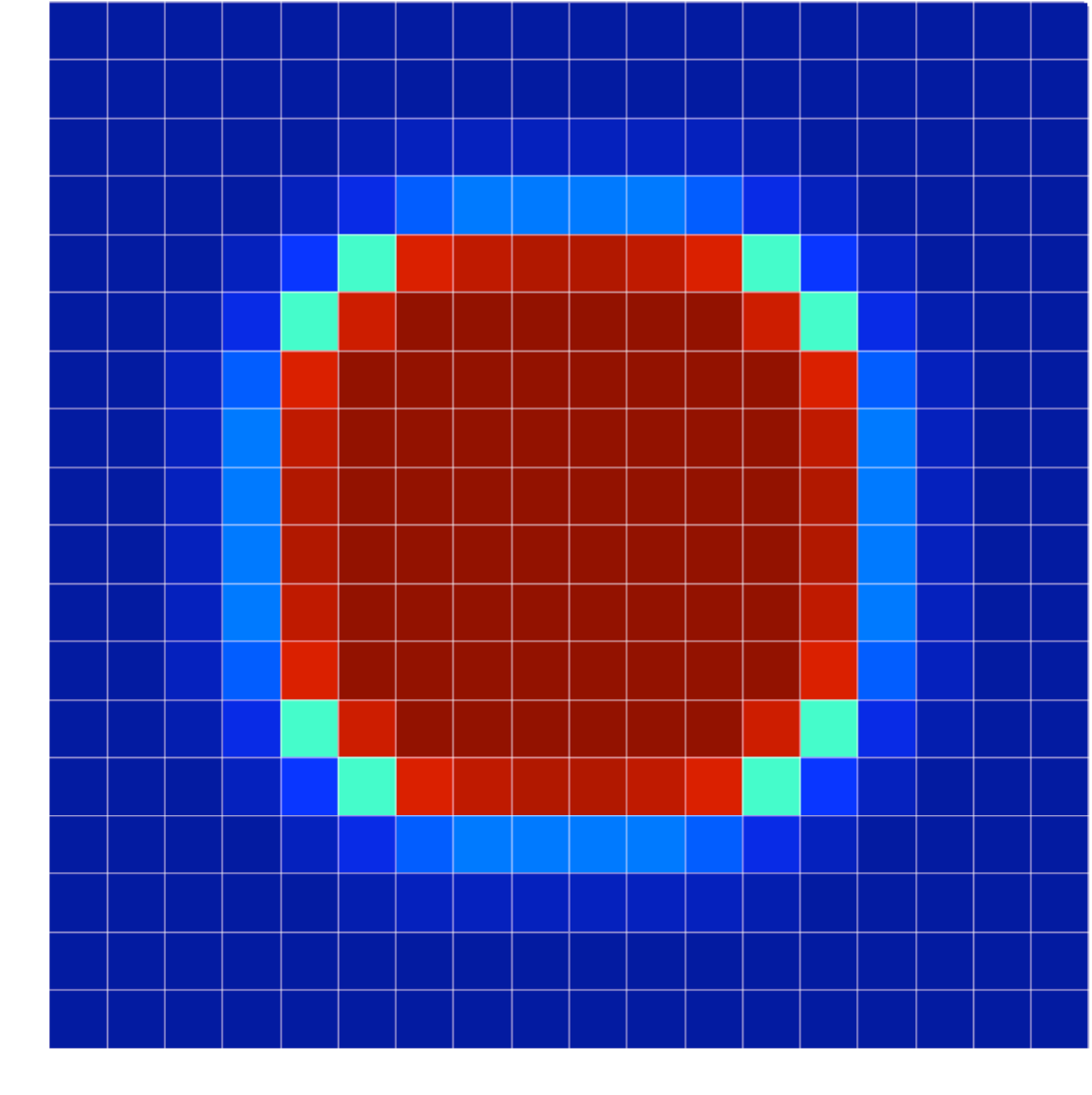}
\includegraphics[width=0.16\textwidth]{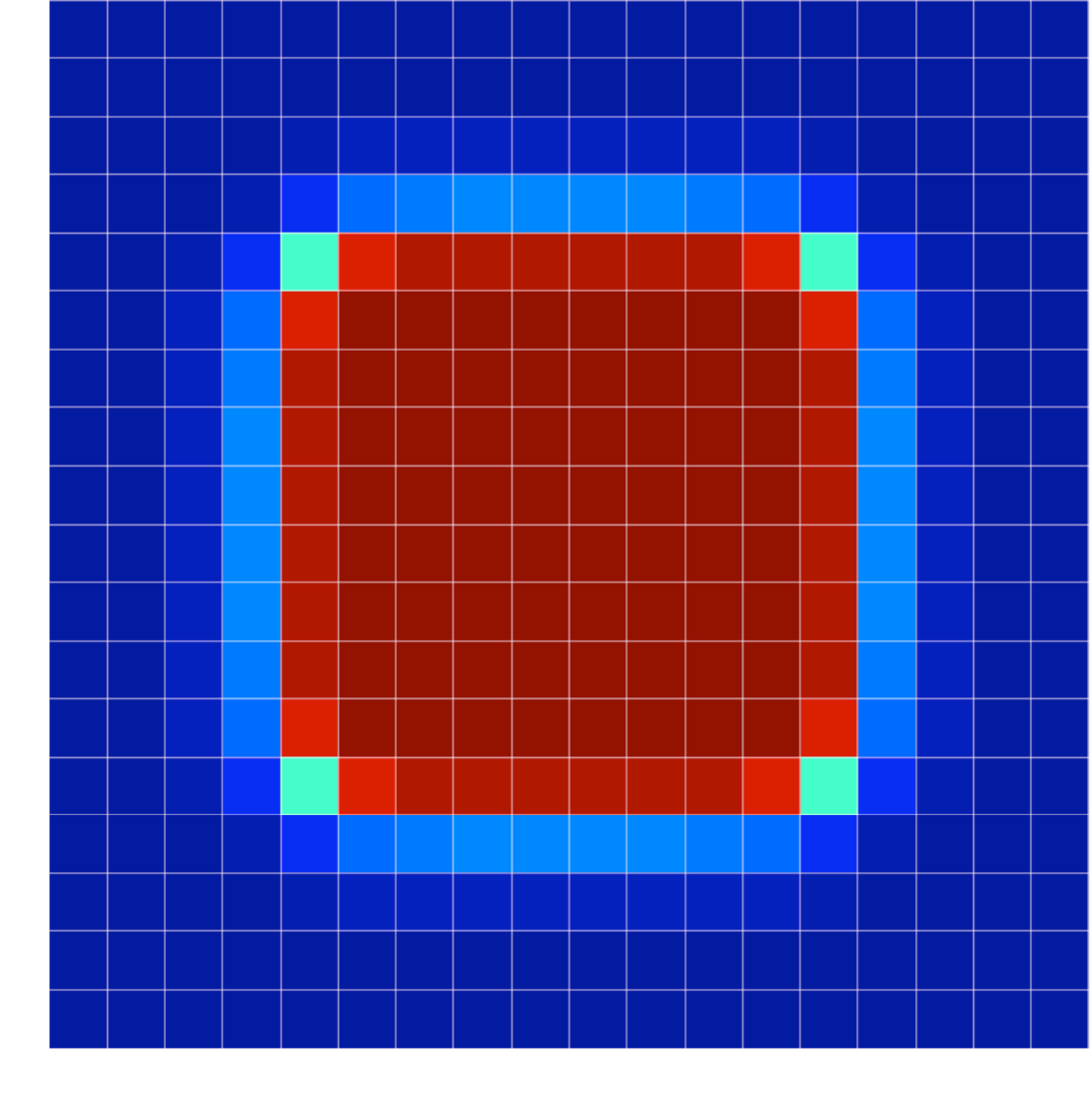}\\
\includegraphics[width=0.16\textwidth]{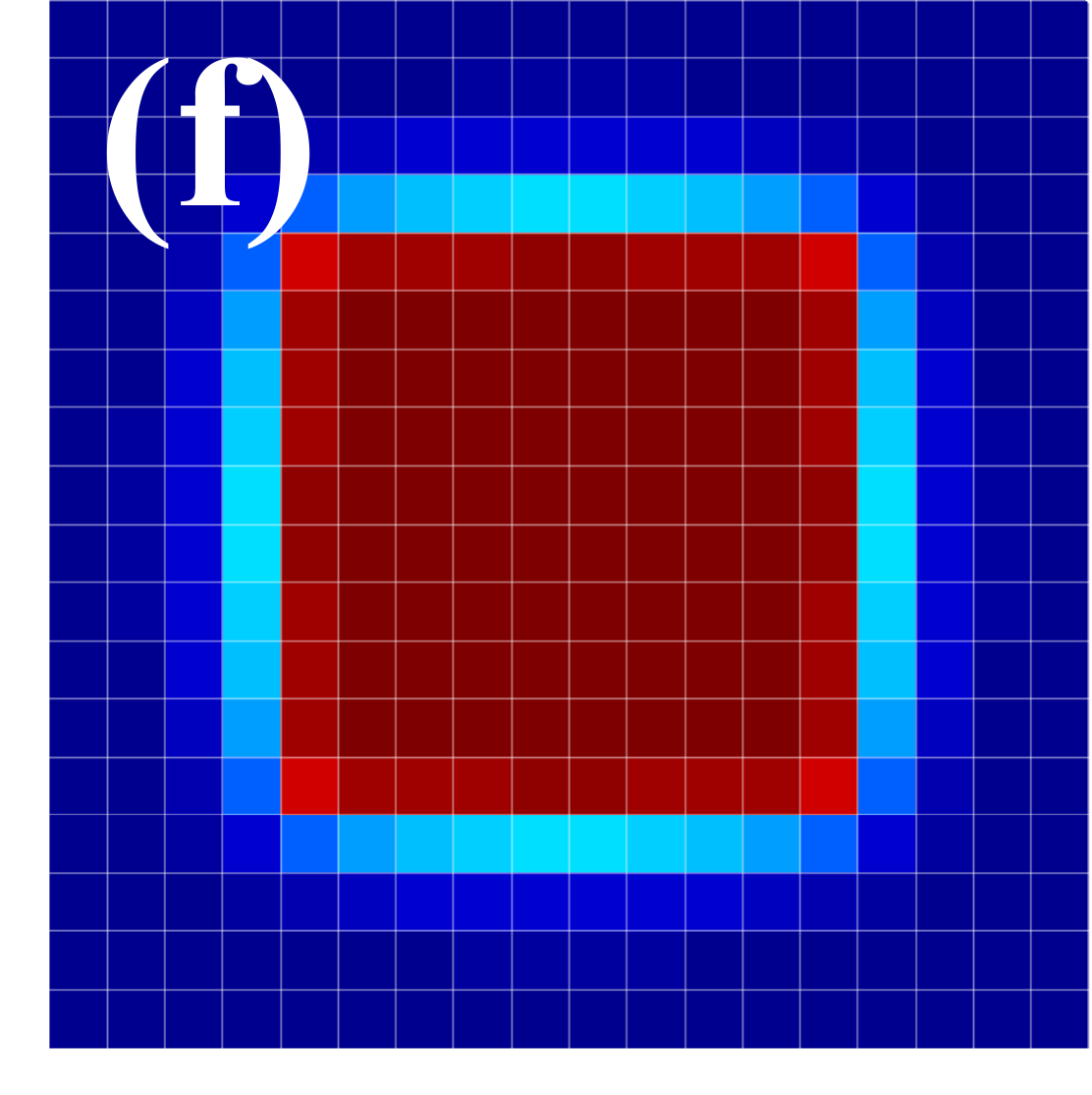}
\includegraphics[width=0.16\textwidth]{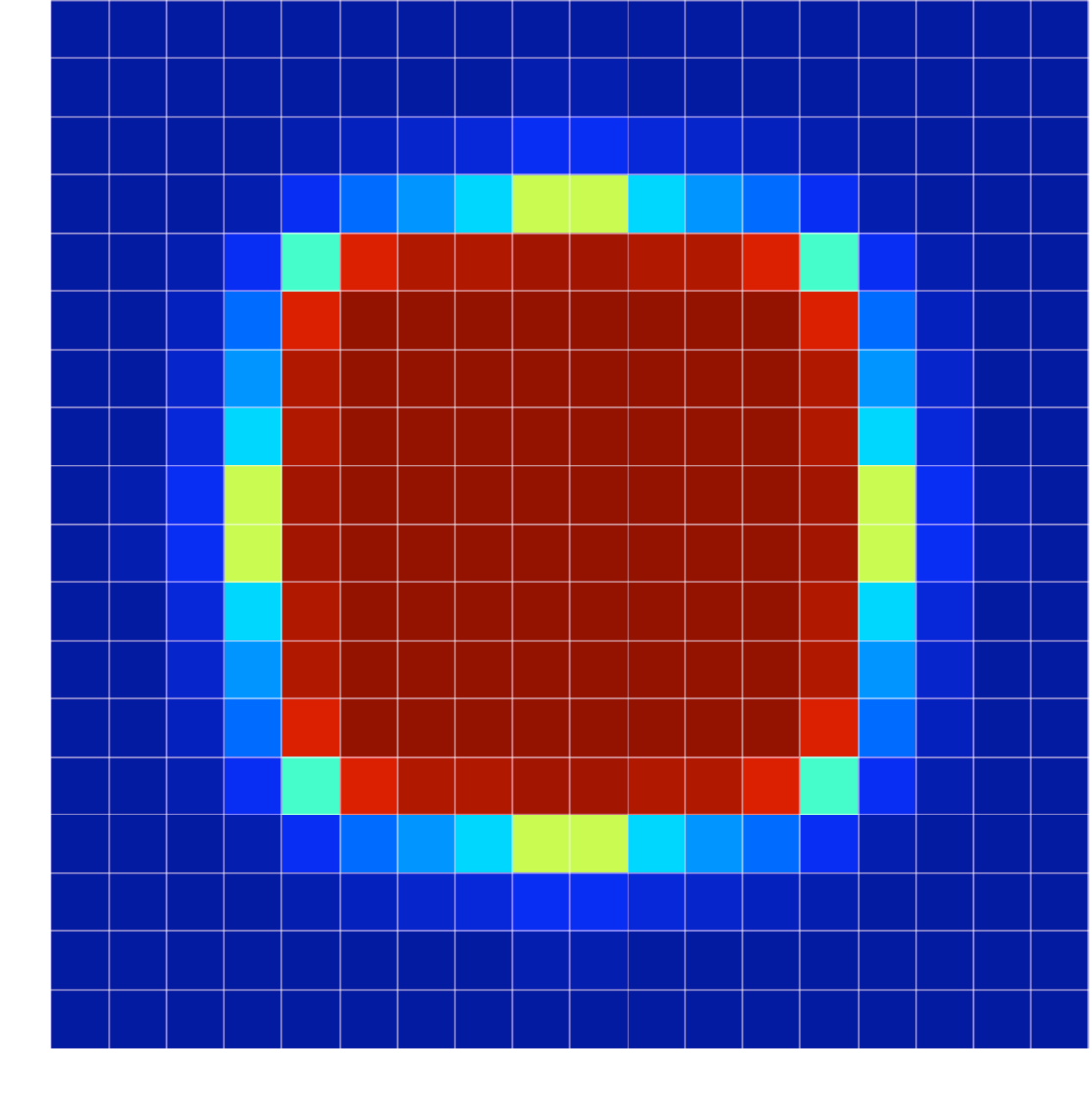}
\includegraphics[width=0.16\textwidth]{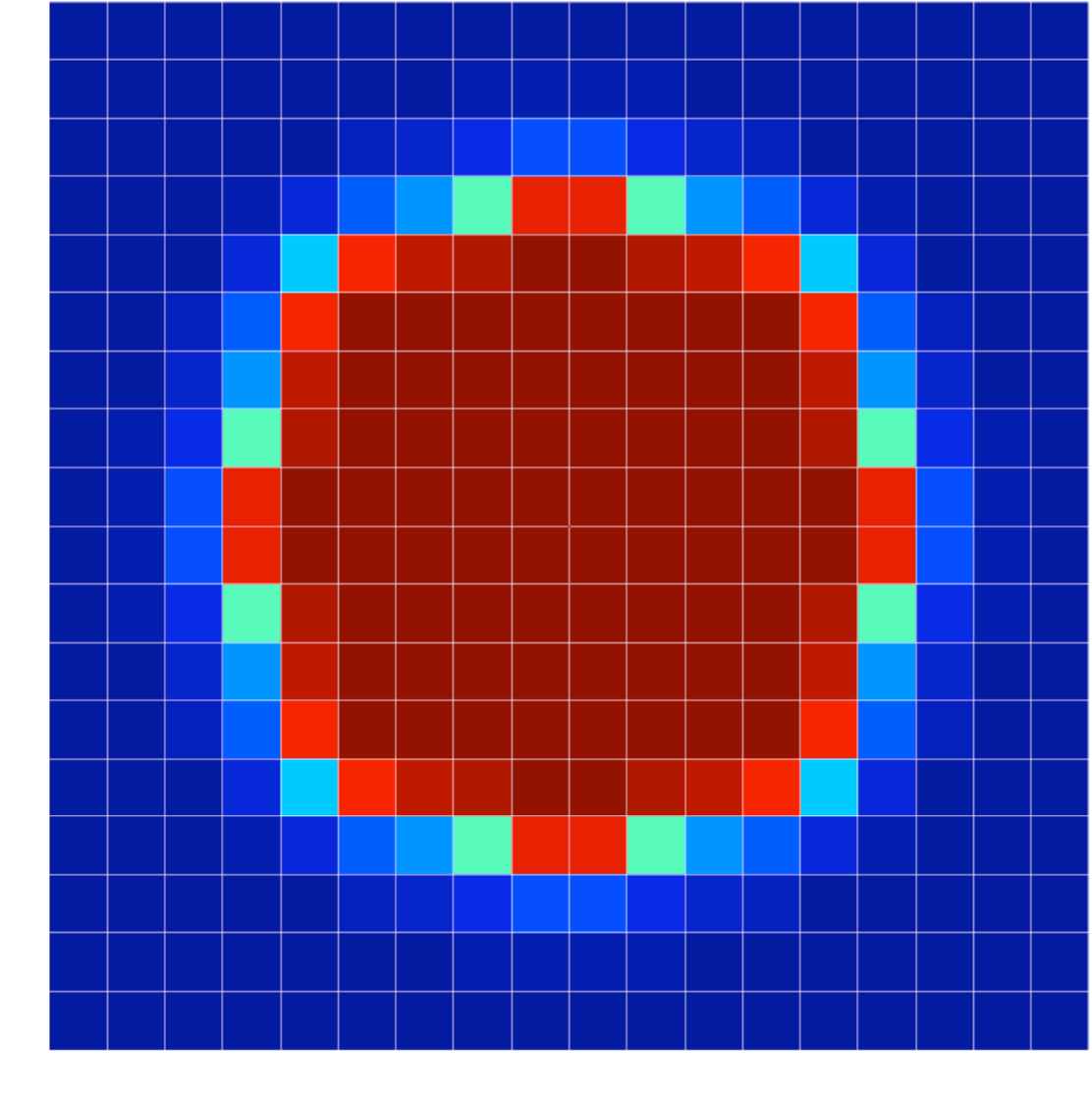}
\includegraphics[width=0.16\textwidth]{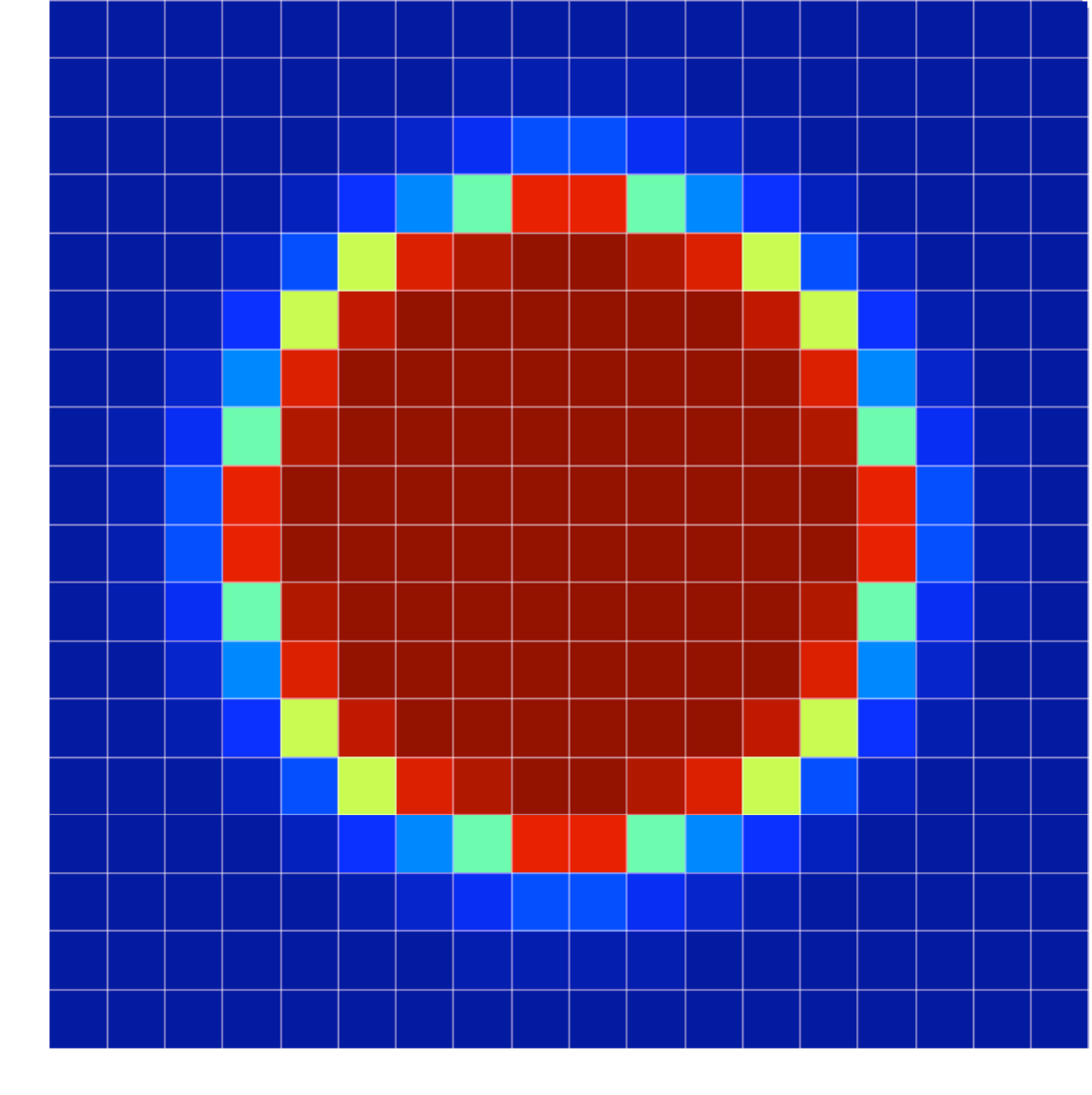}
\includegraphics[width=0.16\textwidth]{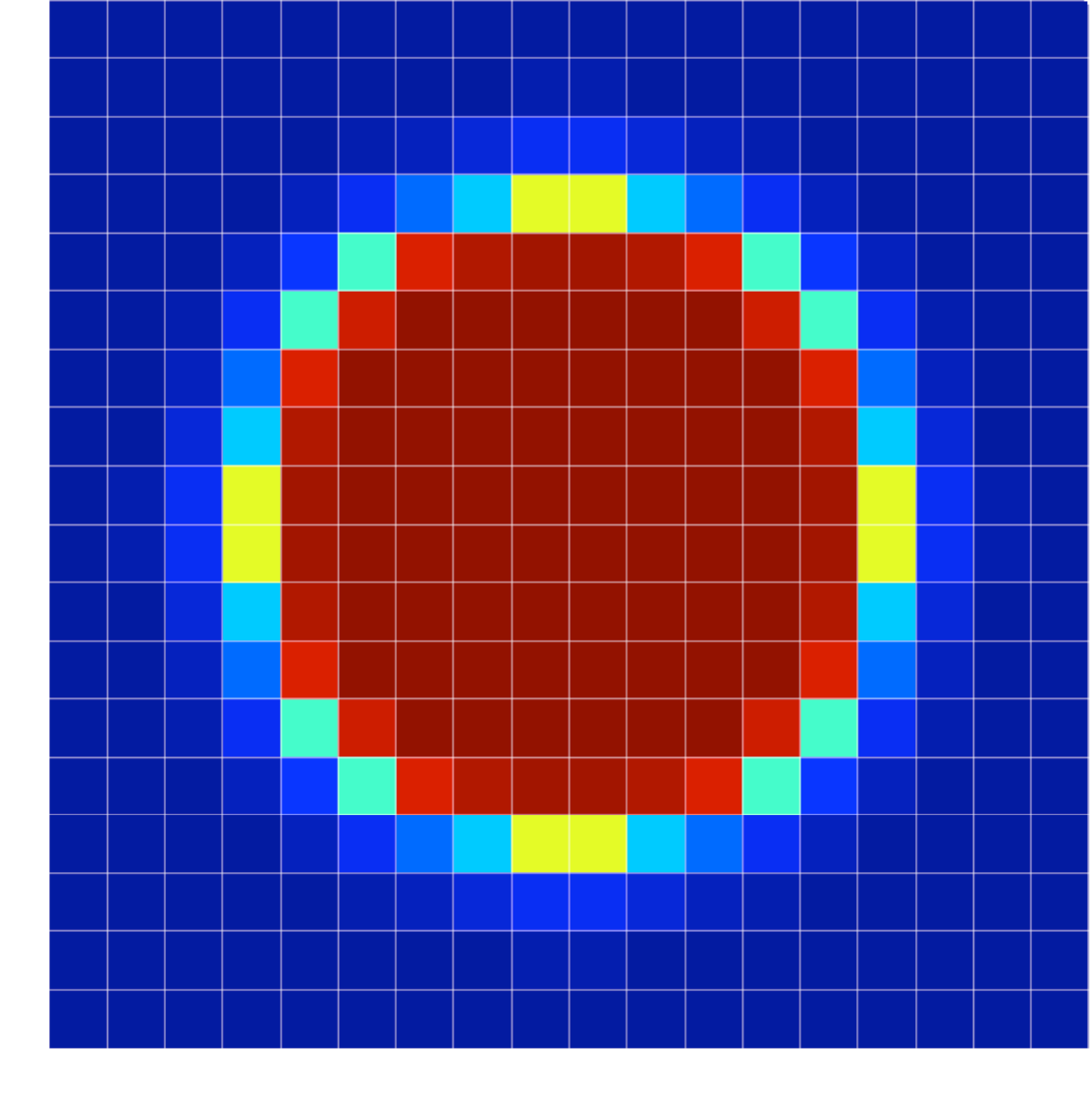}
\includegraphics[width=0.16\textwidth]{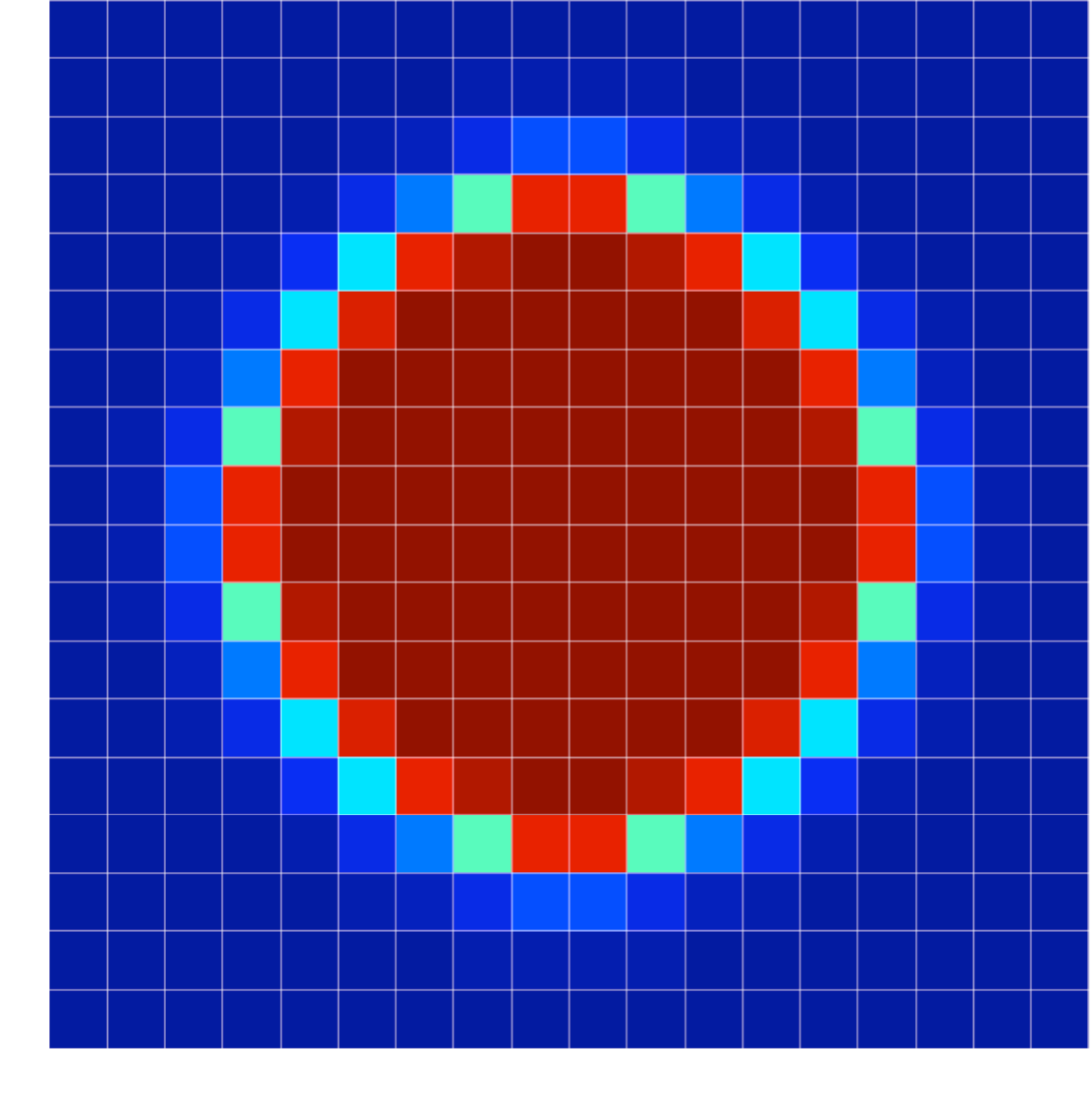}\\
\includegraphics[width=0.16\textwidth]{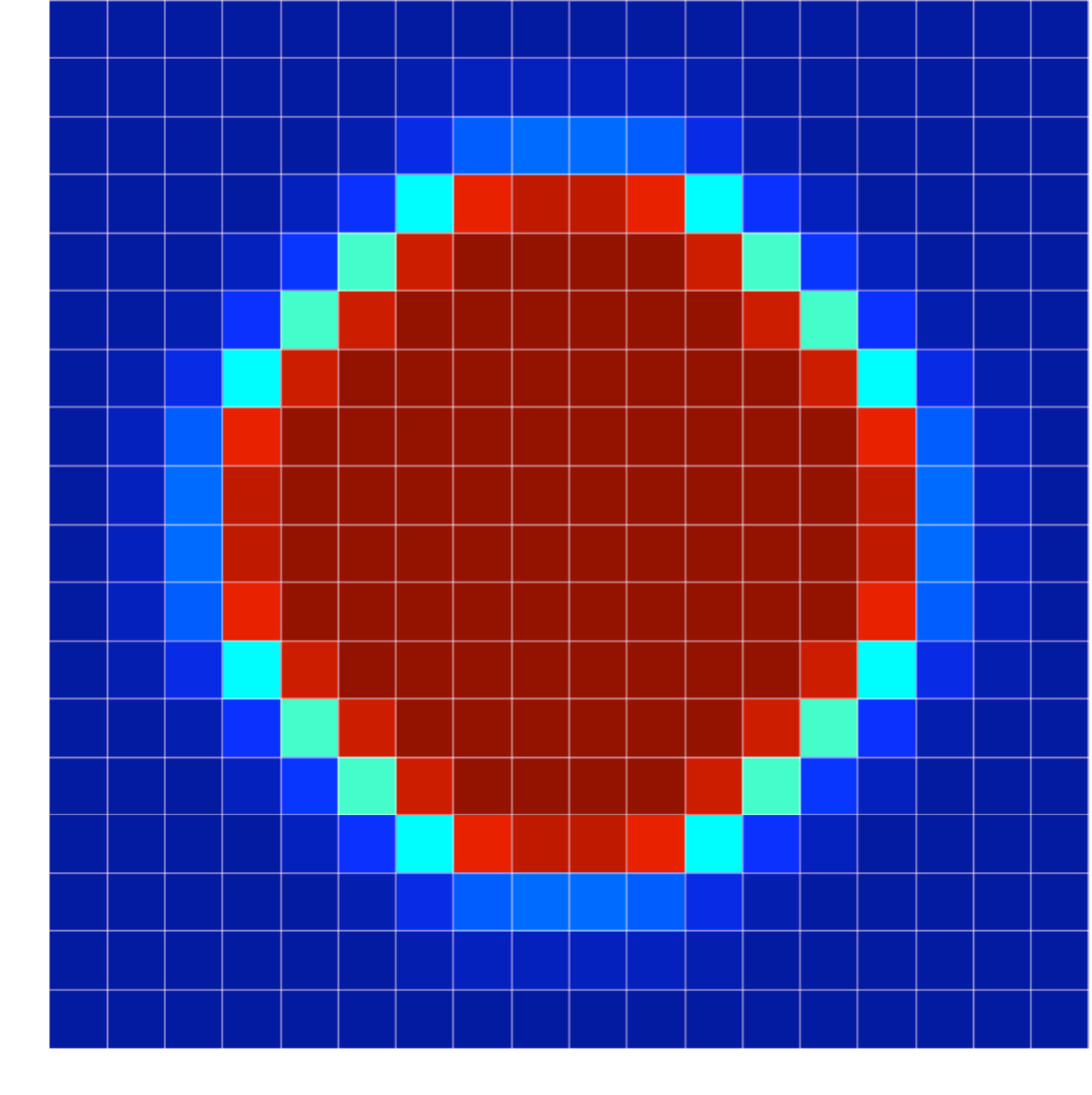}
\includegraphics[width=0.16\textwidth]{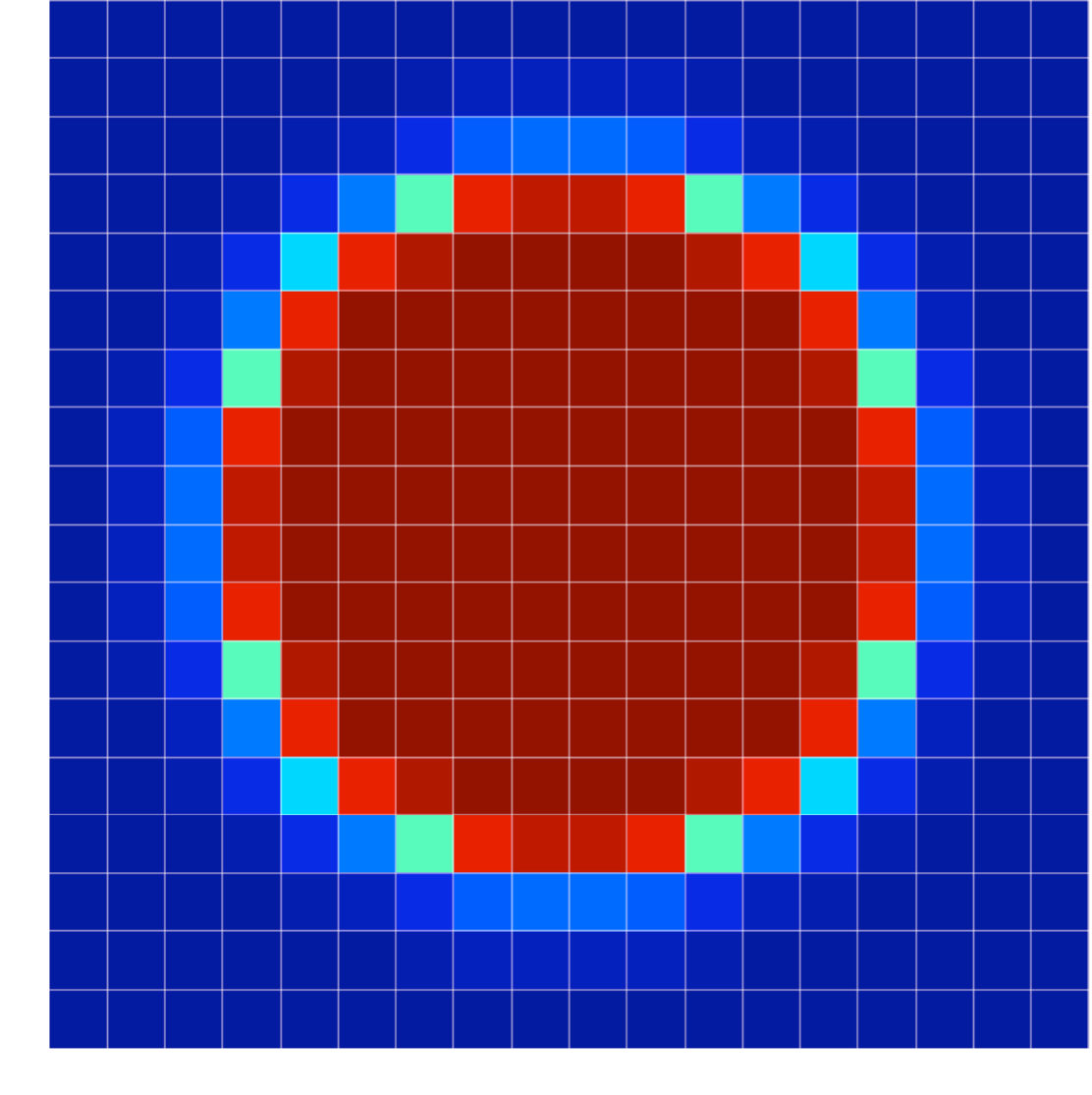}
\includegraphics[width=0.16\textwidth]{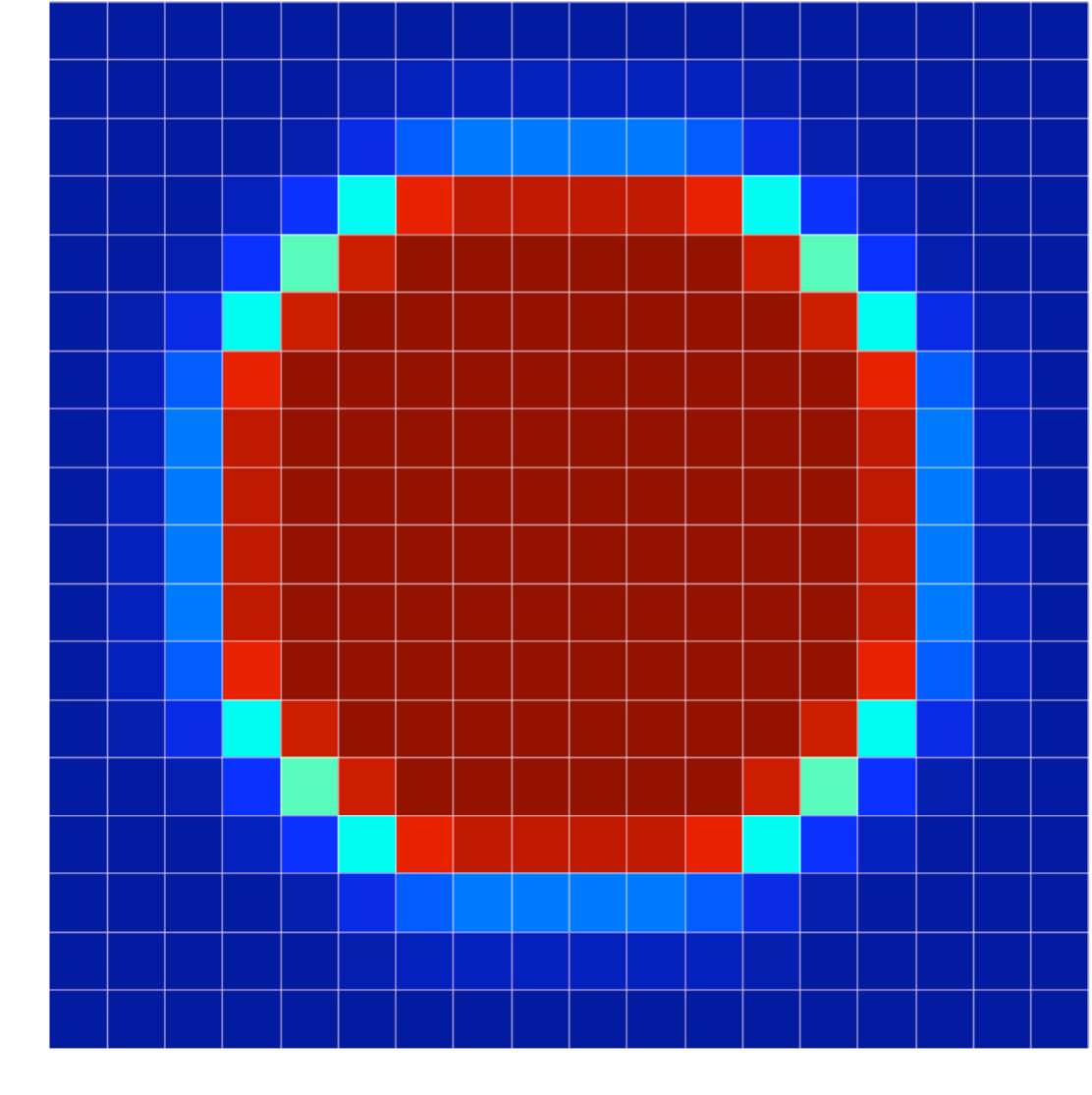}
\includegraphics[width=0.16\textwidth]{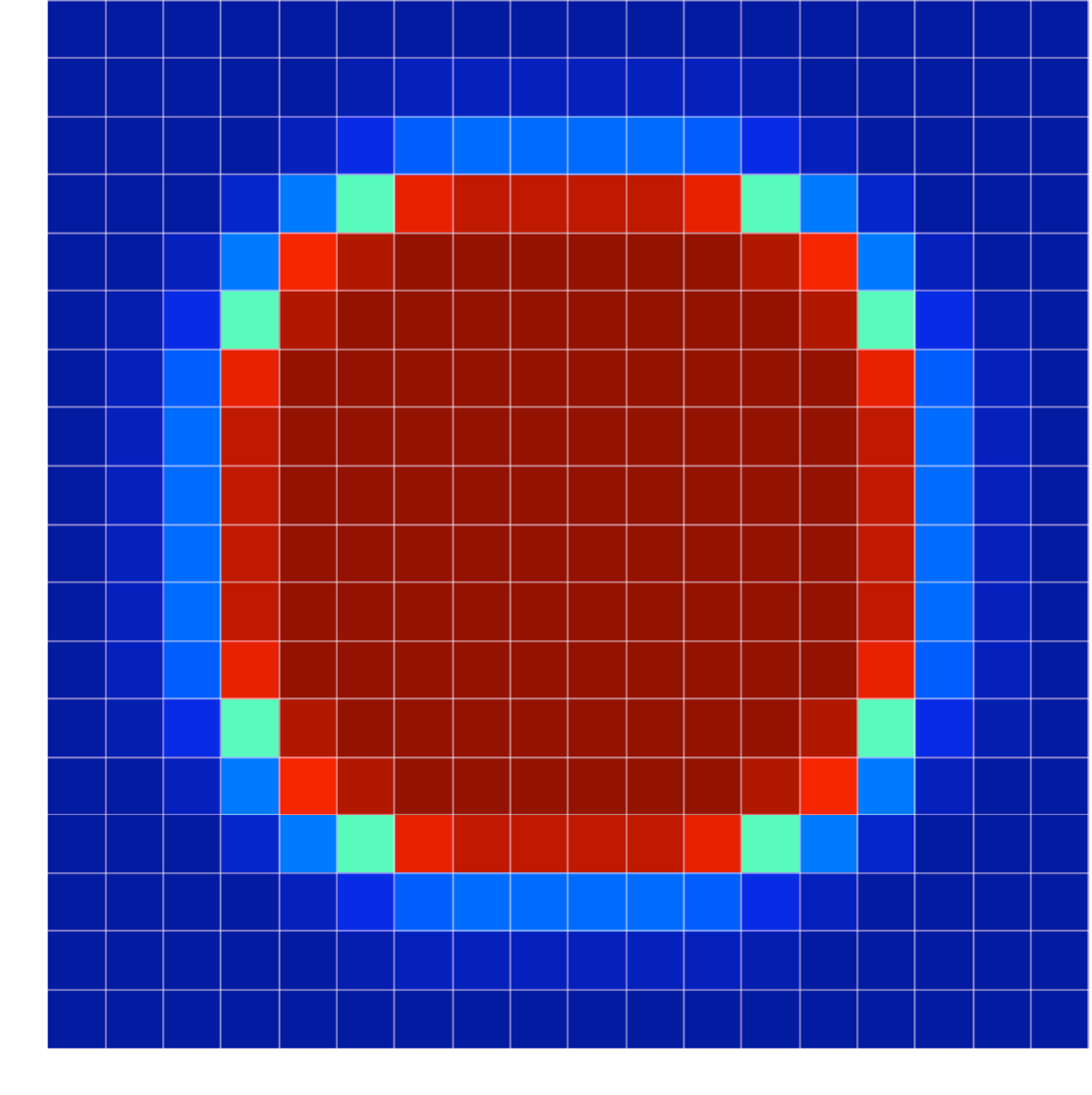}
\includegraphics[width=0.16\textwidth]{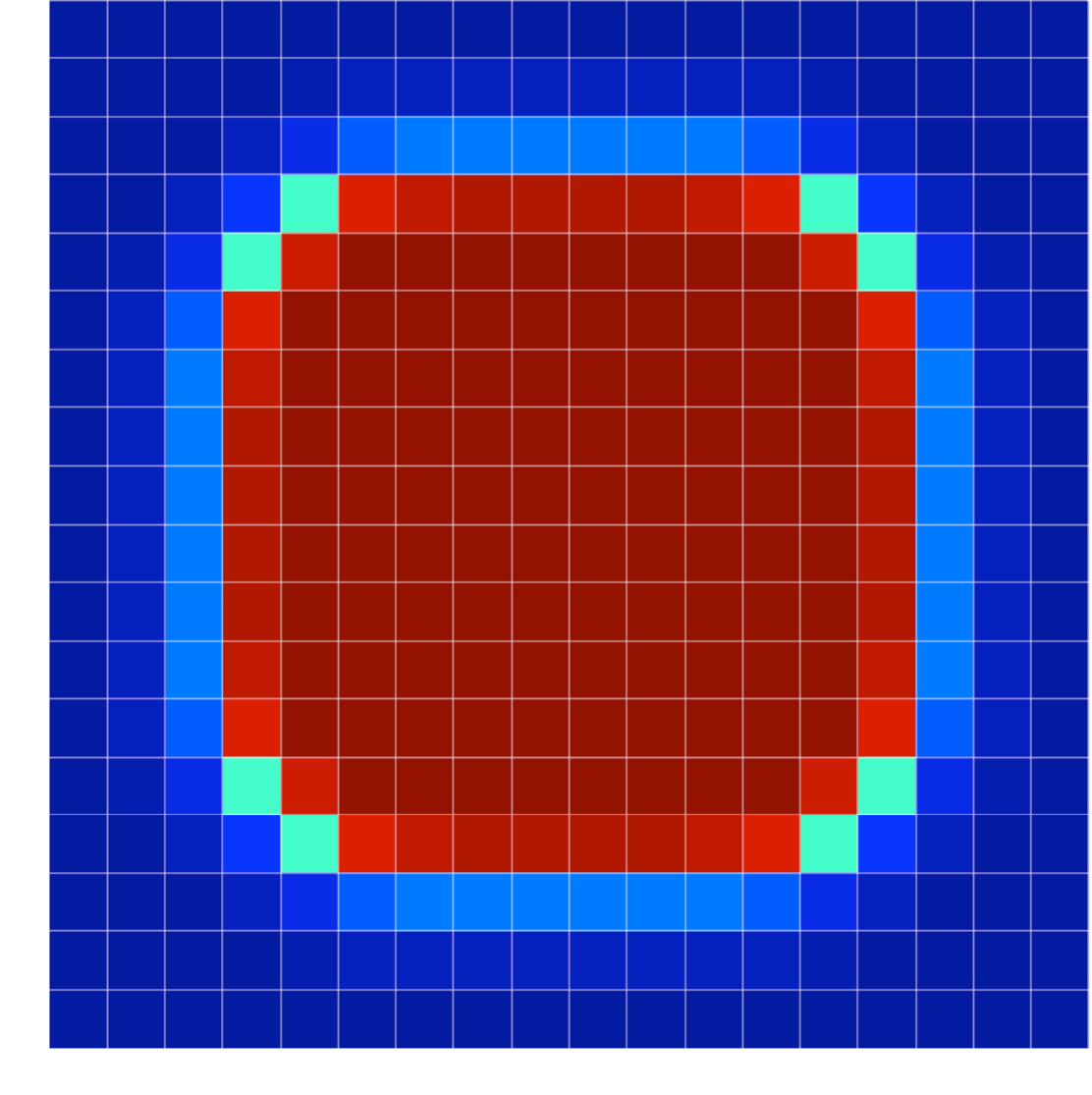}
\includegraphics[width=0.16\textwidth]{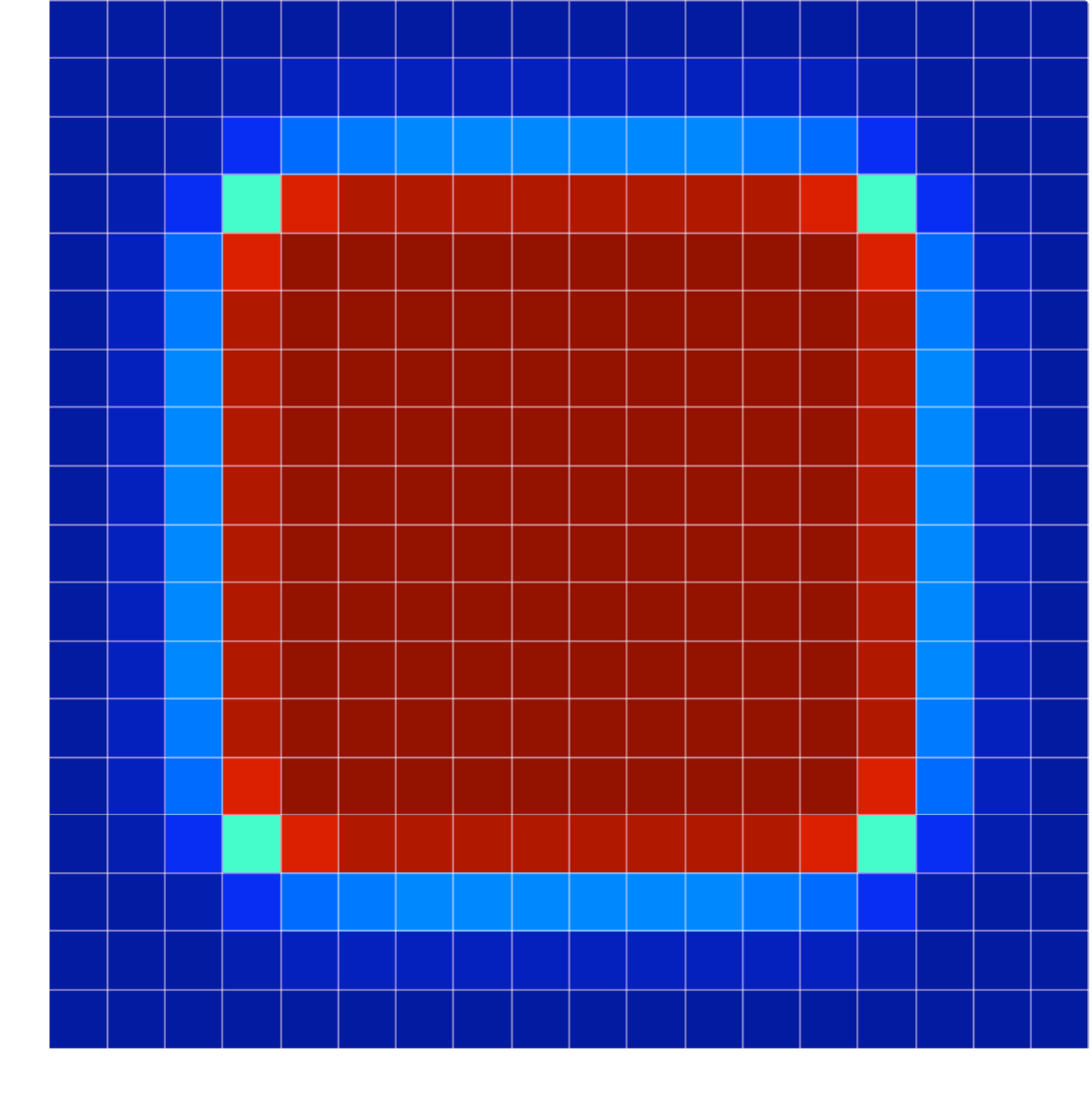}
\caption{The first 30 bond-centred localised states encountered at
saddle-node bifurcation points corresponding to
local maxima of $\mu$, moving along
the snaking curve with $M$ increasing, for fixed $C^+=0.2$.
Labels (a) - (f) correspond to the labelled saddle-node bifurcations
on the snake indicated in figure~\ref{fig:2dsnake}(b).}
\label{fig:bond-centred}
\end{figure}

\begin{figure}[!h]
\centering
\subfigure[]{
\includegraphics[width=\textwidth]{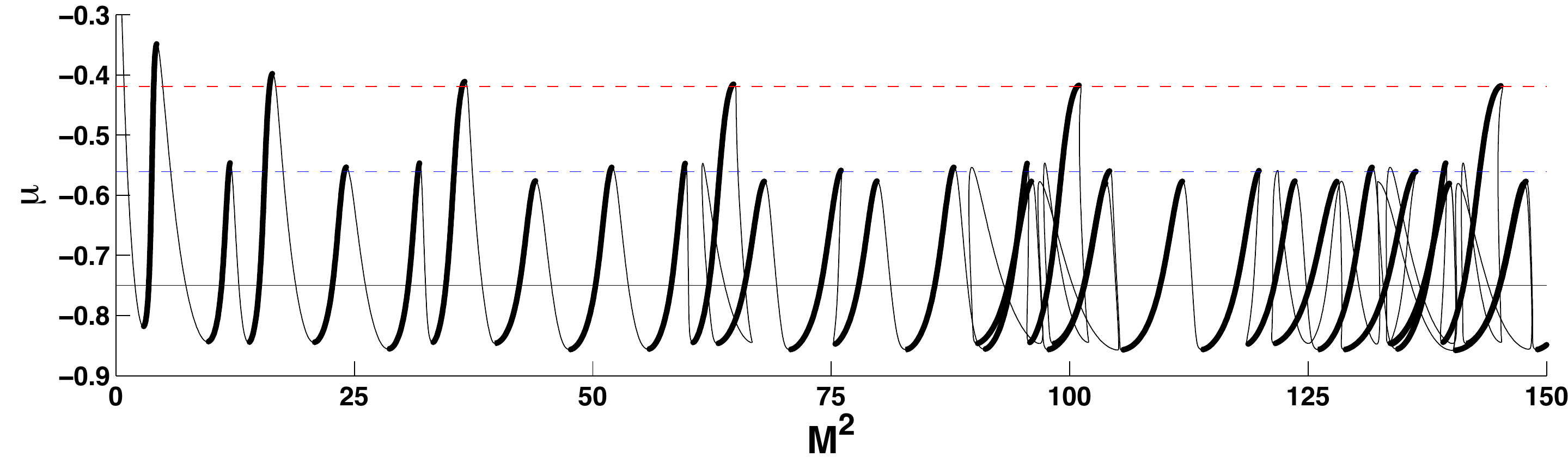}
}
\subfigure[]{
\includegraphics[width=\textwidth]{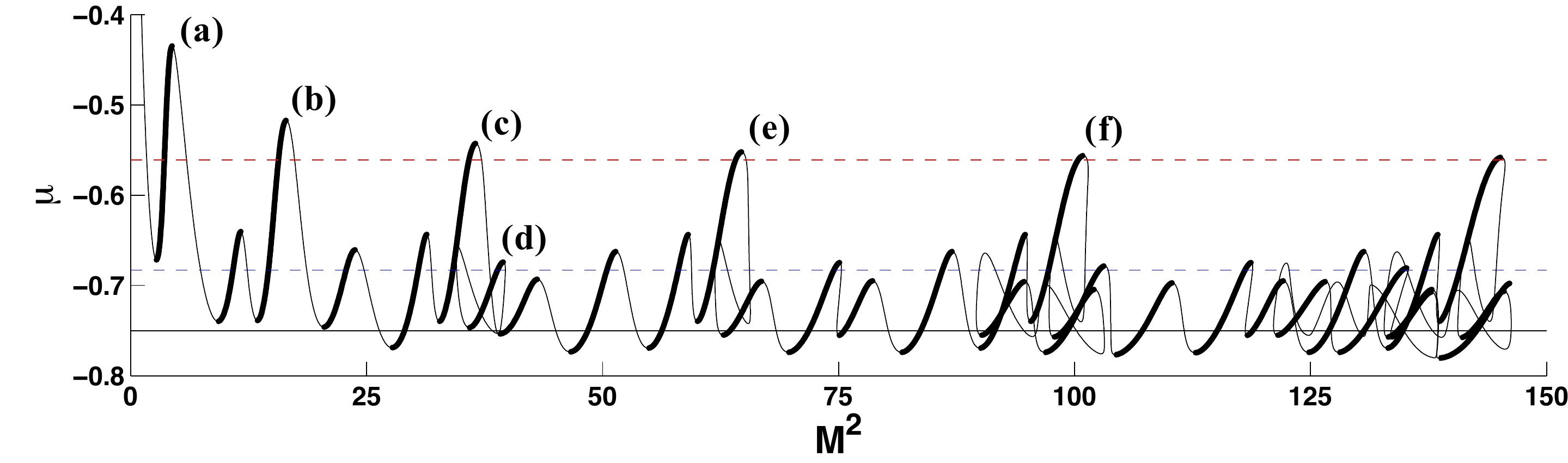}
}
\subfigure[]{
\includegraphics[width=\textwidth]{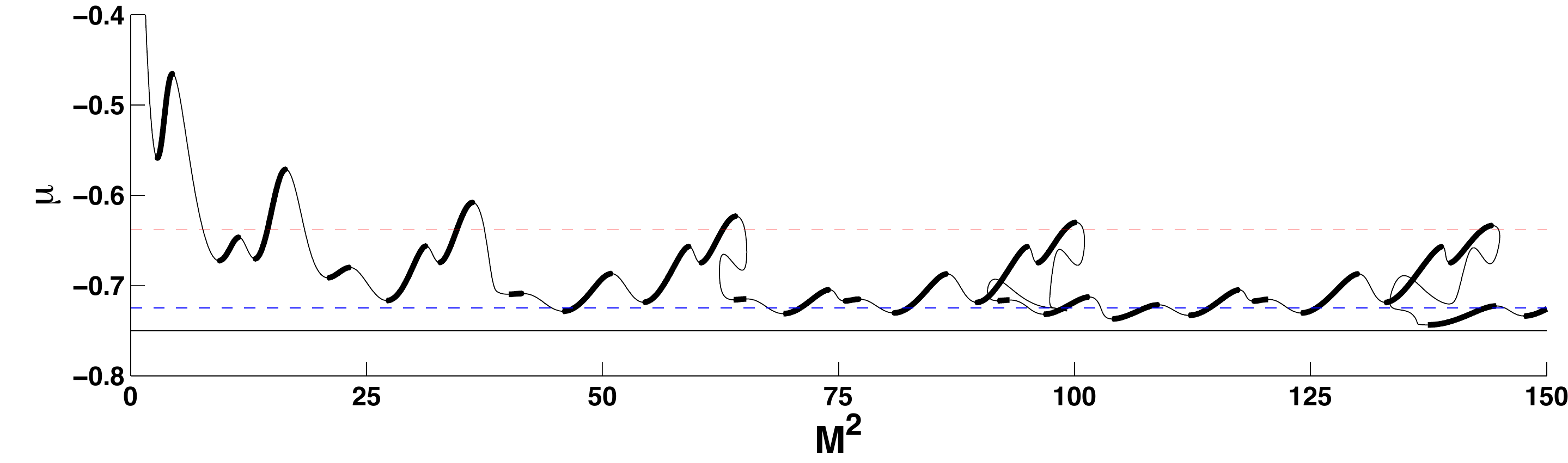}
}
\subfigure[]{
\includegraphics[width=\textwidth]{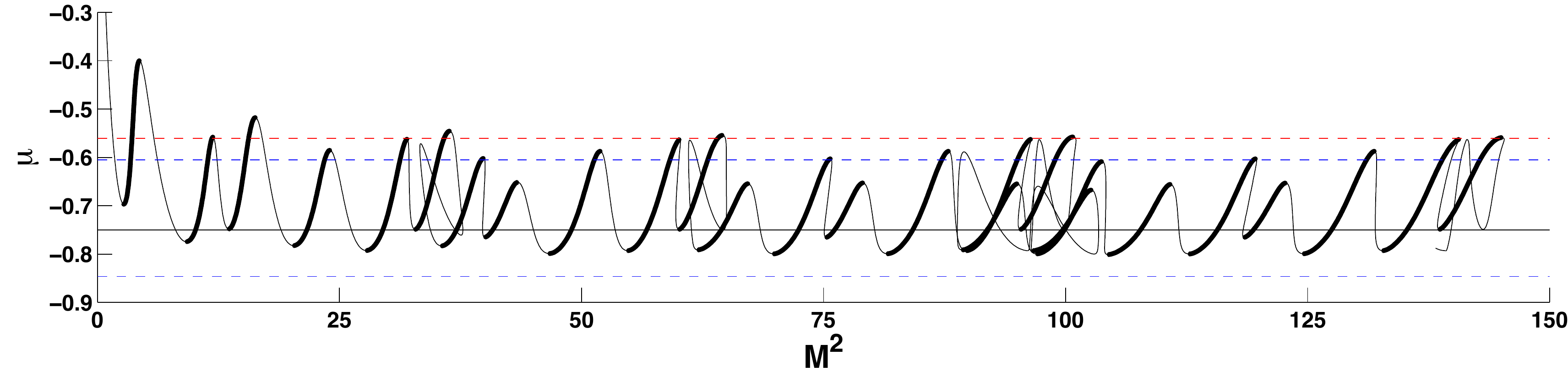}
} 
\caption{Snaking of bond-centred solutions to \eqref{eq:2d}
with (a) $C^+=0.1$, (b) $C^+=0.2$, (c) $C^+=0.3$ (in all cases
$C^\times=0$ and (d) $C^+=0.1$, $C^\times=0.05$.
Note the switchbacks in (b) occur
near $M^2=36$, $64$, $100$ and $144$,
see also figure~\ref{fig:2dsnakecloseups}.
Solid black lines indicate the Maxwell point $\mu=-0.75$.
Red horizontal dashed lines at (a) $\mu\approx-0.41925$,
(b) $\mu\approx -0.56076$, (c) $\mu\approx -0.63832$ and
(d) $\mu\approx -0.56076$
indicate the asymptotic location of saddle-node bifurcations on the
1D snake with coupling strength $C=C^+ + 2C^\times$, far up the 1D snake.
Blue horizontal dashed lines at (a) $\mu\approx-0.56076$,
(b) $\mu\approx-0.68309$, (c) $\mu\approx-0.72479$
and (d) $\mu\approx -0.60495$
indicate the asymptotic location of saddle-node bifurcations on the
1D snake with coupling strength $C=2C^+ + C^\times$. Thin and thick lines
indicate unstable and stable solutions, respectively. Labels (a)
 - (f) in part (b) correspond to the solutions labelled in
figure~\ref{fig:bond-centred}.}
\label{fig:2dsnake}
\end{figure}

The bifurcation structure of localised solutions in 2D is made
complicated by the
existence of at least three snaking curves: a localised
state can be site-centred, bond-centred or a hybrid of the two,
being site-centred in one lattice direction and bond-centred in
the other (for a more complete discussion see \cite{CG06}). Examples of
each of these types of solutions are shown in figure~\ref{fig:local2d}.
For simplicity we will now focus on bond-centred states and
follow the evolution of snaking curves of such states.
Figure~\ref{fig:bond-centred} shows the evolution of bond-centred
localised states along the continuous snaking curve shown in
figure~\ref{fig:2dsnake}(b).

\subsection{Bifurcation structure of localised states}

It is not at all clear theoretically how the well-established
theory for homoclinic snaking in 1D extends to 2D since
there does not appear to be an analogous paradigm to 
the spatial dynamics approach.
However, numerical computations show that
the bifurcation diagram
for localised states contains qualitative similarities
(see Figure~\ref{fig:2dsnake}). We remark that
to generate this figure our computations were
carried out using \AUTO~on a $20\times 20$ grid.
To accelerate the computation we simulated only a quarter
of the domain, using appropriate `reflecting'
boundary conditions to
ensure that the new states generated were bond centred.
This imposed symmetry implies that bifurcations to asymmetric 
`ladder' states could not be detected. Other computations indicate
that such bifurcations exist, and we will discuss the
existence of asymmetric states elsewhere.

The states shown in figure~\ref{fig:bond-centred} lie on the snaking
curve shown in figure~\ref{fig:2dsnake}(b) in which
the bifurcation parameter $\mu$ is plotted on the vertical axis
and the solution measure $M^2$ on
the horizontal axis. $M$ is a scaled version of the $L_2$ norm defined as:
\[
M= \left( \frac{\sum_{n,m} u_{nm}^2}{1+\sqrt{1+\mu}} \right)^{1/2}.
\]
Use of the solution measure $M$ confers the advantage
that at low coupling strengths it measures the (near-integer) number of
cells that are `active' in the localised state. Figure~\ref{fig:2dsnake}(a)
indicates that `high peaks' (e.g. the saddle-nodes labelled (a), (b), (c), (e)
and (f) at which the value of $\mu$
is markedly less negative than at most saddle-nodes)
are closely aligned with even integer values
of $M$. These labels correspond to the labelled
localised states in figure~\ref{fig:bond-centred}
that are in the form of complete $2k \times 2k$ square arrays
of active cells.

\begin{figure}
\centering
\subfigure[]{
\includegraphics[width=0.31\textwidth]{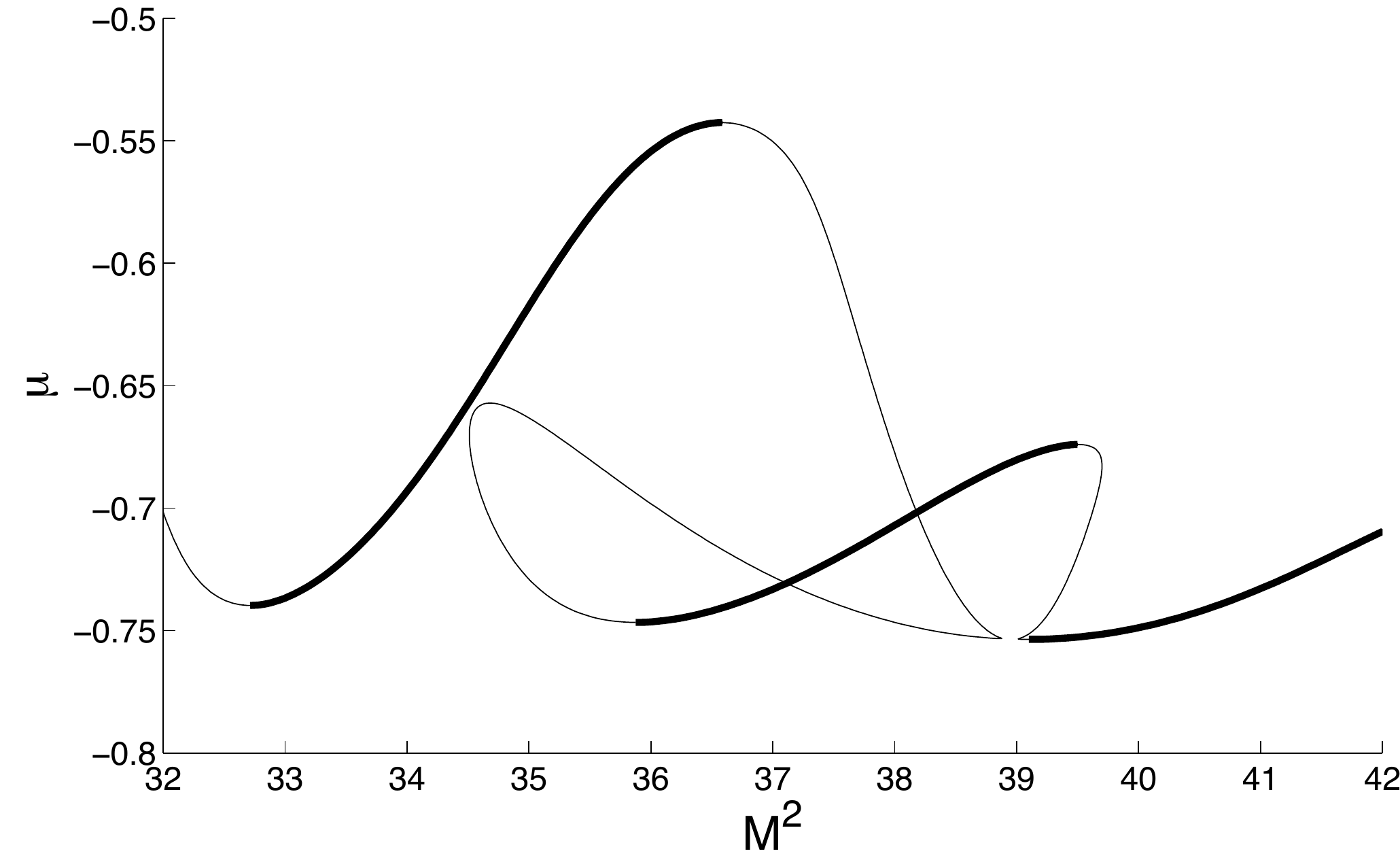}
	}
\subfigure[]{
\includegraphics[width=0.31\textwidth]{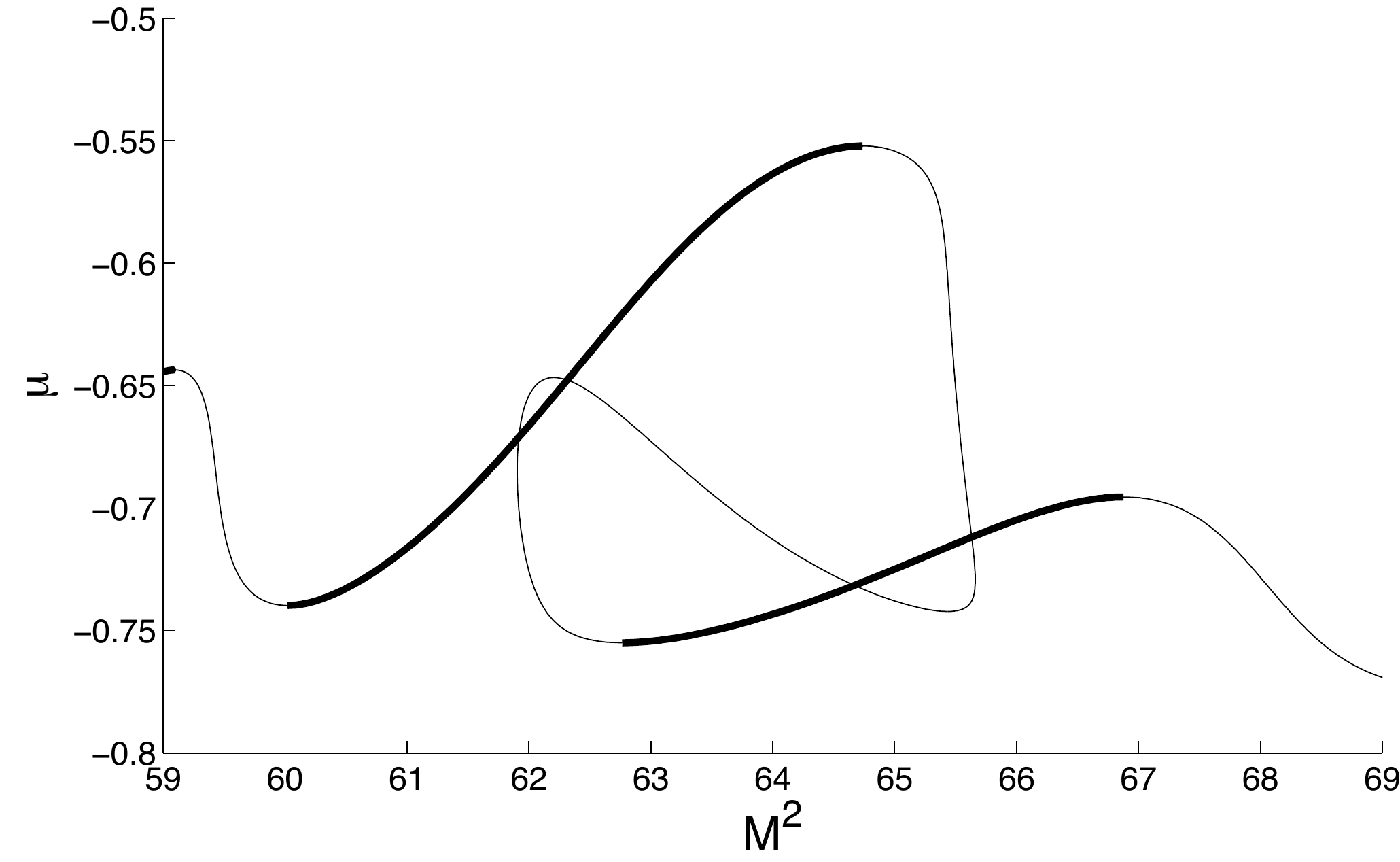}
	}
	\subfigure[]{
\includegraphics[width=0.31\textwidth]{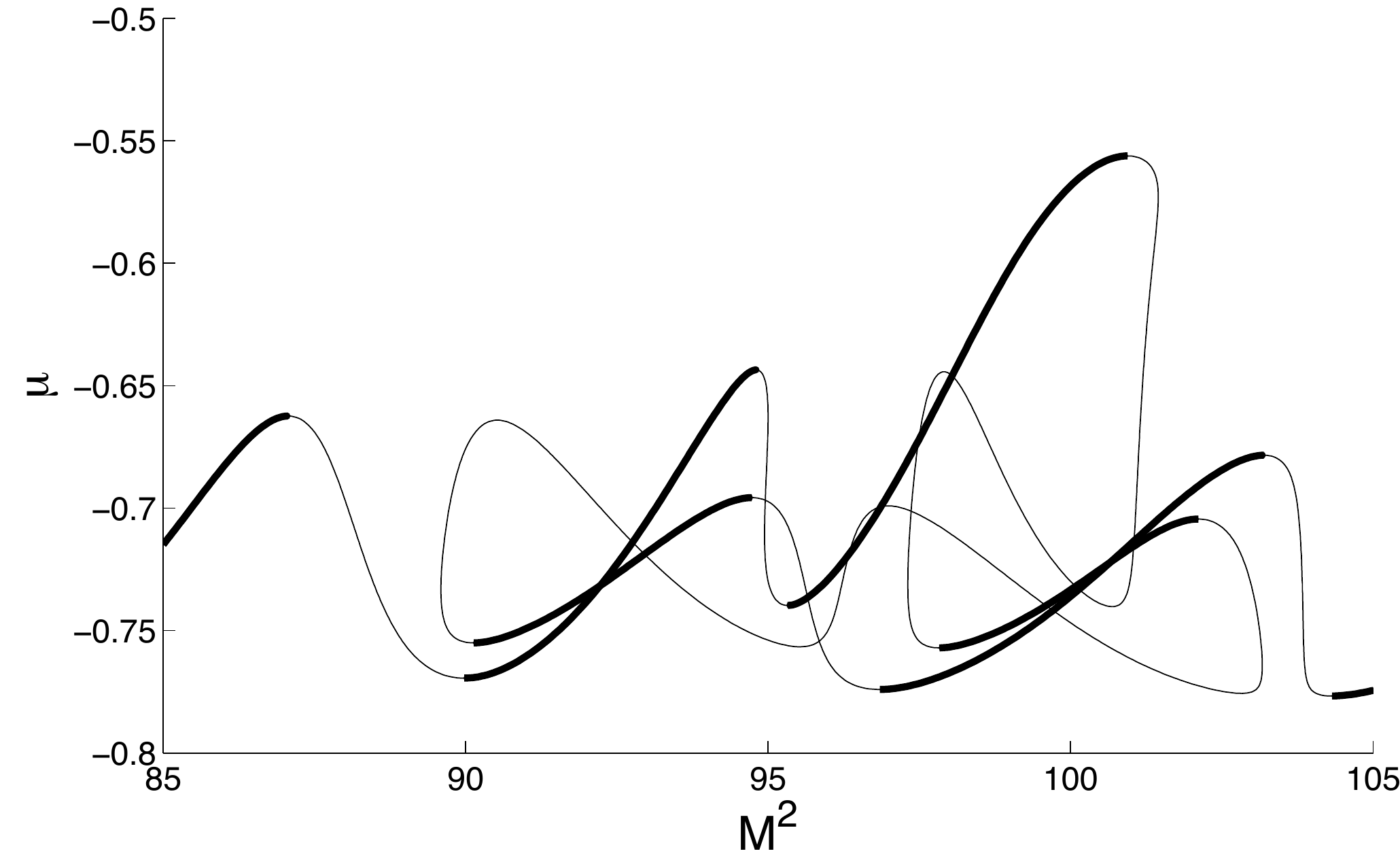}
	}
\caption{(a) Enlargement of figure~\ref{fig:2dsnake}(b) at the first
switchback, near $M^2=36$; (b) a second switchback near $M^2=64$;
(c) a much more complicated switchback near $M^2=100$.}
\label{fig:2dsnakecloseups}
\end{figure}

The evolution of bond-centred localised states as we move up the
snaking curve is shown in figure~\ref{fig:bond-centred}.
Starting from~\ref{fig:bond-centred}(a) in which four cells are
close to $u_+$ and the rest are near zero,
new $u_+$ cells are added to the solution profile
starting in the centre of the faces, and then progressing outwards
towards the corners until the profile is square, at which
point the sequence repeats itself. Numerical continuation suggests
that it is possible to proceed arbitrarily far up the
snaking curve in this manner as long as the coupling coefficient
is sufficiently small. There are similar snaking curves of site-centred
and hybrid localised states which for brevity, and because it
is not central to our discussion of isolas, we will not discuss here.

\subsection{Snaking limits}

The horizontal lines in each part of figure~\ref{fig:2dsnake}
indicate
the locations of saddle-node bifurcations far up the 1D snake for
two values of the coupling coefficient: $C=C^+ + 2C^\times$
(red) and $C=2C^+ + C^\times$ (blue).
These values are determined from the part of the blue curve
in figure~\ref{fig:muC} which lies in $\mu>-0.75$.

At small coupling strengths it is clear
numerically in figure~\ref{fig:2dsnakecloseups}(a) that,
as $M^2$ increases and we move far up the snaking curve,
the width of the snaking region is given overall by $C=C^+$ since
the existence of square $2k \times 2k$ arrays of localised states
is determined by the existence of stationary fronts aligned with
a lattice direction. This is easy to see from the lattice
equations~\ref{eq:2d})
by removing the dependence on the
second coordinate $m$, i.e. setting $u_{nm}=u_n$ we
recover~(\ref{eq:1d})
with $C=C^{+}+2C^\times$.
All other saddle-node bifurcations involve localised
states which are not complete squares and therefore involve
stationary diagonal fronts between $u_+$ and zero. Substituting
the ansatz $u_{nm}=u_{\ell}$ where $\ell=n+m$ into~(\ref{eq:2d})
we obtain
\beq \label{eq:2dreduced}
\dot{u}_\ell = 2C^+(u_{\ell+1}+u_{\ell-1}-2u_\ell) + C^\times(u_{\ell+2}+u_{\ell-2}-2u_\ell) + \mu u_\ell + 2u_\ell^3 - u_\ell^5.
\eeq
Hence, in the case $C^\times=0$ this reduces exactly to~(\ref{eq:1d})
with $C=2C^+$. When $C^\times \neq0$ and the coupling coefficients
are small (and hence the front is sharp compared to the lattice spacing)
we anticipate that $u_{\ell+2}\approx u_{\ell+1} \approx u_+$ on one
side of the front and $u_{\ell-2}\approx u_{\ell-1}\approx 0$ on
the other and so~(\ref{eq:2dreduced}) can be well approximated
by~(\ref{eq:1d}) setting $C=C^{+}+2C^\times$.

Figure 6(d) shows the bond-centred snake with $C^+ = 0.1$ and $C^\times = 0.05$. Qualitatively there is little
difference between this and parts (a) and (b). We observe two
distinct scales in the bifurcation diagram,
with the widths of turns on the snake corresponding roughly to
those in the 1D model with $C = C^+ +2C^\times$
or $C = 2C^+ + C^\times$. We anticipate that for $C^\times$ small
compared with $C^+$ there will be no significant
differences from the $C^\times = 0$ case, but numerical continuation
suggests that larger $C^\times$ can alter the snaking
curve significantly, especially when $C^\times > C^+$.
We should also note that with two independent coupling
parameters there is no equivalent to the staggering transformation
presented for the 1D model.

\subsection{Switchbacks}

As the coupling strength $C^+$ increases the snaking curves
quantitatively depart from the
1D case due to the existence of regions in the bifurcation diagram
where the snaking curve turns back on itself and the $L_2$ norm
(or, equivalently, $M$) of the localised state descreases for a
number of twists before turning back
again and continuing to snake upwards. We refer to such an episode
as a `switchback'.
Examples of switchbacks are shown in figure~\ref{fig:2dsnakecloseups}.
In part (a) of figure~\ref{fig:2dsnake} switchbacks are
observed to begin only at $M^2\approx 64$,
$100$ and $144$.
In (b) an additional switchback appears at $M^2\approx 36$.
In (c) almost all the twists in the switchbacks have disappeared, although
the bifurcation curves near $M^2=64$, $100$ and $144$ show clear
similarities.

\begin{figure}
\centering
\subfigure[$C^+=0.19$]{
\includegraphics[width=0.48\textwidth]{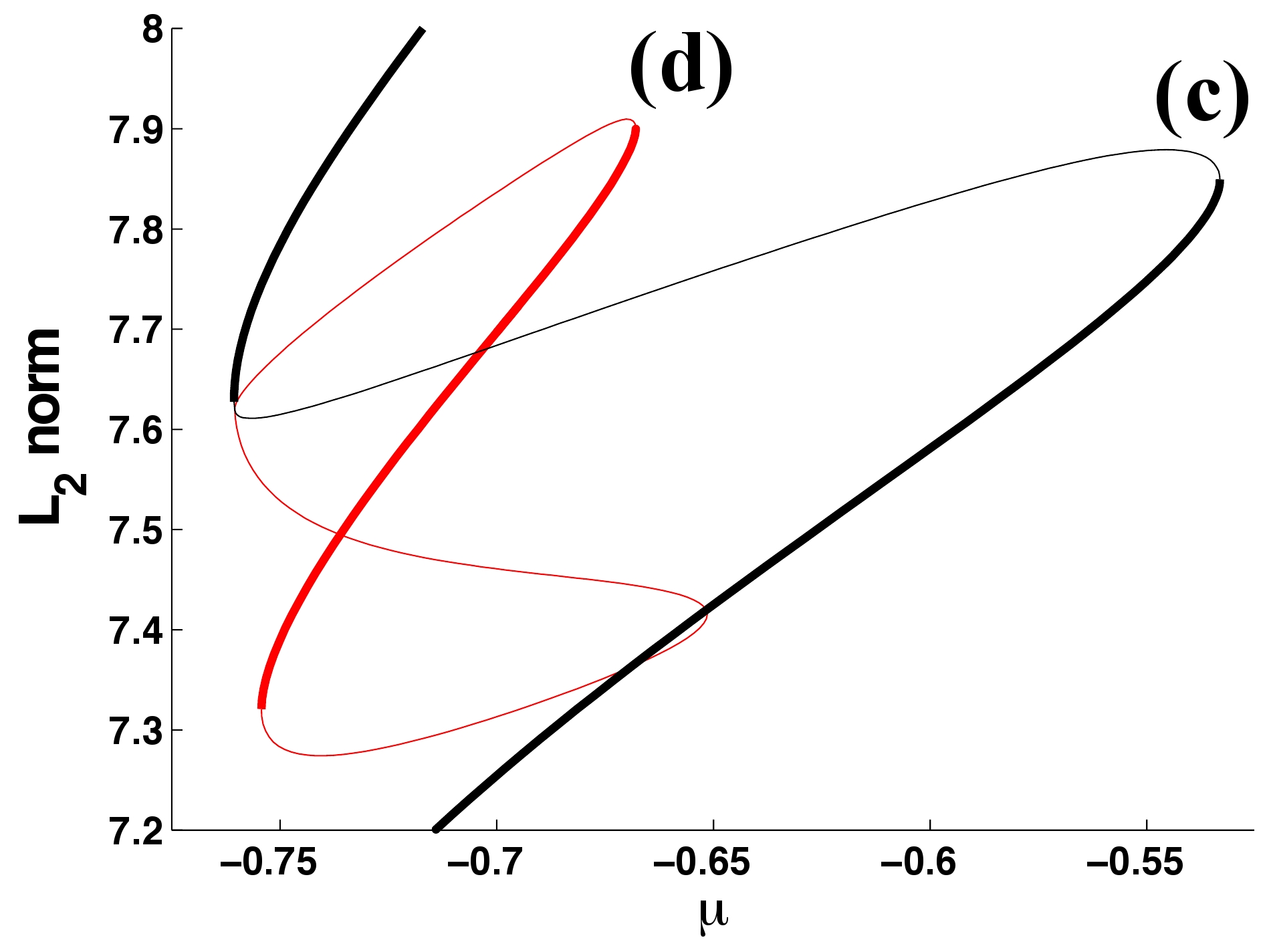}
	}
\subfigure[$C^+=0.21$]{
\includegraphics[width=0.48\textwidth]{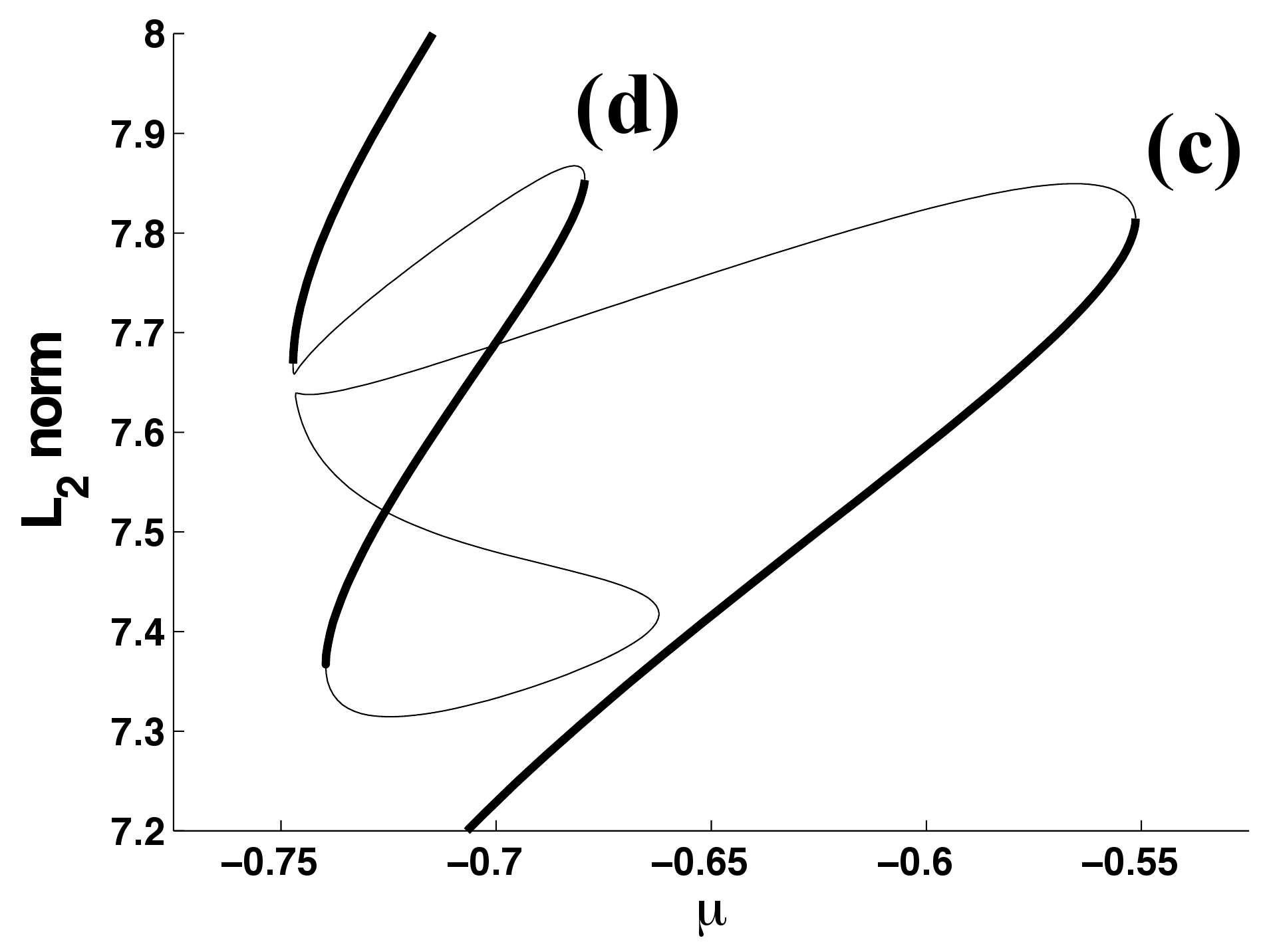}
	}
\subfigure[$C^+=0.27$]{
\includegraphics[width=0.48\textwidth]{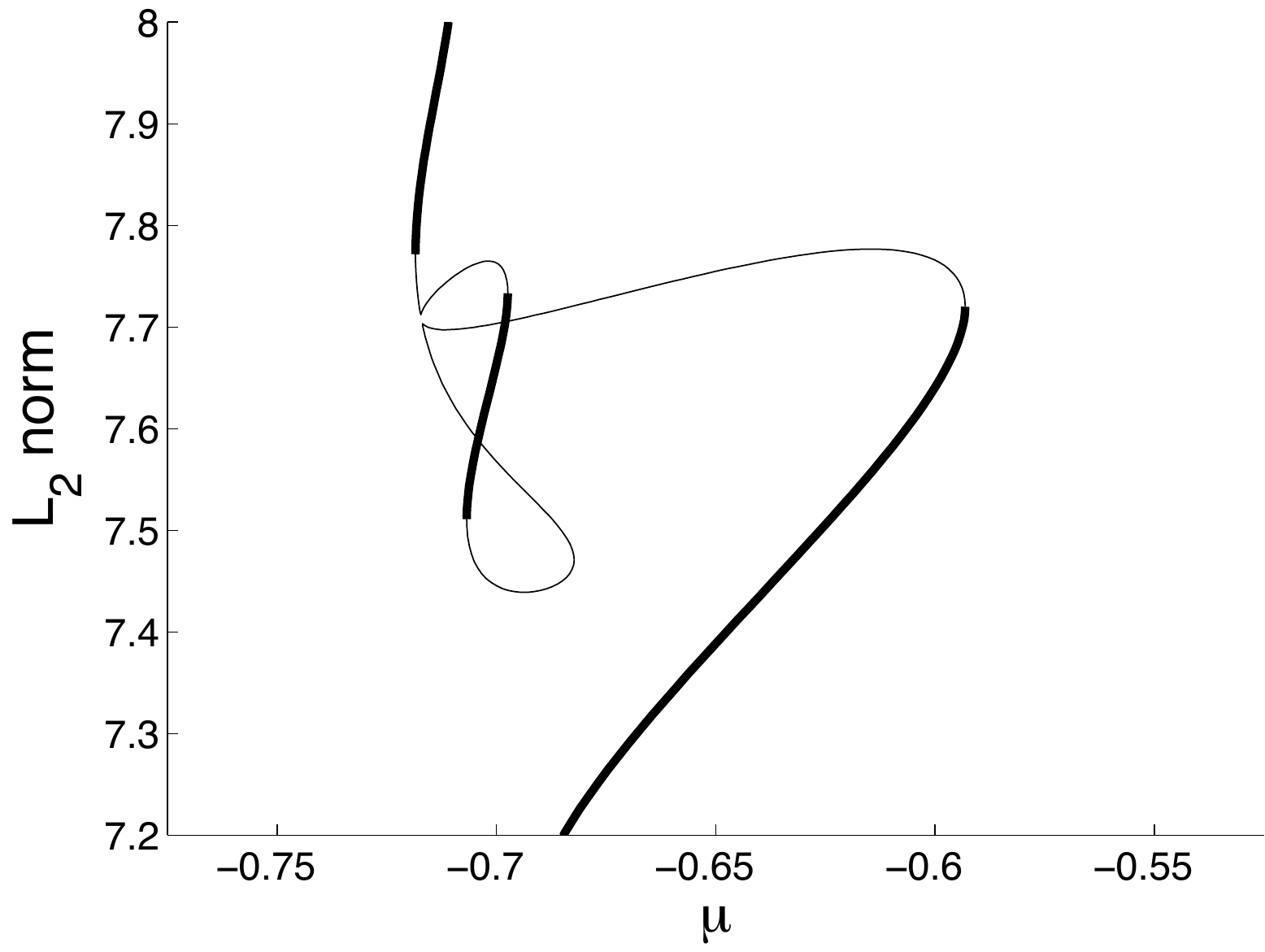}
	}
\subfigure[$C^+=0.28$]{
\includegraphics[width=0.48\textwidth]{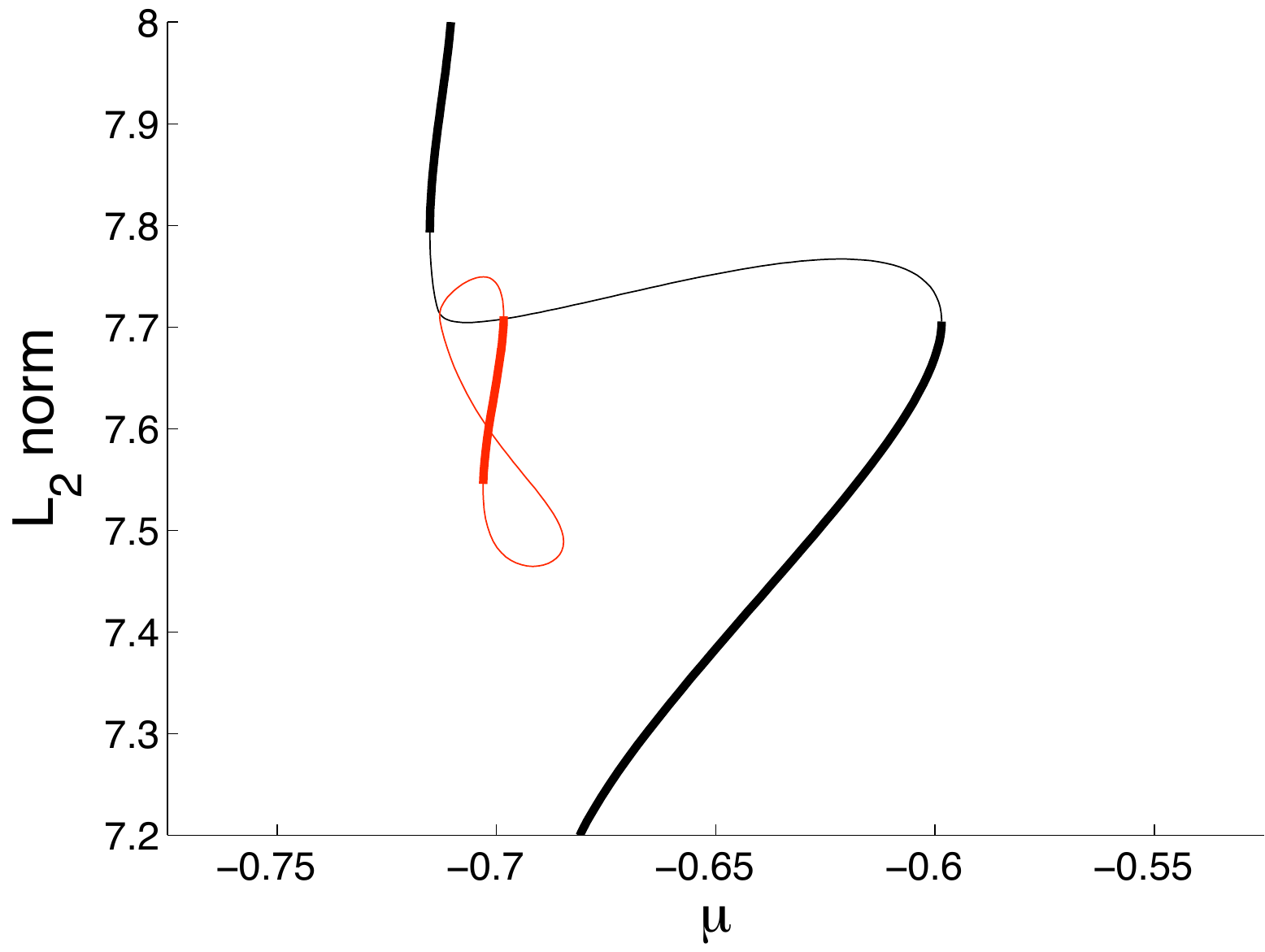}
	}
\caption{(a) The snaking curve (black) and an unconnected isola (red) at $C^+=0.19$, (b) topology change in the snaking curve as we increase the coupling strength, creating a switchback, (c) narrowing of the snaking curve as coupling strength is increased further, (d) eventual detachment of the switchback to form another isola. The labels \textbf{(c)}
and \textbf{(d)} in parts
(a) and (b) correspond to the saddle-nodes labelled in
figure~\ref{fig:2dsnake}(b) and the solution plots shown, respectively,
in figures~\ref{fig:bond-centred}(c) and~\ref{fig:bond-centred}(d).}
\label{fig:switchback}
\end{figure}

\begin{figure}
\centering
\includegraphics[width=0.49\textwidth]{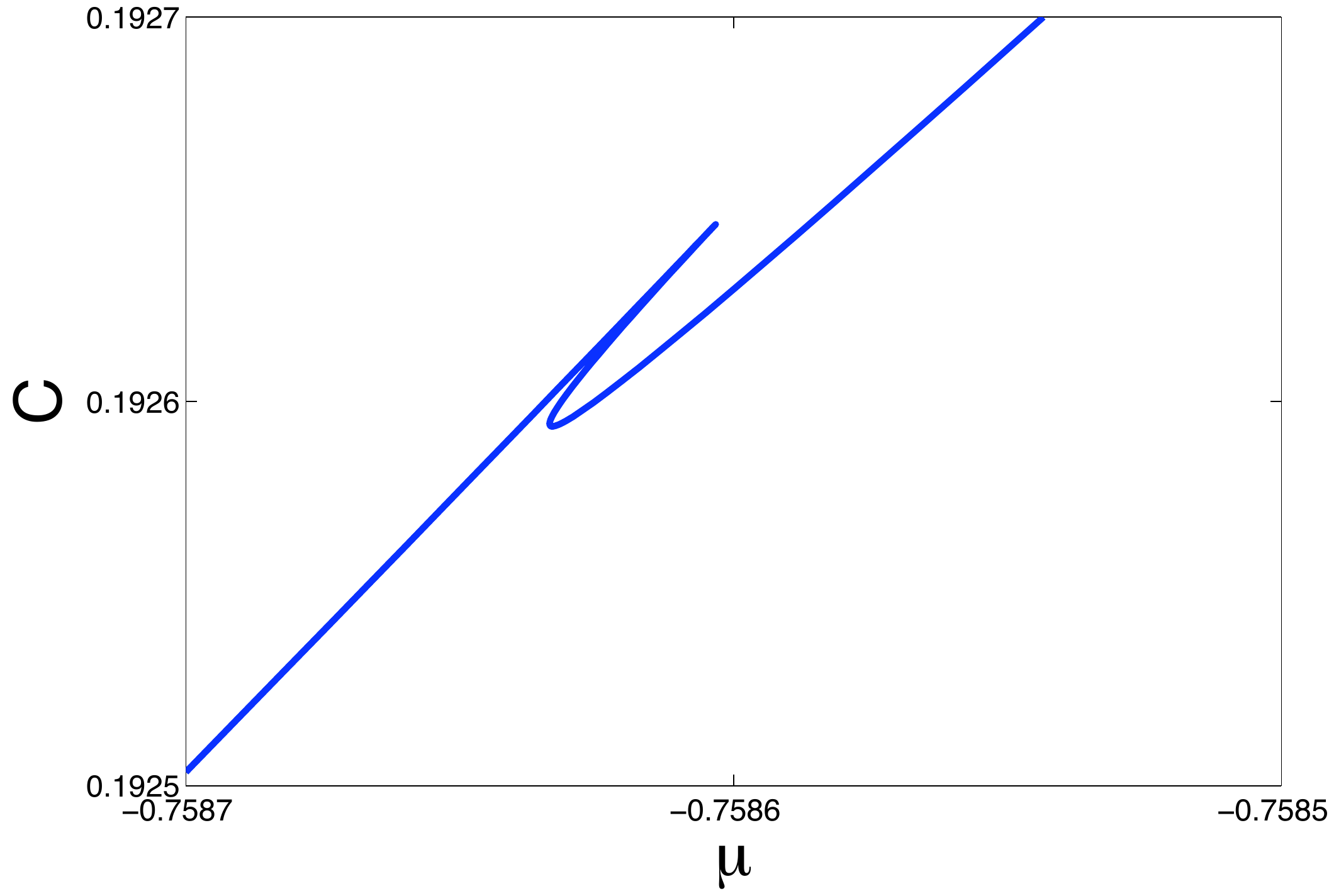}
\includegraphics[width=0.49\textwidth]{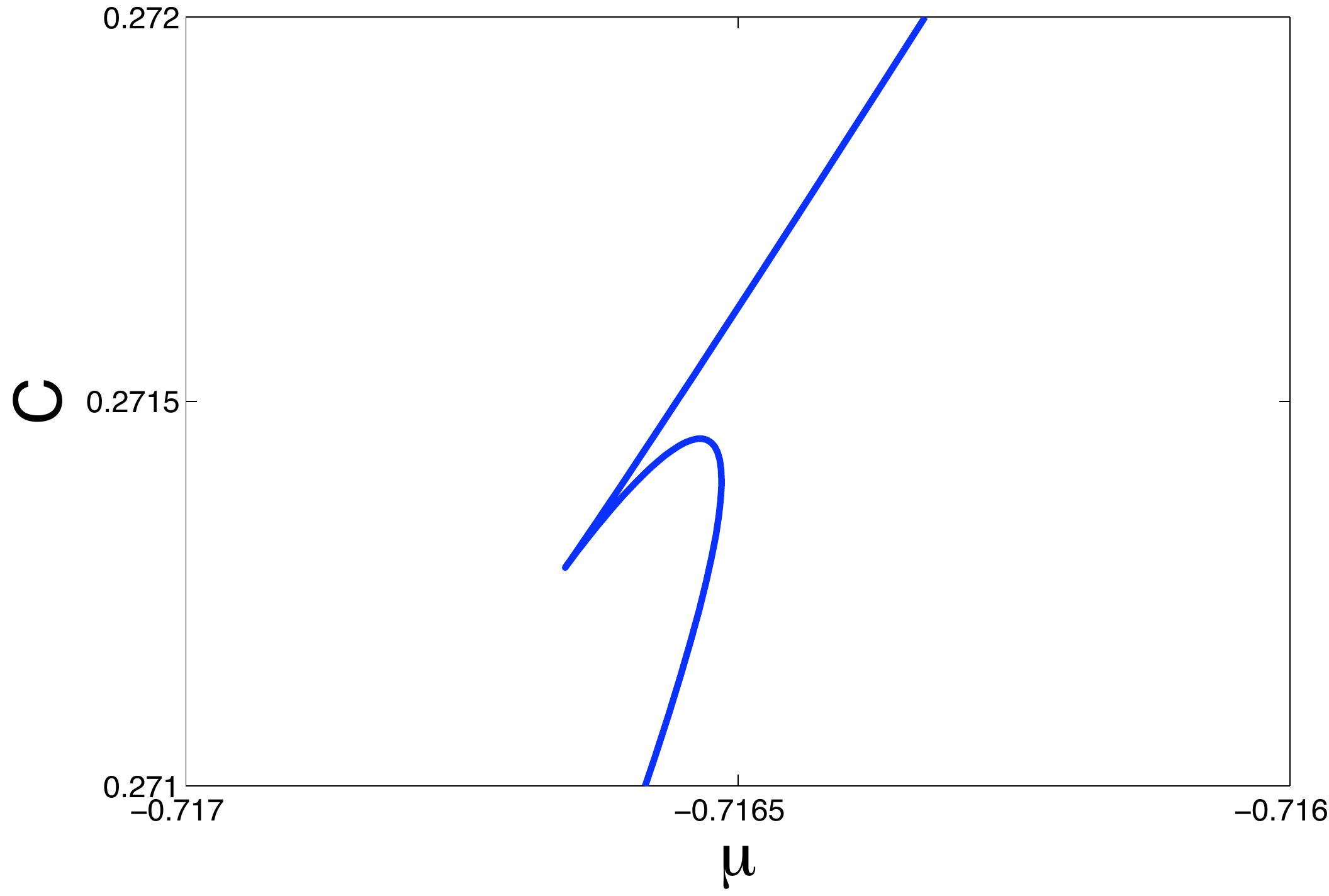}
\caption{The process of isola attachment and detachment in the $(\mu,C)$ plane by following limit points: on the left an isola attaches itself to the snake as $C^+$ is increased, briefly creating a situation with three limit points, two of which disappear in a cusp singularity [corresponding to Fig.~\ref{fig:switchback}(a)--(b)], and on the right the same process in reverse, as an isola detaches at a higher value of $C^+$ [corresponding to Fig.~\ref{fig:switchback}(c)--(d)].}
\label{fig:detachment}
\end{figure}

We find that the appearance and disappearance of
switchbacks as the coupling
strength $C^+$ is varied can be explained by collisions between
solution curves in the bifurcation diagram. When $C^+$ is
small it is important to note that
there are many localised solutions to~\eqref{eq:2d} possessing
square symmetry which do not
appear on the snaking curve, e.g. as shown in
figure~\ref{fig:2dsnake}(a). These additional localised states
exist on isolas in the bifurcation diagram. As $C^+$ is increased
these isolas collide with the snaking branch and form a switchback.
Assuming, as is reasonable at low coupling strength, that the isola
contained a stable solution, the snaking curve will now contain
(at least) one extra stable solution, by which we mean a
localised state now present, but not present on the snake
at lower $C^+$. As $C^+$ is increased still further the snake
narrows. Intriguingly, and unexpectedly,
the switchback may detach itself from the snake as $C^+$
is increased further,
re-forming a disconnected isola.
Figure~\ref{fig:switchback} illustrates
these events with a sequence of bifurcation
diagrams in the $(\mu,L_2)$ plane.
The processes of attachment and detachment are facilitated
by cusp singularities in which pairs of saddle-nodes on the snake collide and
disappear as the coupling strength increases. Figure~\ref{fig:detachment}
shows the cusp bifurcations in the $(\mu,C^+)$ plane
corresponding to the attachment and detachment
of the isola near $M^2=36$ shown in figure~\ref{fig:2dsnakecloseups}(a) and
figure~\ref{fig:switchback}.

In summary these isolas provide a mechanism by which additional
turns can be added to the snake, and they explain
the unintuitive switchback episodes of the snaking curve as the
localised states increase in size. This process of isola
attachment and detachment typically occurs over a remarkably small
range of $C^+$, indicated by the vertical scales in
figure~\ref{fig:detachment}. At larger $L_2$ norms the complete
bifurcation diagram in the $(\mu,C^+)$ plane is
complicated since many isolas attach and detach from
the snake at nearby values of $C^+$.

\section{Summary} \label{sec:conclusions}

In this Letter we have discussed the bifurcation structure of a model
lattice equation, the discrete Allen-Cahn equation with cubic and
quintic nonlinearities, in one and two dimensions. This
equation is of interest in its own right as
a model for spatially discrete pattern formation as well as
being closely related to coupled nonlinear Schr\"odinger equations
which have received substantial recent attention in nonlinear optics.
A subcritical instability of the trivial state, and the resulting
bistability, leads to the creation of localised states which are 
stable over an open interval of the parameter $\mu$, stabilized
by being pinned to the lattice. As remarked on by previous
authors, these localised states
arrange themselves into homoclinic snaking curves. In a finite
domain we demonstrated the existence of a maximum coupling
strength above which there is no modulational secondary instability
which would allow creation of localised states: in systems with
a sufficiently large coupling strength localised states do not appear.

In 1D we showed that breaking discrete translational
symmetry, for example by using Neumann boundary conditions,
allows one of the snakes to persist, but causes the other to
fragment into a sequence of isolas
and bubbles (`S'--shaped curves bifurcating from the persistent snake).
This contrasts with the situation in 2D where isolas
exist even when discrete translational 
symmetry is not broken. These isolas attach and detach 
themselves from the primary snaking curve as the linear 
coupling strength is varied, creating kinks in the snaking curve, 
where the solution norm decreases along a part of the curve before
beginning to increase again, that we
refer to as `switchbacks'. 

Numerical investigations suggest quantitative relations between the
widths of the snaking
curves in 1D and 2D. For wide localised states in 2D the
values of $\mu$ that limit the region over which localised states
exist correspond to the limiting values for fronts aligned
with lattice directions or on the diagonal. This is the natural
interpretation and agrees with the observations and computations
by Lloyd et al. \cite{LSAC08} for localised patches of hexagons
in the 2D Swift--Hohenberg equation.
This correspondence between 1D and 2D is asymptotically exact
in the limit of small coupling coefficients: higher-order corrections
arise at non-zero coupling and from the 
corners of the wide localised state where the fronts meet.

Our results are closely related to the work of 
Carretero-Gonz{\`a}les and Chong et al.~\cite{CG06,CC09} on the 
discrete nonlinear Schr{\"o}dinger equation with cubic and quintic 
nonlinearities. They discussed the time-independent Allen-Cahn 
equation from a spatial dynamics viewpoint, explaining the 
one-dimensional snaking curve as arising from intersections 
of the stable and unstable manifolds of a hyperbolic fixed 
point in a two-dimensional map. They also explored a 
variational approach, minimizing an effective Lagrangian 
to give approximate localised states in one and two dimensions. 

Our work considers additional mechanisms that complicate the usual
snaking picture by causing isolas to appear: these include
the effect of boundary conditions in 1D and exploring
larger localised states, higher up
the snaking curve in 2D where new effects, including the
switchbacks, arise.

\end{document}